\documentclass[preprint,review,3p,times,sort&compress]{elsarticle}

\usepackage{graphics}
\usepackage{graphicx}
\usepackage{subfigure}
\usepackage[numbers]{natbib}

\usepackage{CJKutf8}
\usepackage[utf8]{inputenc}

\usepackage{eqnarray,amsmath}

\usepackage{psfrag}
\usepackage{amssymb}
\usepackage{amsmath}
\usepackage{mathrsfs}

\usepackage{amsthm}
\usepackage{xcolor}
\usepackage{rotating}
\usepackage{lscape}
\usepackage{pdflscape}
\usepackage{booktabs}
\usepackage{multirow}
\usepackage{hyperref}

\usepackage{lineno}
\usepackage{float}
\usepackage{epstopdf}
\usepackage{changes}
\usepackage{lipsum}

\colorlet{Changes@Color}{red}

\usepackage{hyperref}

\usepackage{ragged2e}

\usepackage{amsmath}

\usepackage{lineno,hyperref}
\usepackage{booktabs}
\usepackage{flushend,cuted}
\usepackage{lscape}
\usepackage{float}
\usepackage{fixltx2e}
\usepackage{multirow}
\usepackage{subfigure}
\usepackage{multicol}
\usepackage{graphicx}
\usepackage[singlelinecheck=false]{caption}
\usepackage[font={normalsize}]{caption}
\usepackage{mathrsfs}
\usepackage{epsfig}
\usepackage{epstopdf}
\usepackage{adjustbox}

\usepackage{multirow}
\usepackage{amssymb}

\usepackage{amsmath}
\usepackage{mathrsfs}

\usepackage{array}

\usepackage{ulem}
\usepackage{xcolor}
\usepackage{cancel}
\usepackage{centernot}

\usepackage{changes}
\usepackage{lipsum}

\newcommand{\xdeleted}[1]{}

\colorlet{Changes@Color}{red}
\begin{document}

\begin{frontmatter}



\title{On the flame transfer function models for laminar premixed conical and V- flames considering the stretch effect}


\cortext[cor1]{Corresponding authors.}

\author[af1]{Yu Tian}
\ead{y\_tian@buaa.edu.cn}
\author[af1,af2]{Lijun Yang}
\ead{yanglijun@buaa.edu.cn}
\author[af3]{Aimee S.~Morgans}
\ead{a.morgans@imperial.ac.uk}
\author[af1,af2]{Jingxuan Li\corref{cor1}}
\ead{jingxuanli@buaa.edu.cn}

\address[af1]{School of Astronautics, Beihang University, Beijing 100191, China.}
\address[af2]{Aircraft and Propulsion Laboratory, Ningbo Institute of Technology, Beihang University, Ningbo, 315100, China.}
\address[af3]{Department of Mechanical Engineering, Imperial College London, London, SW7 2AZ, UK.}

\begin{abstract}

This paper investigates a predictive model that considers the impact of stretch on the dynamic response of laminar premixed conical and V- flames; the flame stretch consists of two components: the flame curvature and flow strain. 
The steady  and perturbed flame fronts are determined  via the linearized $G$-equation associated with the flame stretch model.
Parameter analyses of the effects of  Markstein length $\mathcal{L}$, flame radius $R$ and unstretched flame aspect ratio $\beta$ are also  conducted. 
Results show that the flame stretch reduces the steady flame height, with this effect being more significant for larger  Markstein lengths and smaller flame sizes. 
The effects of flame stretch on perturbed flames are evaluated by comparing the flame transfer function (FTF) considering the flame stretch and not. 
For flames of different sizes, the impact of flame stretch on FTF gain can be divided into three regions.
When both $\beta$ and $R$ are relatively small, due to the decrease in steady flame height and the impact of flow strain, the FTF gain increases. 
As $\beta$ and $R$ gradually increase, the FTF gain of the conical flame oscillates periodically while the FTF gain of the V-flame decreases, primarily due to the flame curvature enhancing the flame front disturbance and the wrinkle counteracting effect.
When $\beta$ and $R$ are large, a disruption in wrinkle counteracting effect ensues, leading to a significant increase in FTF gain.
Furthermore, as the actual flame height is reduced, the flame stretch  also  reduces  the FTF phase lag which is related to the disturbance propagation time from the flame root to the tip.

\end{abstract}


\end{frontmatter}

\section{Introduction}
\label{sec-intro}
%
Combustion instability is one of the most significant problems in the development of gas turbines, aero-engines and aerospace engines \citep{Candel_IJA_2009,Lieuwen_2012,Candel_EJMBF_2013}.
It occurs mainly due to the coupling between unsteady heat release and various acoustic perturbations within the combustion chambers \citep{Richards_ASME_1998,Candel_PCI_2002,Paschereit_AIAA_2001,Poinsot_Book_2005}. 
It may cause destructive damage to the engine structure or even failure.
Therefore, it is desirable to reveal the complex mechanisms behind these instabilities. 
When the combustion instability occurs, the flame is subjected to oncoming flow or acoustic disturbances, and flame wrinkles may occur and propagate along the flame, which further leads to perturbations in the heat release rate \citep{Boyer_CNF_1990,Durox_PCI_2009,Acharya_Lieuwen_JPP_2013}. 
The response of heat release rate to the oncoming flow perturbation thus constitutes a transfer relation, generally known as the flame transfer function (FTF) for  linear situations or flame describing function (FDF) for weakly nonlinear situations \citep{Dowling_JFM_1999,Schuller_JFM_2020,Polifke_PECS_2020};   these functions can be coupled in the low order models for the prediction and analysis of combustion instabilities based on the assumption that the flame can be considered compact compared to the wavelength of the dominant acoustic waves \citep{Dowling_JPP_2003,Poinsot_Book_2005,Schuermans_JEGTP_2010,Schuller_JFM_2020}.

The FTF or FDF of  premixed flames  typically can be obtained from experiments \citep{Balachandran_CNF_2005,Noiray_JFM_2008,Durox_PCI_2009,Schuller_CNF_2022}, in which the flow velocity perturbation upstream (typically immediately upstream) the flame is measured by the hot-wire anemometry \citep{Wang_CNF_2021}, multiple microphone methods (MMM) \citep{AEsoy_CNF_2022} or laser Doppler anemometry (LDA) \citep{Durox_PCI_2009}, and the heat release rate perturbation is mostly estimated by the chemiluminescence methods \citep{Schuller_PCI_2002,Li_CNF_2015_me}. 
The numerical simulation, e.g., the large eddy simulation (LES),  can also be applied to simulate the dynamic response of the flame and further the FTF or FDF by imposing a harmonic axial velocity perturbation immediately upstream of the flame \citep{Krediet_PCI_2013,Han_CNF_2015,Han_CNF_2015b,Li_CNF_2017}.  

An alternative approach is to theoretically model the flame response  \citep{Crocco_JARS_1951,Dowling_JFM_1999,Schuller_CNF_2003}, making it possible to  access  the sensitivities of FTF or FDF to different operating conditions or other factors \citep{Schuller_CNF_2003,Schuller_JFM_2020,Lim_IJHE_2021}, to apply these models  to the scheme for the controller design \citep{Dowling_ARFM_2005}  and to reveal the mechanisms of  nonlinear behaviors  of combustion instabilities \citep{Dowling_JFM_1997,Dowling_JFM_1999,Kashinath_Juniper_CNF_2013,Li_JSV_2015}. 
In cases of laminar premixed flames, or even weak turbulent premixed flames, the flame front can be treated as infinitely thin and the heat release rate is approximately proportional to the flame front surface area \citep{Schuller_CNF_2003,Palies_Schuller_PCI_2011}, which makes it possible to reconstruct the normalized heat release rate perturbations by the normalized flame front surface area perturbations \citep{Dowling_JFM_1999,Schuller_CNF_2003}. 
The instantaneous position of an oscillating flame front subjected to oncoming flow perturbations can be captured using the $G$-equation model (or the level-set approach) proposed by Markstein \cite{Markstein_1964}. 
This model typically  ignores the flame inner  structure and the feedback from the dynamic flame to the flow field ahead of the flame \citep{Fleifil_CNF_1996,Schuller_CNF_2003}.

The oncoming flow velocity depends on the geometry of the burner and combustor operating conditions, and other factors,  leading to it being difficult to work out an  appropriate model for all burners and flames. 
A  great many experimental \citep{Durox_PCI_2005,Leitgeb_Schuller_CNF_2013,Durox_PCI_2009,Mejia_CNF_2018} and numerical simulation studies \citep{Preetham_AIAA_2008,Kashinath_Juniper_CNF_2013,Dupuy_Schuller_CNF_2020} have  been conducted  for various representative flames,  such as conical, V-, M- and slit flames,  under various conditions \citep{Schuller_CNF_2003_b,Durox_PCI_2005,Palies_Schuller_PCI_2011,Blanchard_Schuller_POF_2015}, ranging from a single flame to  multiple flames configurations \citep{Durox_PCI_2009}. 
Based on the experimental or numerical results of the flow velocity field ahead of the flame front \citep{Schuller_PCI_2002,Schlimpert_CNF_2015}, the three most often used models of the flow velocity perturbations are proposed \citep{Schuller_AIAA_2012}, typically named
(1) the uniform velocity perturbation model which assumes that the disturbance convection velocity is sufficiently large that the disturbance is the only function of time \citep{Fleifil_CNF_1996,Ducruix_PCI_2000,Lieuwen_JPP_2003}; 
(2) the convective velocity perturbation model which assumes that the convection velocity of the flow disturbance equals the bulk flow velocity \citep{Baillot_CNF_1992,Preetham_AIAA_2004};    
(3) the incompressible velocity perturbation model which takes into account the radial velocity by solving the continuity equation \citep{Baillot_CNF_1992,Preetham_JPP_2008,Cuquel_MCS7_2011}.
Experimental measurements of the dynamic flow field ahead of the flame front showed that the convection velocity of the oncoming perturbations varies with frequencies \citep{Schuller_PCI_2002,Birbaud_CNF_2006,Schlimpert_CNF_2015};
a  convection ratio of the perturbation convection velocity to the bulk time-averaged flow  is thus proposed to account for this difference and frequency dependence in the convective velocity perturbation model   \citep{Birbaud_CNF_2006,Preetham_JPP_2008}.

By imposing expressions for the time-averaged flow and oncoming flow perturbations  into the  $G$-equation, 
it is possible to numerically or analytically resolve the solutions of the flame front and further the FTF or FDF for  premixed  flames, e.g., as that being firstly derived by  \citep{Fleifil_CNF_1996}. 
More oncoming flow velocity models are accounted for and these methods are extended to more kinds of flames  \citep{Dowling_JFM_1999,Ducruix_PCI_2000,Schuller_CNF_2003}, e.g., the V-flame, and are validated by comparing these predicted FTFs or FDFs to those measured experimentally \citep{Schuller_CNF_2003}. 
Lieuwen analyzed the nonlinear dynamics of the premixed conical and V- flames based on the $G$-equation model, and results showed that the nonlinear intensity increases with increasing the perturbation frequency \citep{Lieuwen_PCI_2005}. 
The analytical solutions of the FDF for these two kinds of flames are further derived based on a perturbation method \citep{Preetham_JPP_2008}. Palies et al. \citep{Palies_Schuller_PCI_2011} constructed an analytical model that quantified the linear response of swirling flames submitted to velocity disturbances. 
The $G$-equation model has also been applied to derive the analytical solutions for transversally forced flames \citep{O'Connor_PECS_2015,Acharya_CNF_2012,Smith_CNF_2018,Acharya_JFM_2020}. The FTFs for the laminar premixed two-dimensional slit flame were subjected to simultaneous longitudinal and transverse forced harmonious acoustic disturbances and both transverse and longitudinal time-averaged flow were derived \citep{LiC_CTM_2017,LiC_PCI_2019}. 
These works were further extended to the FDFs for the slit flames \citep{Liu_Fuel_2020,Liu_Fuel_2021}. 

The above research is typically based on the  fact that the flame heat release rate is only related to surface area fluctuations.
However, due to the influence of flow strain and thermal diffusion, the flame speed is actually affected by the flow velocity gradient, leading to the flame stretch rate being different at each position, and this  further affects the entire  heat release rate \citep{Law_PECS_2000}. 
The concept of flame stretch has been widely used in various studies.
Candel's group extended Clavin and Matalon's theories \citep{Clavin_JFM_1982,Matalon_CST_1983} and derived the expression of flame stretch rate \citep{Darabiha_Candel_CNF_1986,Candel_CST_1990}. 
Heat-loss instability was studied from the perspective of thermal diffusion \citep{Joulin_CNF_1979}. 
The relation between  the stretch and flame thickness was constructed \citep{Chung_Law_CNF_1984,Chung_Law_CNF_1988,Goey_CNF_1997,Goey_PCI_2011}. The stretch was also used in the study of turbulent combustion vortices \citep{Somappa_Lieuwen_PCI_2019}. 
In theoretical derivation, the effect of stretch rate upon the flame speed is typically indicated by the Markstein length \citep{Tseng_CNF_1993,Bradley_CNF_1996}, which varies with the equivalent ratio and the Lewis number of premixture \citep{Matalon_CST_1983,Bechtold_Matalon_CNF_2001}. 
The corresponding relation between Markstein length and equivalent ratio of several commonly used hydrocarbon fuels has been studied through experiments \citep{Tang_IJHE_2008}, in which Markstein number of hydrogen and methane increases with the increase of equivalent ratio, while for most other fuels such as propane, the two are negatively correlated.
The Lewis number ($Le$) is a dimensionless number that characterizes the ratio of mass to heat diffusivities in a reacting mixture. When $Le = 1$, the two diffusion coefficients are the same, and the energy balance of the flame is not affected by these diffusion fluxes. However, when $Le \neq 1$, the diffusion fluxes affect the change in energy of the flame itself. For stretched flames, this affects the temperature distribution and flame shape, subsequently impacting the Markstein length \citep{Clavin_PECS_1985,Williams_PECS_2000,Peters_2000,Lieuwen_2012}.
In the $G$-equation model, flame stretch affects both the shape and the FTF of the flame. 
In the experiments, it has been observed that at low disturbance frequencies and amplitudes, the flame front wrinkles with a constant amplitude from the flame root to its tip. 
However, at higher frequencies, a phenomenon referred to as ``filtering'' can be observed, where the flame wrinkling is prominent only at the flame root and decays at downstream flow locations. 
Research by Wang et al. \citep{Wang_CNF_2009} indicated that considering the flame stretch in theoretical models provided a better prediction of this noticeable phenomenon of flame front root wrinkling observed in experiments.
It was shown that the observed decrease in flame wrinkles could be explained by the effect of stretch, and the influencing factors were the Markstein length, flame height and flame aspect ratio \citep{Preetham_AIAA_2006,Preetham_JPP_2010}. 
Further theoretical research showed that the stretch can play a significant role in the FTF  when the perturbation frequency exceeds  a certain value \citep{Wang_CNF_2009}. 
The relation  between the flame wrinkle and stretch was analyzed  from a theoretical perspective~\citep{Shin_Lieuwen_CNF_2012}. 
The FTF of a laminar premixed conical flame only considers the effect of curvature was derived  \citep{Orchini_CNF_2016}. 
In summary, the above studies have shown that considering the flame stretch is of large significance for establishing flame front dynamics models and correcting the theoretical solution of FTF.

It is worth mentioning that the perforated-plate flame has been extensively studied because of its  compact acoustic characteristics and the large cut-off frequency of the FDF \citep{Noiray_JFM_2008,Boudy_Schuller_PCI_2011,Kornilov_PCI_2009,Duchaine_CNF_2011}.
Systematic research has also been carried out in multinozzle configurations to investigate their flame dynamic responses and self-excited combustion instabilities\citep{Lee_CNF_2020,Kang_CNF_2021,Moon_CNF_2022}.
This kind of flame can be regarded as a collection of small size flames, and the radius of each flame is about $1$-$2$ mm \citep{Noiray_JFM_2008,Kornilov_PCI_2009,Duchaine_CNF_2011}, which is equivalent to the Markstein length of typical flames. 
Hence, for flames stabilized downstream of perforated plates, the effect of flame stretch may be of significant importance \citep{Kornilov_CNF_2009,Altay_PCI_2009}.
The gain of the FTF or FDF  exceeds unity in a  specific frequency range \citep{Durox_PCI_2009,Kornilov_PCI_2009}, rendering the existing theoretical models for conical or V- flames that neglect flame stretch inadequate in accurately predicting the flame front dynamic response and FTF for small size flames.
The experimental results further demonstrate that the FTF of an individual small size flame also features similarities with the perforated plate flame.
Furthermore, it was also observed that for flames with a constant cone angle at the flame tip, small size flames exhibit a lower aspect ratio, indicating that the effect of flame stretch on the flame shape, particularly at the flame tip, is more pronounced \citep{Gaudron_CNF_2017}.
It is thus worthy of deriving the analytical solutions of FTFs for the represented flame and evaluating the effect of the flame stretch on the FTFs for different sizes of flames.
This may be a supplementary theory for the FTF model for the  perforated-plate flames.

\begin{figure}[!t]
	\centering
		\subfigure
  		{
\includegraphics[height=4cm]{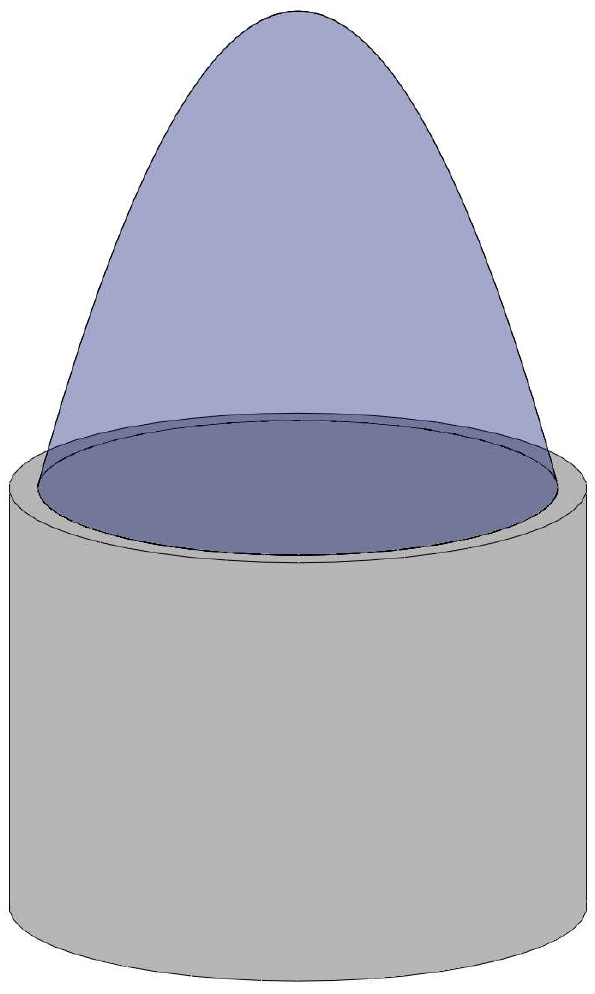}
  		\label{Fig:shape_conical}
  		}\put (-90,10) {\normalsize  $\displaystyle(a)$} 
  		\vspace*{0pt}
  		\hspace*{20pt}	
  		\subfigure
  		{
\includegraphics[height=4.3cm]{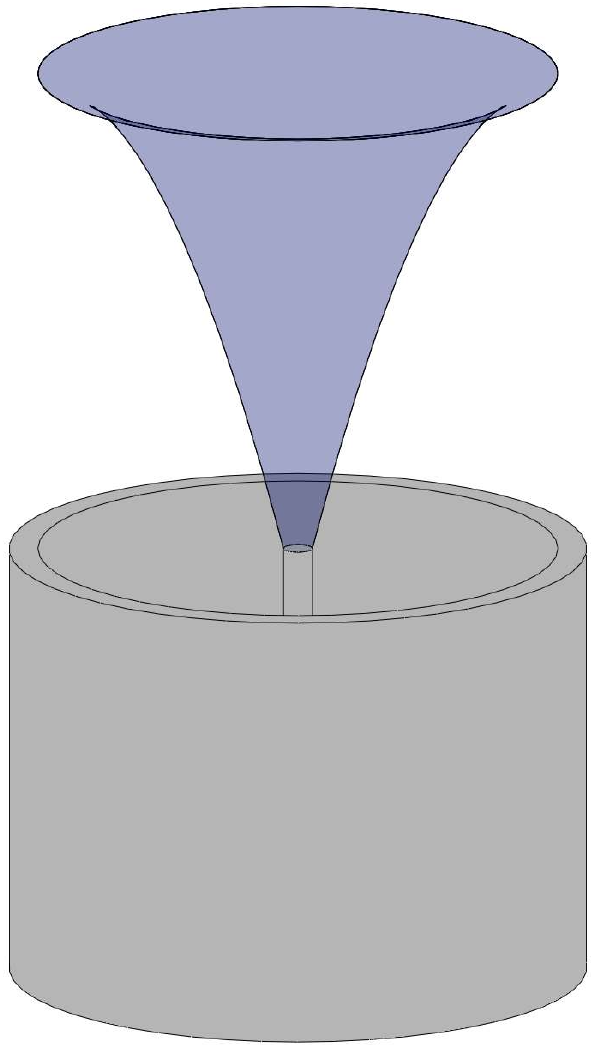}
  		\label{Fig:shape_V}
  		}\put (-90,10) {\normalsize  $\displaystyle(b)$} \\
  		\vspace*{0pt}
  		\hspace*{20pt}	
 	 \caption{Sketches of two types of envisaged laminar premixed flames: (a) conical flame, (b) V-flame.}
	 \label{Fig:Shape}
\end{figure}

The objective of this study is to get the impact of flame stretch effects, comprising flame curvature and flow strain, on the FTF and flame shape for various laminar premixed flames, and to establish a predictive model for flame dynamic response that accounts for flame stretch.
Two kinds of flames are considered: the conical flame and V-flame, as sketched in Fig.~\ref{Fig:Shape}. 
This paper is organized as follows.  
In Section \ref{sec:flame-dynamics}, the models and methods for the FTF of two kinds of flames considering the flame stretch effect are introduced. In Section \ref{sec:flame-FTF}, the FTFs of stretched and unstretched flames are summarized. 
The steady and perturbed results of conical and V- flames are presented and discussed respectively in Section \ref{sec:results}. 
The effects of Markstein length for steady-state flame, flow velocity perturbation model and flame size for FTFs are considered for each flame. 
Conclusions are drawn in the final section. 

\section{Dynamics models of stretched flames}
\label{sec:flame-dynamics}


The flame dynamics model in this study assumes that the flame diffusion time is considered much smaller than the acoustic characteristic time and 
flow thermal expansion and thermo-diffusive effects are ignored. 
The flame radius is much larger than the thickness of the flame front where the chemical reaction occurs, leading to that the flame front (corresponding to the contour $\widetilde{G} = 0$) can be considered as an infinitely thin interface separating the fresh reacting flow (denoted by $\widetilde{G} < 0$) from the burned gases ($\widetilde{G}>0$).
It should be noted that the superscript symbol $\widetilde{()}$ denotes dimensional parameters.
Under the aforementioned assumptions, the flame front is assumed to propagate normally to itself with the velocity $\mathbf{\tilde{u}}\cdot {\mathbf{n}} + \tilde{s}_L$,  where $\mathbf{\tilde{u}}$ is the fresh reacting flow velocity vector,  
$\tilde{s}_L$ is the flame speed and ${\mathbf{n}} = -  {\widetilde{\mathbf{\nabla}} \widetilde{G} }/{| \widetilde{\mathbf{\nabla}} \widetilde{G} |}$ is the normal vector.
Therefore, the $G$-equation describing the flame front location ($\widetilde{G} = 0$) can be expressed as \citep{Schuller_CNF_2003}: 
\begin{equation}\label{Eq:G-equation}
\frac{\partial \widetilde{G}}{\partial \tilde{t}} + \mathbf{\tilde{u}}\cdot \widetilde{\mathbf{\nabla}} \widetilde{G} = \tilde{s}_L \left| \widetilde{\mathbf{\nabla}} \widetilde{G} \right|
\end{equation}

\begin{figure}[h]
	\centering
		\subfigure
  		{
\includegraphics[height=5cm]{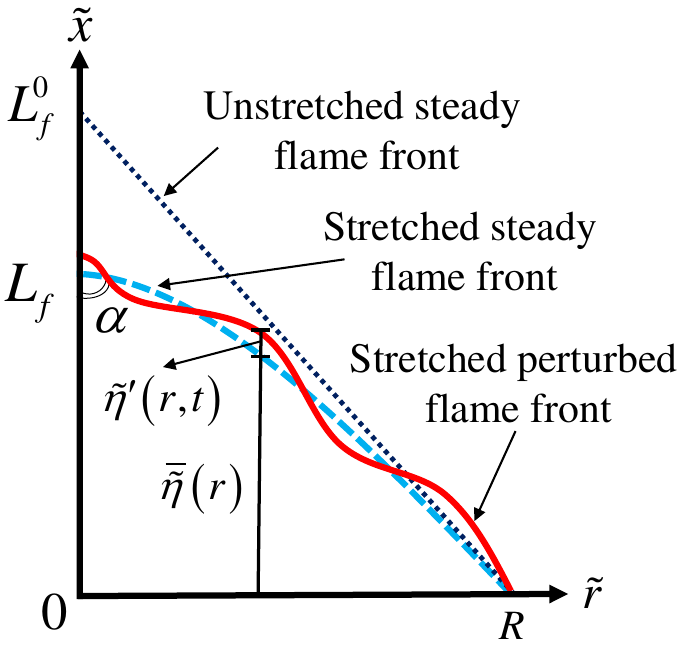}
  		\label{Fig:cal_conical}
  		}
  		\vspace*{0pt}
  		\hspace*{10pt}	
 	 \caption{Flame geometry used in the determination of dynamic flame front  and FTF. $R$: the radial distance from the tip to the root of the flame, referred to the flame radius, $L_{f}^0$: unstretched steady flame height, $L_{f}$: stretched steady flame height, $\alpha$: inclined angle of the local steady flame front.}
	 \label{Fig:cal}
	 \vspace*{00pt}
\end{figure}

In this study, the two types of flames under investigation, namely the conical and V- flames, are both axially symmetric. This property allows for the simplification of the actual three-dimensional flames into a two-dimensional $\tilde{r}-\tilde{x}$ coordinate system, where $\tilde{r}$ corresponds to the radial direction and $\tilde{x}$ indicates the axial direction  as sketched in Fig.~\ref{Fig:cal}. 
Similarly, the oncoming velocity $\mathbf{\tilde{u}}$ can also be divided into two directions, where $\tilde{{u}}_{x}$ and $\tilde{{u}}_{r}$ denote axial and radial flow velocity ahead of the flame front, respectively. 
For the sake of simplicity, it is also possible to express the flame front as $\widetilde{G} = \tilde{x} - \widetilde{\eta}(\tilde{r},\tilde{t}) = 0$, where $\widetilde\eta$ represents the axial position of the flame front at different times and radii. 
Then the $G$-equation can be changed to: 
\begin{equation}\label{Eq:G}
-\frac{\partial {\widetilde{\eta}}}{\partial \tilde{t}}+\tilde{{{u}}}_{x}-\tilde{{{u}}}_{r}\frac{\partial \widetilde{\eta}}{\partial \tilde{r}}
	=\tilde{s}_L \left(1+\left(\frac{\partial \widetilde{\eta}}{\partial \tilde{r}}\right)^2 \right)^{\frac{1}{2}}
\end{equation}

In Eq.~\eqref{Eq:G}, the flame speed $\tilde{s}_L$, taking into account the effect of flame stretch, can be expressed as:
\begin{equation}\label{Eq:Sl}
\tilde{s}_{L}=
\tilde{s}_{L}^{0}\left(1-\widetilde{\mathcal{L}}\widetilde{\mathcal{K}}\right)-\widetilde{\mathcal{L}}\widetilde{\mathcal{S}}
\end{equation}
where
$\tilde{s}_{L}^{0}$ indicates the flame speed of an unstretched flame,  $\widetilde{\mathcal{L}}$  represents the Markstein length, $\widetilde{\mathcal{K}}={\widetilde{\mathbf{\nabla}}}\cdot {\mathbf{n}}$ denotes the flame front curvature and $\widetilde{\mathcal{S}}=-{\mathbf{n}}\cdot{\widetilde{\mathbf{\nabla}}} {\tilde{\mathbf{u}}} \cdot {\mathbf{n}}$ is the flow strain, and the combination of these two terms constitutes the flame stretch \citep{Candel_CST_1990,Matalon_JFM_1982}. 
%

It should be noted that the time-averaged values and perturbations are represented by $\overline{()}$ and $()^\prime$ in this paper respectively. 
Therefore, the oncoming flow velocity of the flame can also be written in the same form of ${\tilde{\mathbf{u}}} = {\bar{\tilde{\mathbf{u}}}} + {\tilde{\mathbf{u}}}^\prime$.
In this work, only the axial time-averaged flow ${\bar{\tilde{u}}}_x \equiv const.$  is taken into account and the radial time-averaged flow equals ${\bar{\tilde{u}}}_r \equiv 0$. 
Specifically, taking the Fourier transform of the perturbation axial flow velocity $\tilde{{u}}_x^\prime$, one can obtain its amplitude in the frequency domain, which corresponds to $\hat{\tilde{u}}_x$.
The three types of flow velocity perturbation models prescribed introduction section are all accounted for, and they have the expressions as follows:
\begin{enumerate}[{(1)}]
\item Uniform velocity perturbation model.
\begin{equation}\label{Eq:K_0}
\tilde{{u}}_x^\prime 
=\hat{\tilde{u}}_x \exp \left(-\mathrm{i}\omega \tilde{t}\right),
\quad 
\tilde{u}_r^\prime =0
\end{equation}

It is noted that only the linear situation is accounted for,  indicating that the flow perturbation amplitude $\hat{\tilde{u}}_x$ is much smaller than the time-averaged value $\bar{\tilde{u}}_x$.

\item convective velocity perturbation model. 
\begin{equation}\label{Eq:K_1S}
\tilde{u}_x^\prime 
=\hat{\tilde{u}}_x \exp \left(\mathrm{i} k\tilde{x}-\mathrm{i}\omega \tilde{t}\right),
\quad 
\tilde{u}_r^\prime =0
\end{equation}
Herein, $k=\omega/\tilde{{u}}_c$. $\tilde{{u}}_c$  is the propagation velocity of perturbation. Note that this model covers the situation that the convection velocity does not equal the time-averaged flow velocity \citep{Kashinath_Juniper_CNF_2013,Jiang_AIAA_2022}. 

\item  incompressible velocity perturbation model. 
\begin{equation}\label{Eq:K_1C}
\tilde{u}_x^\prime 
=\hat{\tilde{u}}_x \exp \left(\mathrm{i} k\tilde{x}-\mathrm{i}\omega \tilde{t}\right),
\quad 
\tilde{u}_r^\prime=-\frac{1}{2} \mathrm{i} k \tilde{r}_{f} \hat{\tilde{u}}_x \exp \left(\mathrm{i} k\tilde{x} - \mathrm{i}\omega \tilde{t}\right)
\end{equation}
where,   $\tilde{r}_{f}=\tilde{r}$ for conical flame, and $\tilde{r}_{f}=R-\tilde{r}$ for V-flame. In this model, the flow is assumed incompressible. The radial velocity is obtained by solving the continuity equation.
\end{enumerate}

One then introduces the non-dimensional parameters as follows:
\begin{align}
& \eta = \frac{\widetilde{\eta}}{R},
\quad && x = \frac{\tilde{x}}{R},
\quad  && r = \frac{\tilde{r}}{R}, 
\quad && \beta=\frac{L_{f}^0}{R}, 
&& \mathbf{u} = \frac{\tilde{\mathbf{u}}}{\bar{\tilde{u}}_x}, 
\\ \nonumber
& t = \frac{{\tilde{t} \bar{\tilde{u}}_x}}{R},
\quad &&\mathcal{L}=\frac{\widetilde{\mathcal{L}}}{R}, 
\quad && \mathcal{K}= \widetilde{\mathcal{K}}{R}, 
\quad && \mathcal{S}= \widetilde{\mathcal{S}} \frac{R}{\bar{\tilde{u}}_x}, 
&& s_L = \frac{\tilde{s}_L}{\bar{\tilde{u}}_x} 
\\ \nonumber
 & {s}_{L}^{0} = \frac{\tilde{s}_{L}^{0}}{\bar{\tilde{u}}_x} = \frac{1}{\sqrt{1+{{\beta }^{2}}}} , 
\quad && \gamma = \frac{\beta}{1+{\beta}^2} ,
\quad &&{\omega}_{\ast} = \frac{\omega R}{\bar{\tilde{u}}_x \gamma} , 
\quad && K =\frac{\bar{\tilde{u}}_x}{\tilde{{u}}_c} = \frac{k R }{{\omega}_{\ast} \gamma},  
\quad &&   \mathbf{\nabla} = \left( \frac{\partial }{\partial r}, \frac{\partial }{\partial x} \right)
\end{align}

Therefore, the dimensionless Eq.~\eqref{Eq:G} becomes:
\begin{equation}\label{Eq:G_end_non}
-\frac{\partial {{\eta}}}{\partial {t}}+{{{u}}}_{x}-{{{u}}}_{r}  \frac{\partial {\eta}}{\partial {r}}
	={s_L} \left(1+ \left(\frac{\partial {\eta}}{\partial {r}}\right)^2 \right)^{\frac{1}{2}}
\end{equation}

The three types of flow velocity perturbation models can be unified  as:
\begin{equation}\label{Eq:u_non}
{u}_x
=1+ \varepsilon \exp \left(\mathrm{i} {\gamma}{\omega}_{\ast} \left(Kx-t\right)\right) ,
\quad 
{u}_r=-\mathrm{i} B K {\gamma}{\omega}_{\ast}{r}_{f}\varepsilon \exp \left(\mathrm{i} {\gamma}{\omega}_{\ast} \left(Kx-t\right)\right)
\end{equation}
%
where, $\varepsilon \ll 1$ is the perturbation parameter, for the small perturbations and linear limit considered in this paper, it can be assumed that $\varepsilon = {\hat{\tilde{u}}_x}/{\bar{\tilde{u}}_x}$. The flow velocity perturbation model is determined by two coefficients, $B$ and $K$, while $B=0,K=0$ for uniform velocity perturbation model; $B=0,K=1$ for convective velocity perturbation model; and $B=\frac{1}{2},K=1$ for incompressible velocity perturbation model.

Furthermore, it is possible to assume  that all perturbations have a time dependence in the form $()^\prime = \varepsilon \hat{()}\exp \left(-\mathrm{i}{\gamma}{\omega}_{\ast} t\right)$ under the linear perturbation condition, where $\hat{()}$ denotes the Fourier component. That is:
\begin{equation}\label{Eq:eta_non}
\eta
=\bar{\eta}+ \varepsilon \hat{\eta}\exp \left(-\mathrm{i} {\gamma}{\omega}_{\ast} t\right) ,
\quad 
s_L
=\bar{s}_L+ \varepsilon {\hat{s}_L}\exp \left(-\mathrm{i} {\gamma}{\omega}_{\ast} t\right)
\end{equation}

The time-averaged properties of Eq.~\eqref{Eq:G_end_non} satisfy the relations:
\begin{equation}\label{Eq:G_zero}
\bar{s}_L = \sin\alpha
\end{equation}
where, $\sin\alpha = \left(1+\left({\mathrm{d} \bar{\eta}}/{\mathrm{d} r}\right)^2 \right)^{-\frac{1}{2}} $ is  defined as the geometrical inspection. 
It should be noted that, due to the stretch effect, the time-averaged flame speed $\bar{s}_L^{}$ relies on the position, the local flame inclined angle $\alpha(r)$ is thus also a function of $r$. 
When the stretch is not accounted for,  $\alpha(r)$ becomes a constant value $\alpha_0 = \sin^{-1}s_L^0$.    The flame height when the stretch is not considered equals $L_f^0 = R/\tan\alpha_0$. 

The first-order term of Eq.~\eqref{Eq:G_end_non}, which represents the perturbation term of the $G$-equation, can be expressed as:
\begin{equation}\label{Eq:G_line}
\mathrm{i}{\gamma}{\omega}_{\ast} \hat{\eta} +  \exp \left( \mathrm{i} {\gamma}{\omega}_{\ast} Kx\right) - \mathrm{i} B K {\gamma}{\omega}_{\ast}{r}_{f} \cot\alpha \exp  \left( \mathrm{i} {\gamma}{\omega}_{\ast} Kx\right) = \frac{\hat{s}_L}{\sin\alpha}- \bar{s}_L  \frac{\mathrm{d} \hat{\eta}}{\mathrm{d} r} \cos\alpha
\end{equation}
where, $\cot\alpha = -\mathrm{d}\bar{\eta}/\mathrm{d}r $. 
It should be noted that the above equation degenerates to that in \citep{Ducruix_PCI_2000} when $s_L$ remains constant.

Following previous studies \citep{Humphrey_IJSCD_2014,Steinbacher_CNF_2019}, the conical flame is anchored at the edge of the combustor, while the V-flame is anchored at the combustor center stabilizer.
Within the coordinate system defined in this study, the anchor points of both flames are located at $r = 1$ ($\tilde{r} = R$ shown in Fig.~\ref{Fig:cal}). Therefore, ${r}_{f}={r}$ for conical flame, and ${r}_{f}=1-{r}$ for V-flame in Eq.~\eqref{Eq:G_line}. For these two kinds of flame, the location $r = 1$ represents the edge and the center of the combustor, respectively.
The boundary condition for the flame anchor point is:
\begin{equation}\label{Eq:boundary1}
\bar{\eta}\left( r=1 \right)=0,
\hat{\eta}\left( r=1 \right)=0
\end{equation}

Additionally, both flames are axisymmetric, so the symmetry axis of the conical and V- flames are located at $r = 0$ and $r = 1$, respectively.
For the conical flame, $r = 0$ represents the flame tip, and it is necessary to ensure a smooth transition at this point.
Therefore, the boundary condition for the tip of the conical flame is:
\begin{equation}\label{Eq:boundary_conical}
\frac{\mathrm{d}\bar{\eta}_c}{\mathrm{d}r}\left( r=0 \right)=0,
\frac{\mathrm{d}\hat{\eta}_c}{\mathrm{d}r}\left( r=0 \right)=0
\end{equation}

While for the V-flame, $r = 0$ represents the flame tail. It is assumed that the flame edge is free to move only in the $x$-direction with perturbations and that all information cannot propagate into or out of the flame \citep{Wang_CNF_2009,Preetham_JPP_2010}. Therefore, the boundary condition for the edge of the V-flame is:
\begin{equation}\label{Eq:boundary_V}
\frac{\mathrm{d}^2\bar{\eta}_v}{\mathrm{d}r^2}\left( r=0 \right)=0,
\frac{\mathrm{d}^2\hat{\eta}_v}{\mathrm{d}r^2}\left( r=0 \right)=0
\end{equation}

In the following  analysis, the time-averaged and perturbed flame speed considering the flame stretch can be expanded as:
\begin{equation}\label{Eq:Sl_ave}
\bar{s}_{L}=
{s}_{L}^{0}\left(1-{\mathcal{L}}{\overline{\mathcal{K}}}\right) - \mathcal{L}{\overline{\mathcal{S}}}= \frac{1-{\mathcal{L}}{\overline{\mathcal{K}}}}{\sqrt{1+{{\beta }^{2}}}}- \mathcal{L}{\overline{\mathcal{S}}}
\end{equation}
\begin{equation}\label{Eq:Sl_per}
\hat{s}_{L}=
-{s}_{L}^{0}{\mathcal{L}}\hat{\mathcal{K}}-{\mathcal{L}}\hat{\mathcal{S}}
= -\frac{{\mathcal{L}}\hat{\mathcal{K}}}{\sqrt{1+{{\beta }^{2}}}}-{\mathcal{L}}\hat{\mathcal{S}}
\end{equation}

Under the perturbation, the normal vector  $\mathbf{n}$ becomes: 
\begin{equation}\label{Eq:n}
\mathbf{n}=-\frac{\nabla {G}}{\left| \nabla {G} \right|}=[{{n}}_{x},{{n}}_{r}]
\end{equation}
where
\begin{equation}
n_x = - \left(1+\left(\frac{\mathrm{d} {\eta}}{\mathrm{d} r}\right)^2 \right)^{-\frac{1}{2}}, \quad
n_r =  \frac{\mathrm{d}\eta}{\mathrm{d}r} \left(1+ \left(\frac{\mathrm{d} {\eta}}{\mathrm{d} r}\right)^2 \right)^{-\frac{1}{2}}
\end{equation}

The flame curvature and the flow strain can be expressed as:
\begin{equation}\label{Eq:K_c}
\begin{split}
{{{\mathcal{K}}}}=\mathbf{\nabla}\cdot \mathbf{n} &= \frac{\partial n_r}{\partial r} + \frac{n _r}{r}  + \frac{\partial n_x}{\partial x} \\
\end{split}
\end{equation}
\begin{equation}\label{Eq:S_c}
\mathcal{S}= -\mathbf{n}\cdot\nabla \mathbf{u}\cdot \mathbf{n} = -n_x^2 \frac{\partial u_x}{\partial x} -n_x n_r \left(\frac{\partial u_x}{\partial r}+\frac{\partial u_r}{\partial x}\right) -n_r^2 \frac{\partial u_r}{\partial r} + \frac{1}{r}\frac{\partial \left(r u_r\right)}{\partial r} + \frac{\partial u_x}{\partial x}
\end{equation}

The full expressions of the time-averaged and perturbed flame curvatures for the conical and V-flames are summarized as:
\begin{equation}\label{Eq:K_ave}
\overline{\mathcal{K}}=
\frac{\mathrm{d}^2\bar{\eta}}{\mathrm{d}r^2} \sin^3 \alpha - \frac{1}{r}\cos \alpha
\end{equation}
\begin{equation}\label{Eq:K_per}
\hat{\mathcal{K}}=
\frac{\mathrm{d}^2\hat{\eta}}{\mathrm{d}r^2} \sin^3\alpha
+ 3 \frac{\mathrm{d}^2\bar{\eta}}{\mathrm{d}r^2}  \frac{\mathrm{d} {\hat{\eta}}}{\mathrm{d} r}  \cos\alpha \sin^4\alpha + \frac{1}{r}    \frac{\mathrm{d} \hat{\eta}}{\mathrm{d} r} \sin^3\alpha
\end{equation}

It is worth noting that for the V-flame at the boundary, i.e., when $r = 0$, in order to ensure the solvability of the Eqs.~\eqref{Eq:K_ave} and \eqref{Eq:K_per}, it is required that $\cos \alpha (r = 0) = 0$ and ${\mathrm{d} \hat{\eta}}(r = 0)/{\mathrm{d}r = 0}$, thus the boundary condition represented by Eq.~\eqref{Eq:boundary_V} becomes:
\begin{equation}\label{Eq:boundary_V2}
\frac{\mathrm{d}\bar{\eta}_v}{\mathrm{d}r}\left( r=0 \right)=0,
\frac{\mathrm{d}\hat{\eta}_v}{\mathrm{d}r}\left( r=0 \right)=0
\end{equation}

Due to the uniform time-averaged flow assumption, the time-averaged flow strain remains zero, $\overline{\mathcal{S}}= 0$. The full expressions for the flow strain of the perturbed flames are summarized as:
\begin{equation}\label{Eq:S_per_c}
\hat{\mathcal{S}}_c=
K {\gamma}{\omega}_{\ast} \exp \left( \mathrm{i} {\gamma}{\omega}_{\ast}Kx\right)\left(\mathrm{i} B \cos^2\alpha -  B K {\gamma}{\omega}_{\ast} r \sin\alpha\cos\alpha - \mathrm{i}\sin^2\alpha - \mathrm{i}\left( 2B - 1\right)\right)
\end{equation}
\begin{equation}\label{Eq:S_per_V}
\hat{\mathcal{S}}_V=
K {\gamma}{\omega}_{\ast} \exp \left( \mathrm{i} {\gamma}{\omega}_{\ast} Kx\right)\left(-\mathrm{i} B \cos^2\alpha -  B K {\gamma}{\omega}_{\ast} \left( 1-r\right) \sin\alpha\cos\alpha - \mathrm{i}\sin^2\alpha - \mathrm{i}\left( 2B - 1\right)\right)
\end{equation}
The full expressions of the time-averaged and perturbed laminar flame speed of the conical and V- flames can be obtained by substituting expressions of the flame curvature and flow strain (Eqs.~\eqref{Eq:K_ave},\eqref{Eq:K_per},\eqref{Eq:S_per_c} and \eqref{Eq:S_per_V}) into Eqs.~\eqref{Eq:Sl_ave} and \eqref{Eq:Sl_per}; these are then substituted into  Eqs.~\eqref{Eq:G_zero} and \eqref{Eq:G_line} to expand the  governing equations of the zero and first order terms, respectively.
By numerically resolving these equations, one can get the time-averaged ($\bar{\eta}$, $\frac{\mathrm{d}\bar{\eta}}{\mathrm{d}r}$, $\frac{{{\mathrm{d}}^{2}}\bar{\eta}}{\mathrm{d}{{r}^{2}}}$, $\alpha$) and perturbed quantities (${\hat{\eta}}$, $\frac{\mathrm{d}\hat{\eta}}{\mathrm{d}r}$, $\frac{{{\mathrm{d}}^{2}}\hat{\eta}}{\mathrm{d}{{r}^{2}}}$) related to the dynamic flame front.

\section{Flame transfer function modelling}
\label{sec:flame-FTF}

The flame transfer function, representing the response of flame heat release rate perturbation to the velocity perturbation immediately upstream of the flame,  can be defined as:
\begin{equation}\label{Eq:FTF}
\mathcal{F}=\frac{\widetilde{\dot{Q}}^{\prime}/{\overline{\widetilde{\dot{Q}}}}}{\tilde{{u}}_x^\prime(x=0)/\bar{\tilde{{u}}}_x}
\end{equation}
where,  ${\tilde{{u}}_x^\prime}(x=0)/\bar{\tilde{{u}}}_x= \varepsilon  \exp \left(-\mathrm{i} {\gamma}{\omega}_{\ast} t\right)$ denotes the normalized axial velocity perturbation at the inlet of the flame. 
$\overline{\widetilde{\dot{Q}}}$  and 
$\widetilde{\dot{Q}}^\prime$    are the time-averaged and perturbed heat release rates, respectively.  
 $\widetilde{\dot{Q}}$ can be determined by:
\begin{equation}\label{Eq:Q}
\widetilde{\dot{Q}} = \int_{\widetilde{A}} \tilde{\rho} \tilde{s}_{L} \tilde{h}_r \mathrm{d}{\widetilde{A}}
\end{equation}
where, $\widetilde{A}$  is the flame surface area. $\tilde{\rho}$ and $\tilde{h}_r$ represent the unburned gas density and enthalpy per mass, respectively, and are considered constants  for a low-Mach number flow rate and laminar premixed flame \citep{Schuller_CNF_2003, Wang_CNF_2009}.

\subsection{Flame transfer functions for stretched flames}
When the flame curvature and flow strain are both considered, besides the flame surface area $\widetilde{A}$, the laminar flame speed $\tilde{s}_{L}$ also oscillates. The heat release rate of the conical and V- flames can be expressed as:
\begin{equation}\label{Eq:Q_c}
\widetilde{\dot{Q}}_c = 2\pi \tilde{\rho} \tilde{h}_r\int_{0}^{R}{\tilde{s}_L \left(1+\left(\frac{\partial \widetilde{\eta}}{\partial \tilde{r}}\right)^2 \right)^{\frac{1}{2}}}\tilde{r}\mathrm{d}{\tilde{r}}
\end{equation}
\begin{equation}\label{Eq:Q_V}
\widetilde{\dot{Q}}_V = 2\pi \tilde{\rho} \tilde{h}_r\int_{R_h}^{R_b}{\tilde{s}_L \left(1+\left(\frac{\partial \widetilde{\eta}}{\partial \tilde{r}}\right)^2 \right)^{\frac{1}{2}}}\left(1-\tilde{r}\right)\mathrm{d}{\tilde{r}}
\end{equation}
where, $R_b - R_h = R$, $R_b$ and $R_h$ are the burner outlet radius and the combustor stabilizer radius, respectively.
It should be noted that the term   ${\tilde{s}_L \left(1+\left(\frac{\partial \widetilde{\eta}}{\partial \tilde{r}}\right)^2 \right)^{\frac{1}{2}}}$ is  exactly the r.h.s of the governing equation (Eq.~\eqref{Eq:G}).
It is thus possible to simplify the calculation and substitute the zero and   first order terms of  Eq.~\eqref{Eq:G} into Eqs.~\eqref{Eq:Q_c} and \eqref{Eq:Q_V}. The dimensionless expression for the FTF of the conical and V- flames are:
\begin{equation}\label{Eq:Q_ratio_non}
\mathcal{F}_c
= 2 \int_{0}^{1}{\left(\frac{\hat{s}_L}{\sin\alpha}- \bar{s}_L  \frac{\mathrm{d} \hat{\eta}}{\mathrm{d} r} \cos\alpha\right){r}}\mathrm{d}{{r}}
\end{equation}
\begin{equation}\label{Eq:Q_ratio_non_V}
\mathcal{F}_V
=2{\int_{r_h}^{r_b}{\left(\frac{\hat{s}_L}{\sin\alpha}- \bar{s}_L  \frac{\mathrm{d} \hat{\eta}}{\mathrm{d} r} \cos\alpha\right)\left(b-r\right)}\mathrm{d}{{r}}}
\end{equation}
where, $r_b - r_h = 1$, $r_b$ and $r_h$ are the dimensionless burner outlet radius and the combustor stabilizer radius, respectively.

\subsection{Flame transfer functions for unstretched flames}
\label{subsec:FTF_unstretched}
In order to identify and highlight the effect of flame stretch, the FTFs of the two kinds of flame neglecting these effects are also introduced and partly derived. 
It should be noted that some of them are already derived from previous research. 
Schuller et al. \citep{Schuller_CNF_2003} derived the FTFs for the laminar premixed  conical and V- flames under uniform and convective velocity perturbation models. 
Cuquel et al. \citep{Cuquel_MCS7_2011} further obtained the FTFs of conical flame by accounting for  the incompressible velocity perturbation model. 
Based on similar methods, the FTFs for the rest flames by accounting for the three kinds of flow velocity perturbation models are also derived. 
These results are summarized in Table \ref{TableFTF_ana}. 

\begin{table}[h]
\small \centering
\caption{Governing equations for different flames.}
\label{TableFTF_ana}
\begin{tabular}{p{1.5cm} p{1cm} p{13cm}<{\centering}}
\hline
& &Unstretched flame transfer functions \\
\hline
\rule{0pt}{20pt}
\multirow{5}{*}{Conical} 
& UVP\citep{Schuller_CNF_2003}
&$\frac{2}{{\omega}^{2}_{\ast}}\left( 1-\exp \left( \mathrm{i}{\omega}_{\ast} \right)+\mathrm{i}{\omega}_{\ast} \right)$\\
\rule{0pt}{20pt}
& CVP\citep{Schuller_CNF_2003}
&$\frac{2}{{\omega}_{\ast}^{2}\left(1-\xi\right) }\left( 1-\exp \left( \mathrm{i}{\omega}_{\ast} \right)+\frac{\exp \left( \mathrm{i}{\omega}_{\ast}\xi  \right)-1}{\xi } \right)$\\
\rule{0pt}{20pt}
& IVP\citep{Cuquel_MCS7_2011}
&$-\frac{1}{\mathrm{i}{\omega}_{\ast}\left(1-\xi\right)}\left[ \left( 2-\mathrm{i}{\omega}_{\ast}\xi +\frac{\xi}{1-\xi}  \right)\left( \frac{\exp \left( \mathrm{i}{\omega}_{\ast}\xi \right)-1}{\mathrm{i}{\omega}_{\ast}\xi }-\frac{\exp \left( \mathrm{i}{\omega}_{\ast} \right)-1}{\mathrm{i}{\omega}_{\ast}} \right)+\left( \exp \left( \mathrm{i}{\omega}_{\ast}\xi \right)-\frac{\exp \left( \mathrm{i}{\omega}_{\ast}\xi  \right)-1}{\mathrm{i}{\omega}_{\ast}\xi } \right) \right]$\\
\rule{0pt}{0pt}
 & & \\
\hline
\rule{0pt}{20pt}
\multirow{8}{*}{V-} 
& UVP\citep{Schuller_CNF_2003}
&$\frac{2}{{\omega}_{\ast}^{2}}\left( \exp \left( \mathrm{i}{\omega}_{\ast} \right)-1-\mathrm{i}{\omega}_{\ast}\exp (\mathrm{i}{\omega}_{\ast}) \right)$\\
\rule{0pt}{20pt}
& CVP\citep{Schuller_CNF_2003}
& $\frac{2}{{\omega}_{\ast}^{2}\left(1-\xi\right) }\left( \exp \left( \mathrm{i}{\omega}_{\ast} \right)-1-\frac{\exp \left( \mathrm{i}{\omega}_{\ast}\xi  \right)-1}{\xi } \right)+\frac{2\mathrm{i}}{{\omega}_{\ast}\left(1-\xi\right) }\left( \exp \left( \mathrm{i}{\omega}_{\ast}\xi  \right)-\exp \left( \mathrm{i}{\omega}_{\ast} \right) \right)$\\
\rule{0pt}{40pt}
& IVP 
&$\begin{aligned} &\frac{2-\mathrm{i}{\omega}_{\ast}\xi +\frac{\xi}{1-\xi} }{\mathrm{i}{\omega}_{\ast}\left(1-\xi\right) }\left( \exp \left( \mathrm{i}{\omega}_{\ast} \right)-\exp \left( \mathrm{i}{\omega}_{\ast}\xi  \right)+\frac{\exp \left( \mathrm{i}{\omega}_{\ast}\xi \right)-1}{\mathrm{i}{\omega}_{\ast}\xi }-\frac{\exp \left( \mathrm{i}{\omega}_{\ast} \right)-1}{\mathrm{i}{\omega}_{\ast}} \right)...\\
&-\frac{1}{\mathrm{i}{\omega}_{\ast}\left(1-\xi\right) }\left( \mathrm{i}{\omega}_{\ast}\exp \left( \mathrm{i}{\omega}_{\ast}\xi  \right)-\exp \left( \mathrm{i}{\omega}_{\ast}\xi  \right)+\frac{\exp \left( \mathrm{i}{\omega}_{\ast}\xi  \right)-1}{\mathrm{i}{\omega}_{\ast}\xi } \right)\end{aligned}$\\
\rule{0pt}{0pt}
 & & \\
\hline
\end{tabular}
\begin{flushleft}where, UVP: uniform velocity perturbation model; CVP: convective velocity perturbation model; IVP: incompressible velocity perturbation model. $\xi = \frac{{\beta }^2}{{1+{{\beta }^{2}}}} $. \end{flushleft}
\end{table}

\section{Results and discussion}
\label{sec:results}

\begin{figure}[h]
	\centering
		\subfigure
  		{
\includegraphics[height=4.5cm]{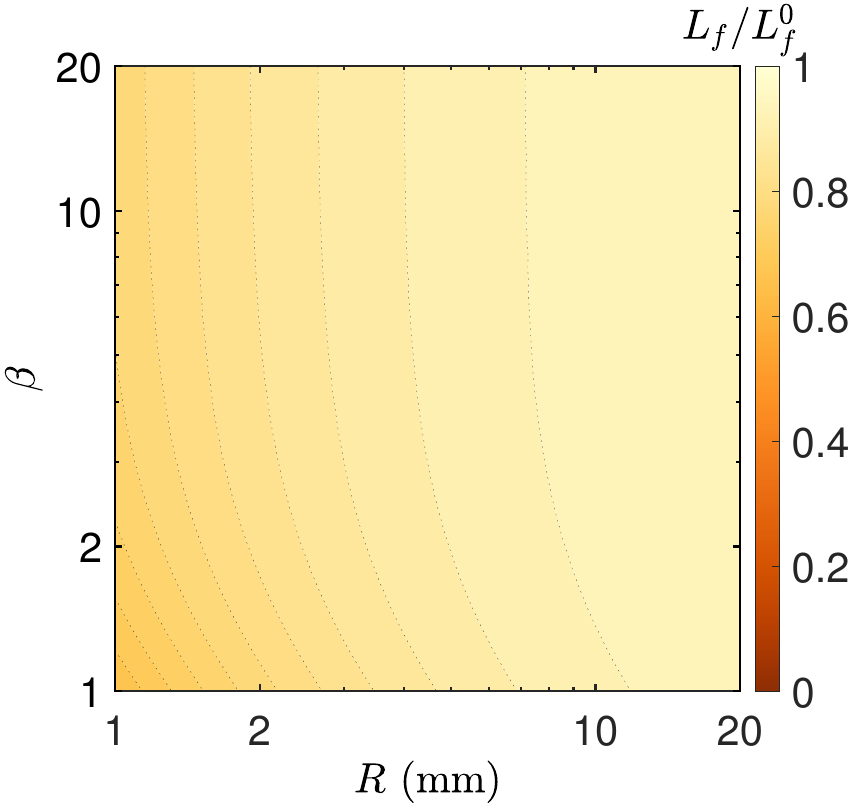}
  		\label{Fig:Steady_1}
  		}\put (-140,10) {\normalsize  $\displaystyle(a)$} 
  		\vspace*{0pt}
  		\hspace*{10pt}
  		\subfigure
  		  		{
\includegraphics[height=4.5cm]{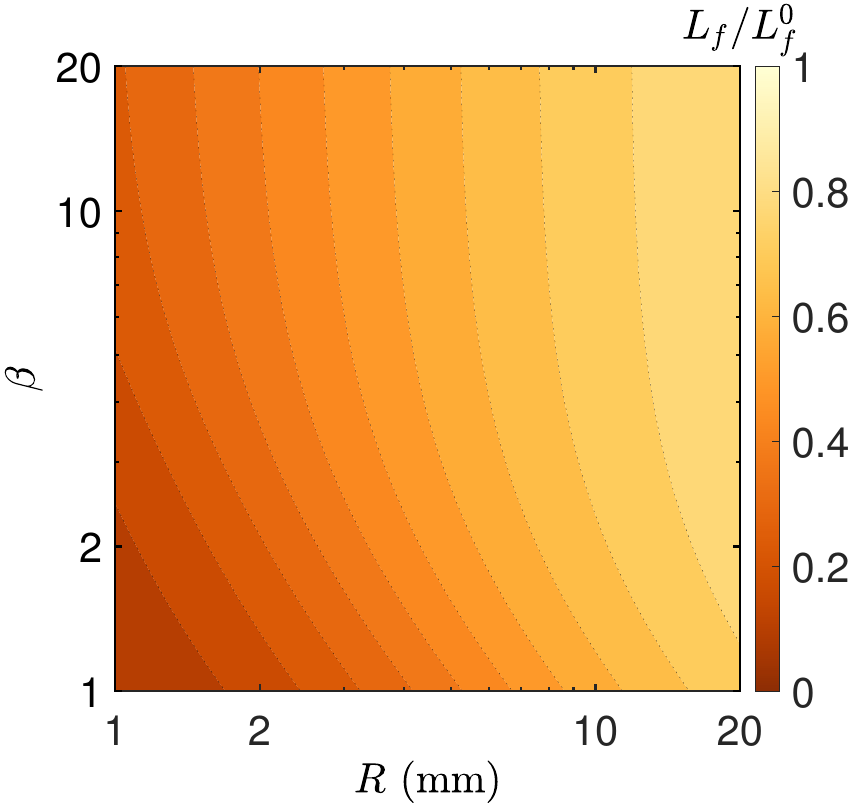}
  		\label{Fig:Steady_2}
  		}\put (-140,10) {\normalsize  $\displaystyle(b)$} 
  		\vspace*{0pt}
  		\hspace*{10pt}
  		\subfigure
  		  		{
\includegraphics[height=4.5cm]{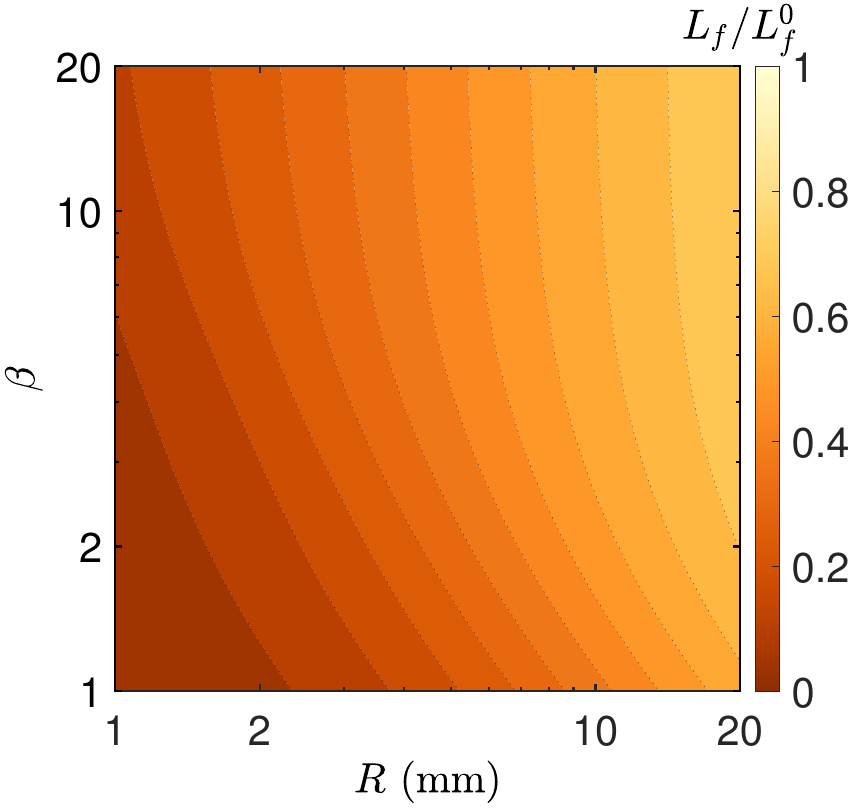}
  		\label{Fig:Steady_3}
  		}\put (-140,10) {\normalsize  $\displaystyle(c)$} \\
  		\vspace*{0pt}
  		\hspace*{0pt}	
 	 \caption{Evolutions of the flame height ratio $L_f/L_{f}^0$ as functions of  the flame radius $R$ and the unstretched flame aspect ratio $\beta = L_{f}^0/R$. The Markstein lengths and equivalent ratios for the conical / V- flames are (a) $ \widetilde{\mathcal{L}}=0.075$ mm, $\phi=1.4$, (b) $ \widetilde{\mathcal{L}}=1$ mm, $\phi=1$ and (c) $ \widetilde{\mathcal{L}}=2.1$ mm, $\phi=0.6$.}
	 \label{Fig:Steady}
	 \vspace*{00pt}
\end{figure}

This section delves into the implications of flame stretch on the dynamics of both steady-state and perturbed flames. 
The propane-air premixed flame is considered; the Markstein length $\widetilde{\mathcal{L}}$ varies with changing the equivalence ratio $\phi$ and Lewis number $Le$.  
According to the experimental results \citep{Tang_IJHE_2008}, when the equivalent ratios of propane-air premixed flames are 1.4, 1 and 0.6, the corresponding Markstein lengths $\widetilde{\mathcal{L}}$  are 0.075 mm, 1 mm and 2.1 mm, respectively. 
Thus the flame stretch is more important for the lean propane-air premixed flame.

\subsection{Steady flame results}
\label{subsec:steady}

Due to the presence of stretch, the flame speed $s_L$ does not remain constant for a given operating condition. For the envisaged flames, the steady flame front is not a straight  line, and features the shape of an arc with a certain curvature (as sketched in Fig.~\ref{Fig:Shape}). 
For steady flames, the flow strain $\overline{\mathcal{S}}= 0$, so the flame stretch consists entirely of time-averaged flame curvature $\overline{\mathcal{K}}$.  
By substituting Eq.~\eqref{Eq:Sl_ave} into Eq.~(\ref{Eq:G_zero}), the steady flame front shapes considering flame curvature can be obtained.
For the sake of clarity, the variation trend of flame shape with and without considering curvature can be demonstrated by the flame height ratio $L_f/L_{f}^0$, where $L_f$ and $L_{f}^0$ are the heights when the curvature is considered and not, respectively. 
The specific trends are further highlighted by the contours of the flame height ratio $L_f/L_f^0$ for different Markstein lengths and flame sizes as shown in Fig.~\ref{Fig:Steady}. 
It is noted that considering the curvature will lead to a decrease in the flame height, and the smaller value of  $L_f/L_{f}^0$ corresponds to a larger effect of curvature on the flame shape. 
For each subfigure in Fig.~\ref{Fig:Steady}, $L_f/L_f^0$ typically decreases with decreasing the flame radius $R$ and unstretched flame aspect ratio $\beta$, indicating that these effects are more sound for short and small flames.  
Additionally, Fig.~\ref{Fig:Steady}(a) to (c) sequentially represents the cases with increasing Markstein length $ \widetilde{\mathcal{L}}$.
When the Markstein length becomes large (e.g., for lean propane/air flames), the reduction in the value of $L_f/L_{f}^0$ from unity is even obvious for large $R$ and $\beta$. 
Overall, the steady flame shapes are influenced by the combined effects of $R$, $\beta$ and $\widetilde{\mathcal{L}}$. Decreasing $R$ and $\beta$ and increasing $\widetilde{\mathcal{L}}$ will result in a more pronounced decrease in flame height due to the consideration of flame curvature.

\subsection{Perturbed flame front and flame transfer function}
\label{subsec: Perturbed }

In this section, effects of the flame stretch on FTFs for the conical and V- flames are discussed based upon comparisons between  results when the flame stretch is considered or not. 
As the oncoming flow disturbance distributions vary for different configurations, it is thus not possible to illustrate the effect of the flame stretch based on only one oncoming flow disturbance model.    
One thus considers three models. 
Furthermore, the individual contributions of flame curvature and flow strain to the flame stretch expressions are investigated separately.

In the steady flame results, it has been observed that the effect of stretch on the flame is more significant for a larger  Markstein length.  In the following sections, the Markstein length $\widetilde{\mathcal{L}}=1$ mm (equivalence ratio $\phi=1$) is taken as an example for the analyses of FTFs considering different  flow velocity perturbation models and flame geometries.
%

\subsubsection{Conical flame }
\label{subsubsec:conical}

\subsubsection*{A. The effect of flame stretch}

\begin{figure}[!t]
	\centering
		\subfigure
  		{
\includegraphics[height=5.5cm]{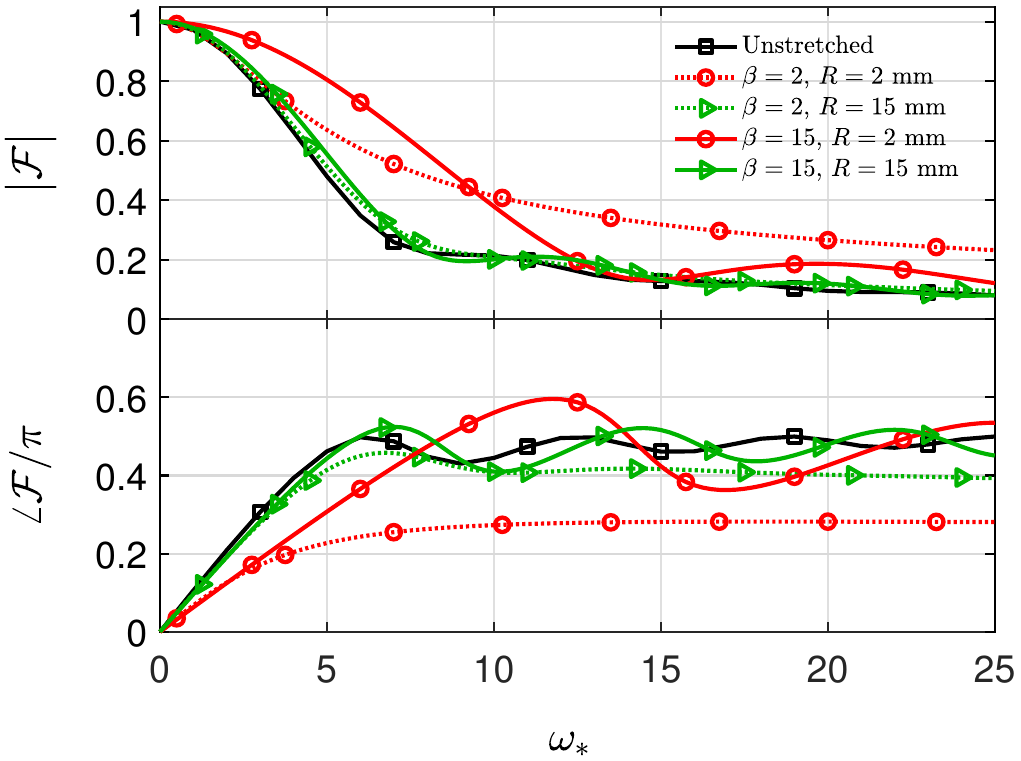}
  		}\put (-200,5) {\normalsize  $\displaystyle(a)$} \\
  		\vspace*{0pt}
  		\hspace*{0pt}
  		\subfigure
  		  		{
\includegraphics[height=5.5cm]{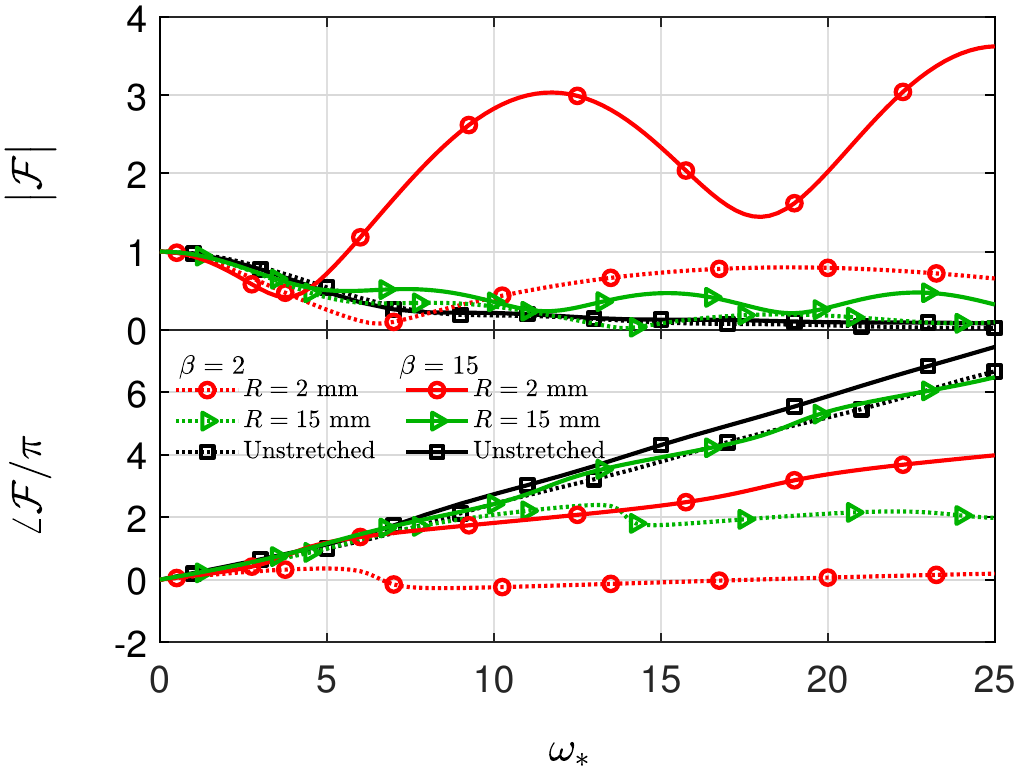}
  		}\put (-200,5) {\normalsize  $\displaystyle(b)$} 
  		\vspace*{0pt}
  		\hspace*{0pt}
  		\subfigure
  		  		{
\includegraphics[height=5.5cm]{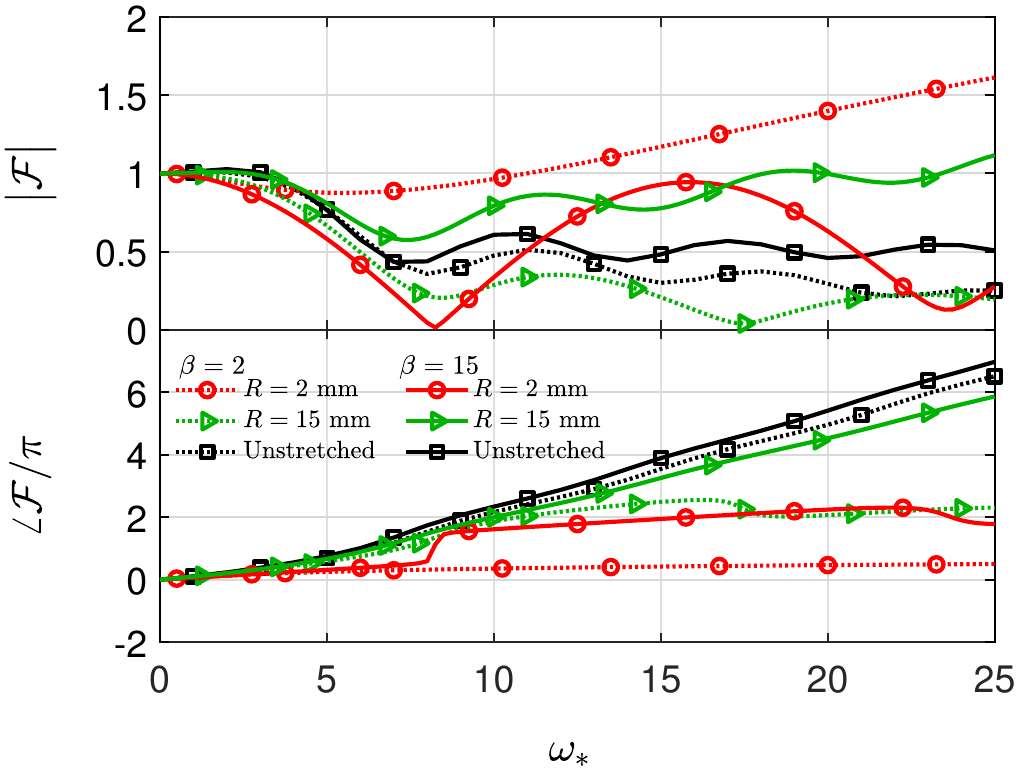}
  		}\put (-200,5) {\normalsize  $\displaystyle(c)$} \\
  		\vspace*{0pt}
  		\hspace*{0pt} 
  		 	 \caption{FTFs of the conical flames for different flame geometries when the flame stretch is considered or not; the (a) uniform velocity perturbation model, (b) convective velocity perturbation model and (c) incompressible velocity perturbation model are considered, respectively.  }
	 \label{Fig:FTF_cflame_stretch}
	 \vspace*{00pt}
\end{figure}

One firstly considers the effect of flame stretch based on three oncoming flow perturbation models for the conical flames.
Figure~\ref{Fig:FTF_cflame_stretch} shows the FTFs for different unstretched flame aspect ratios $\beta$ and flame radii $R$  when the stretch is  accounted for or not. 
The top and bottom of each subfigure show the evolutions of the FTF gain and phase with increasing frequency. Herein, for the sake of comparisons, the normalized frequency $\omega_\ast$ is used. 
For the uniform velocity perturbation model shown in Fig.~\ref{Fig:FTF_cflame_stretch}(a), it is possible to notice that all these FTFs feature shapes of low pass filters. 
The FTF when $R=15$ mm is close to that when the stretch is not accounted for, indicating that the stretch can be neglected for the FTF for the large flame radius. 
When $R=2$ mm, for small $\beta$, the FTF gain is larger than that neglecting the stretch, and the phase lag is smaller.
While for large $\beta$, the reduction of FTF gain is slow, but it approaches the result without considering flame stretch at high frequencies.
For the convective velocity perturbation model shown in Fig.~\ref{Fig:FTF_cflame_stretch}(b), in the case of large $\beta$ and small $R$, the gain increases and fluctuates with frequency, eventually surpassing unity.
While for the incompressible velocity perturbation model shown in Fig.~\ref{Fig:FTF_cflame_stretch}(c), the gain exhibits a pronounced rise with increasing frequency when both $\beta$ and $R$ are small.

\begin{figure}[!t]
	\centering
		\subfigure
  		{
\includegraphics[height=4.5cm]{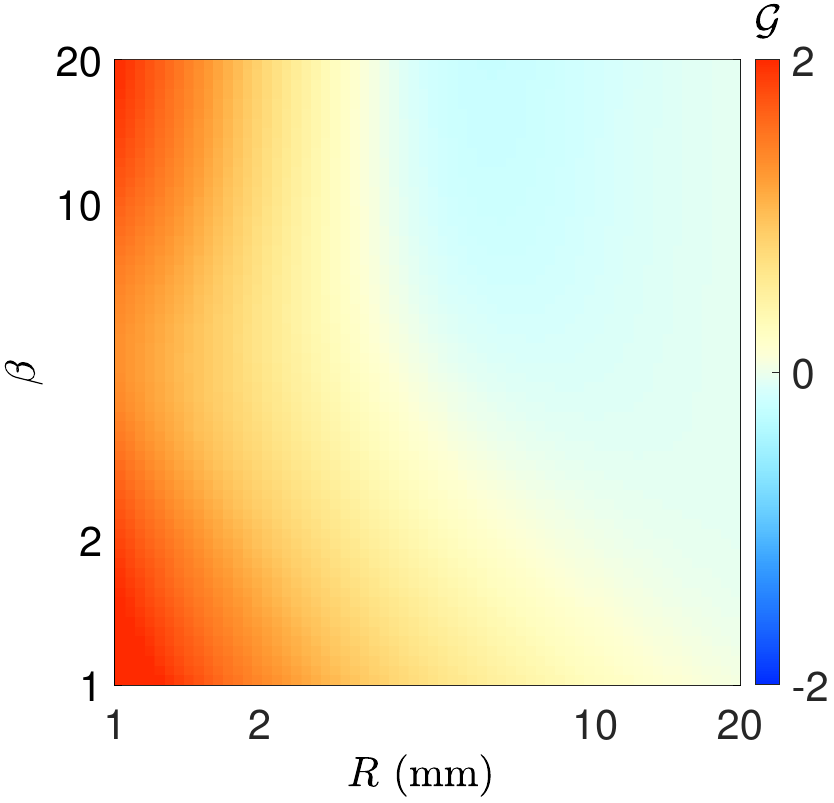}
  		}\put (-140,0) {\normalsize  $\displaystyle(a.1)$} 
  		\vspace*{0pt}
  		\hspace*{0pt}
  		\subfigure
  		  		{
\includegraphics[height=4.5cm]{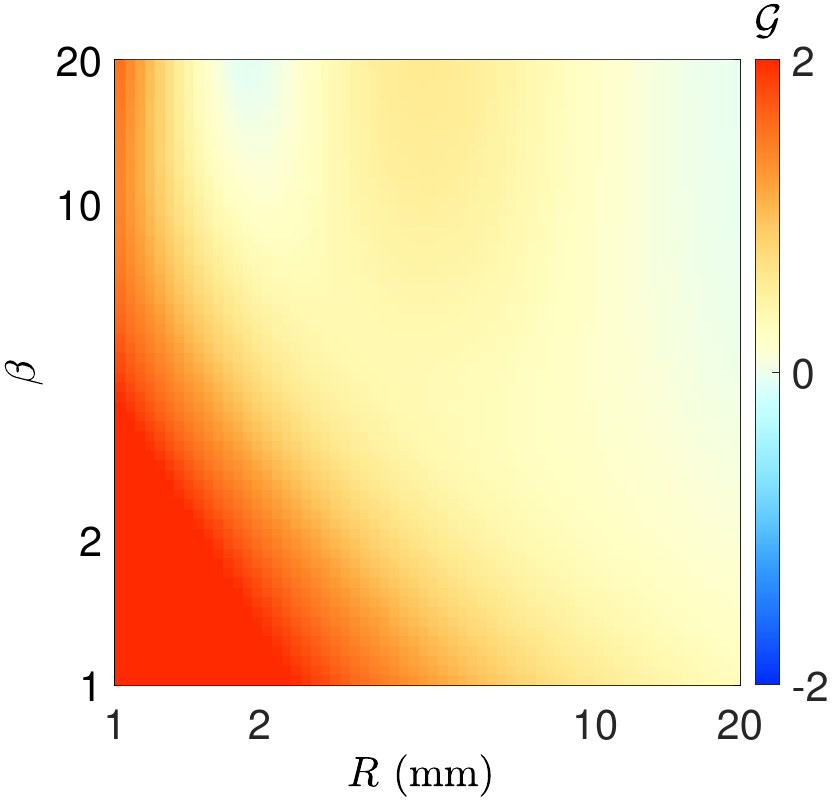}
  		}\put (-140,0) {\normalsize  $\displaystyle(a.2)$} 
  		\vspace*{0pt}
  		\hspace*{0pt}
  		\subfigure
  		  		{
\includegraphics[height=4.5cm]{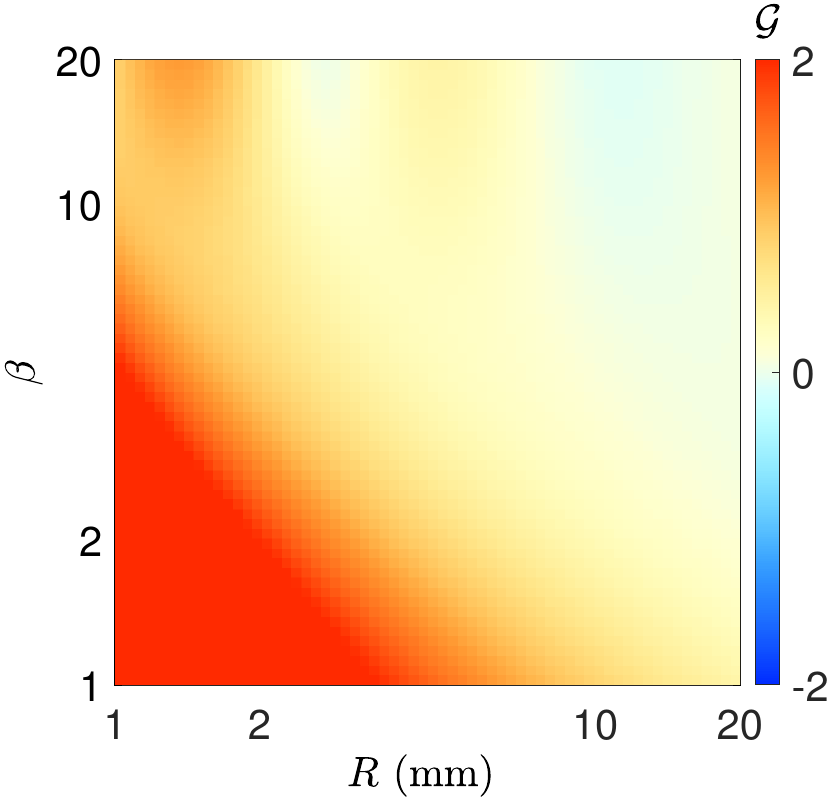}
  		}\put (-140,0) {\normalsize  $\displaystyle(a.3)$} \\
  		\vspace*{0pt}
  		\hspace*{0pt}
  		\subfigure
  		{
\includegraphics[height=4.5cm]{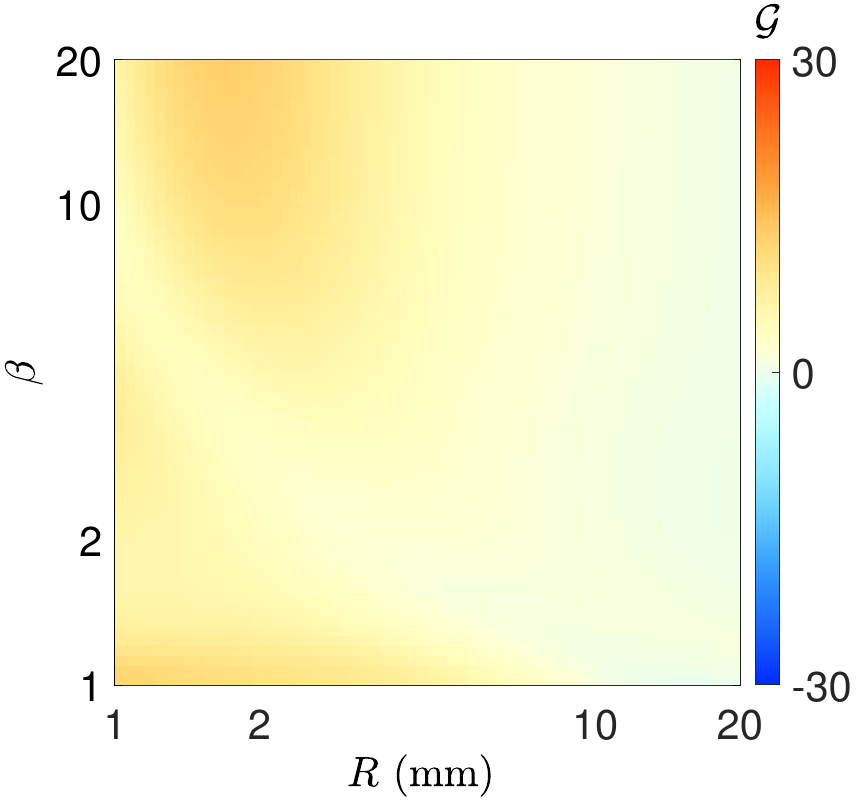}
  		}\put (-140,0) {\normalsize  $\displaystyle(b.1)$} 
  		\vspace*{0pt}
  		\hspace*{0pt}
  		\subfigure
  		  		{
\includegraphics[height=4.5cm]{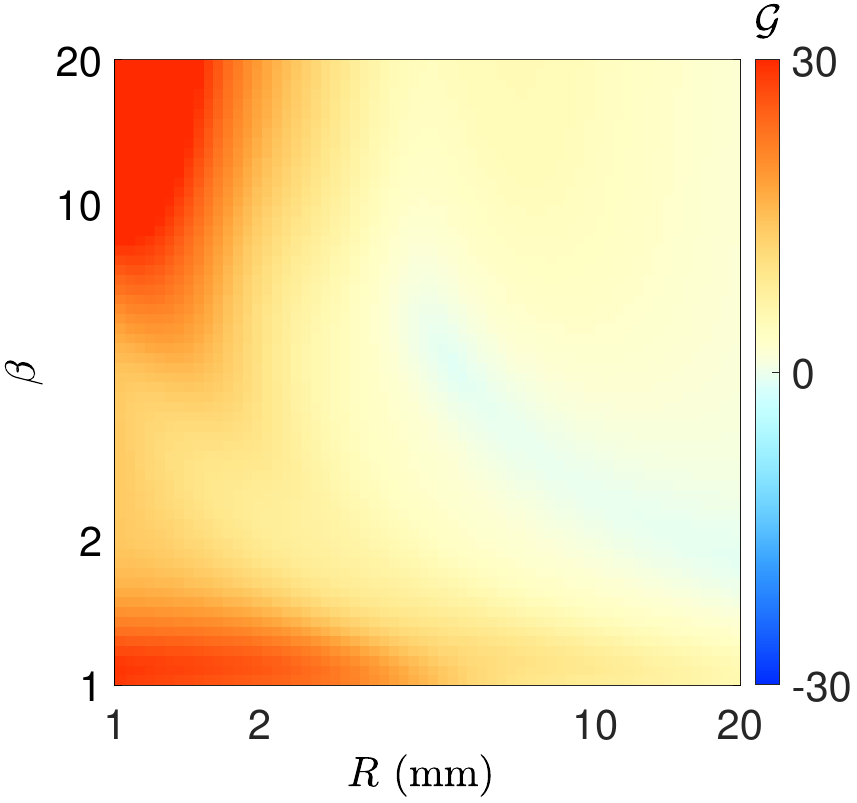}
  		}\put (-140,0) {\normalsize  $\displaystyle(b.2)$} 
  		\vspace*{0pt}
  		\hspace*{0pt}
  		\subfigure
  		  		{
\includegraphics[height=4.5cm]{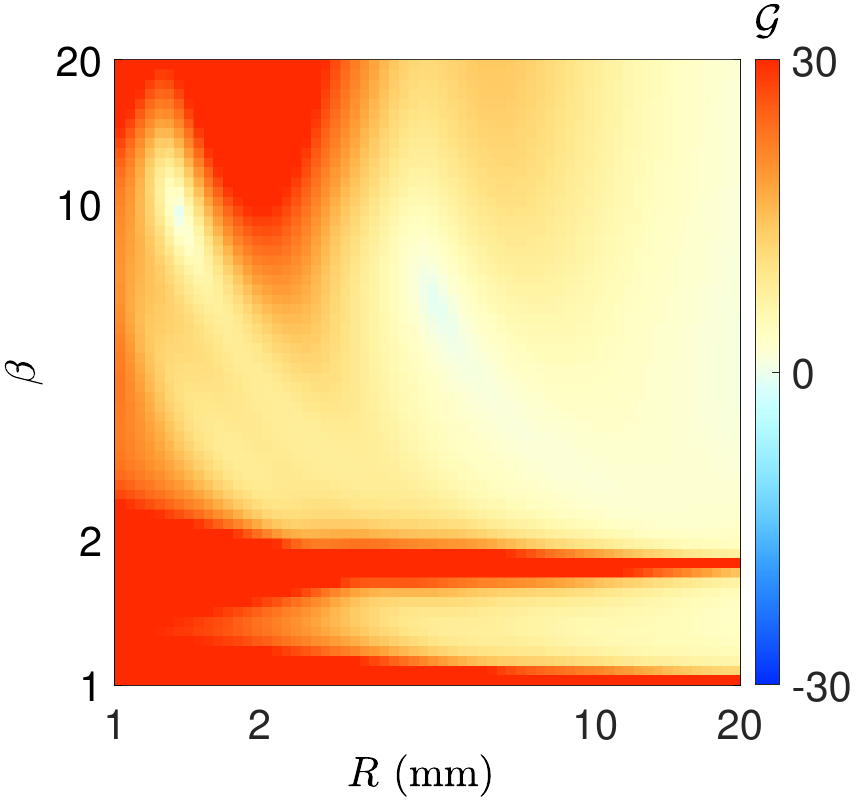}
  		}\put (-140,0) {\normalsize  $\displaystyle(b.3)$} \\
  		\vspace*{0pt}
  		\hspace*{0pt}
  		\subfigure
  		{
\includegraphics[height=4.5cm]{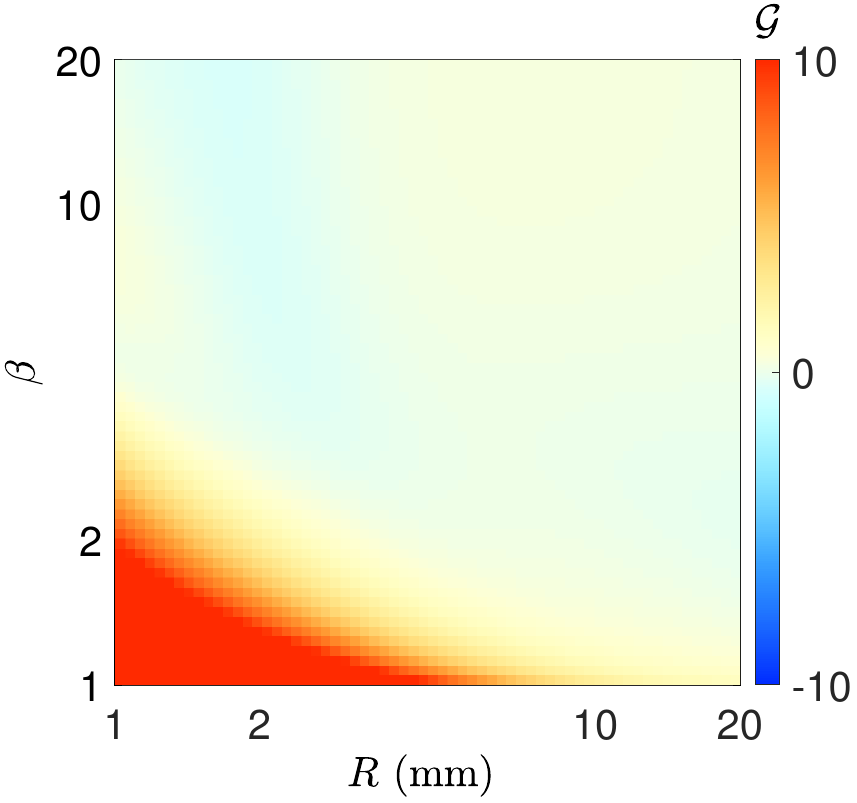}
  		}\put (-140,0) {\normalsize  $\displaystyle(c.1)$} 
  		\vspace*{0pt}
  		\hspace*{0pt}
  		\subfigure
  		  		{
\includegraphics[height=4.5cm]{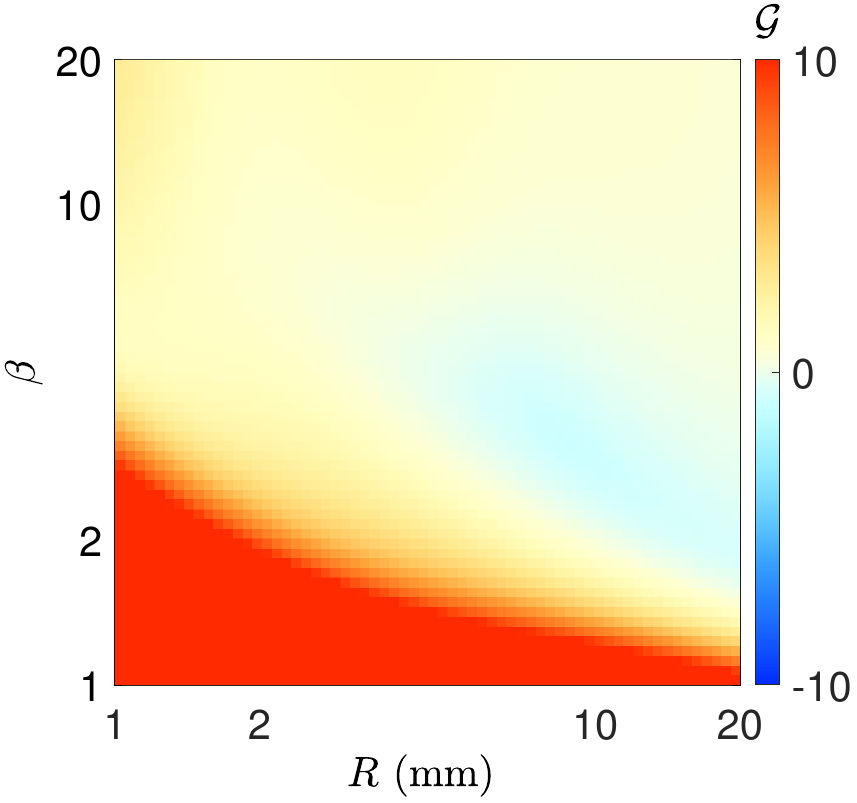}
  		}\put (-140,0) {\normalsize  $\displaystyle(c.2)$} 
  		\vspace*{0pt}
  		\hspace*{0pt}
  		\subfigure
  		  		{
\includegraphics[height=4.5cm]{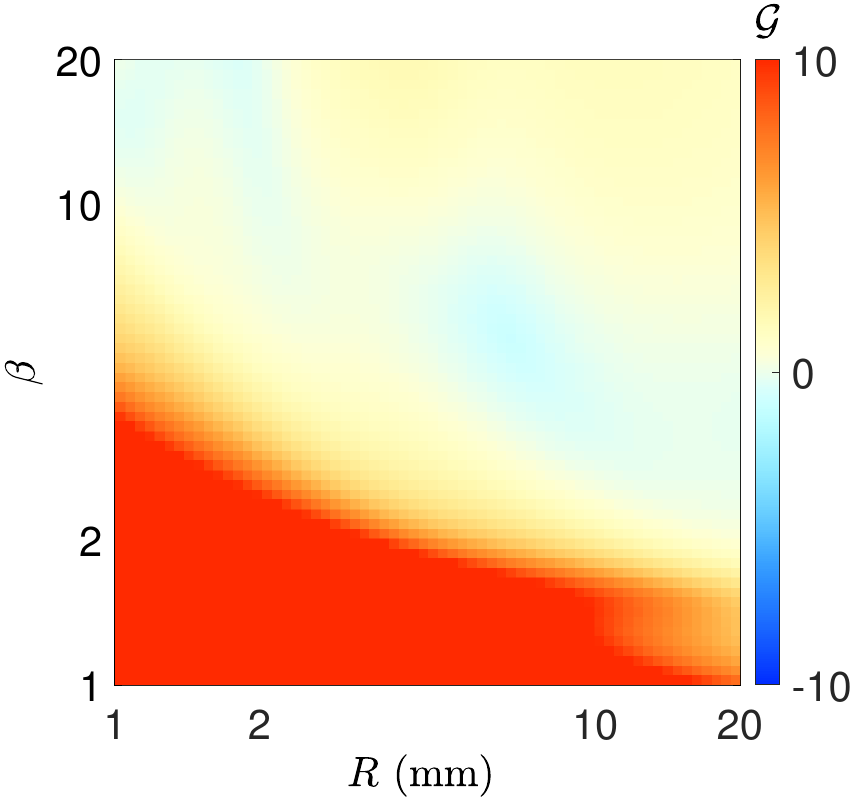}
  		}\put (-140,0) {\normalsize  $\displaystyle(c.3)$} \\
  		\vspace*{0pt}
  		\hspace*{0pt}
 	 \caption{Contours of the FTF gain difference ratios $\mathcal{G}$  of conical flames as functions of  $\beta$ and $R$ considering the flame stretch effect for the (a) uniform velocity perturbation model, (b) convective velocity perturbation model and (c) incompressible velocity perturbation model. The normalized frequencies are (1) ${\omega}_{\ast}$ = 10, (2) ${\omega}_{\ast}$ = 15 and (3) ${\omega}_{\ast}$ = 25. }
	 \label{Fig:FTF_conical_stretch_ratio}
	 \vspace*{00pt}
\end{figure}

In order to quantify the FTF differences considering the flame stretch (or the flame curvature and flow strain in the following sections) or not, it is possible to define the FTF gain difference ratio:
\begin{equation}\label{Eq:gainratio}
\mathcal{G} = \frac{|\mathcal{F}_{with}| - |\mathcal{F}_{none}|}{|\mathcal{F}_{none}|}
\end{equation}
where $\mathcal{F}_{with}$ and $\mathcal{F}_{none}$ indicate the FTFs when the flame stretch (curvature or flow strain) is accounted for or not, respectively.
Figure~\ref{Fig:FTF_conical_stretch_ratio}
shows the FTF gain difference ratio $\mathcal{G}$ for the conical flames considering flame stretch under the three kinds of flow
perturbation models.
The larger absolute value of the  FTF gain difference ratio $\mathcal{G}$ indicates  a larger effect of the flame stretch. 
As shown in these contours, the same trends as those shown in Fig.~\ref{Fig:FTF_cflame_stretch} can be observed: large deviations occur for smaller flame radius $R$ and unstretched flame aspect ratio $\beta$. 
The FTF gains $|\mathcal{F}_{with}|$ are typically larger than $|\mathcal{F}_{none}|$ neglecting the stretch effect, which is primarily attributed to the significant decrease in the steady state flame height at these flame geometries shown in Fig.~\ref{Fig:Steady}. 
At large frequencies, the region with  large  gain difference ratio $\mathcal{G}$ expands for small size flames, as the denominator of  $\mathcal{G}$ -- the gain  $|\mathcal{F}_{none}|$ remains small. 
To elucidate the gain characteristics of the convective and incompressible velocity perturbation models at high $\beta$ depicted in Fig.~\ref{Fig:FTF_conical_stretch_ratio}(b) and (c), respectively, one proceeds to analyze the distinct influences of flame curvature and flow strain.

\subsubsection*{B. The effect of flame curvature}

\begin{figure}[!t]
	\centering
		\subfigure
  		{
\includegraphics[height=5.5cm]{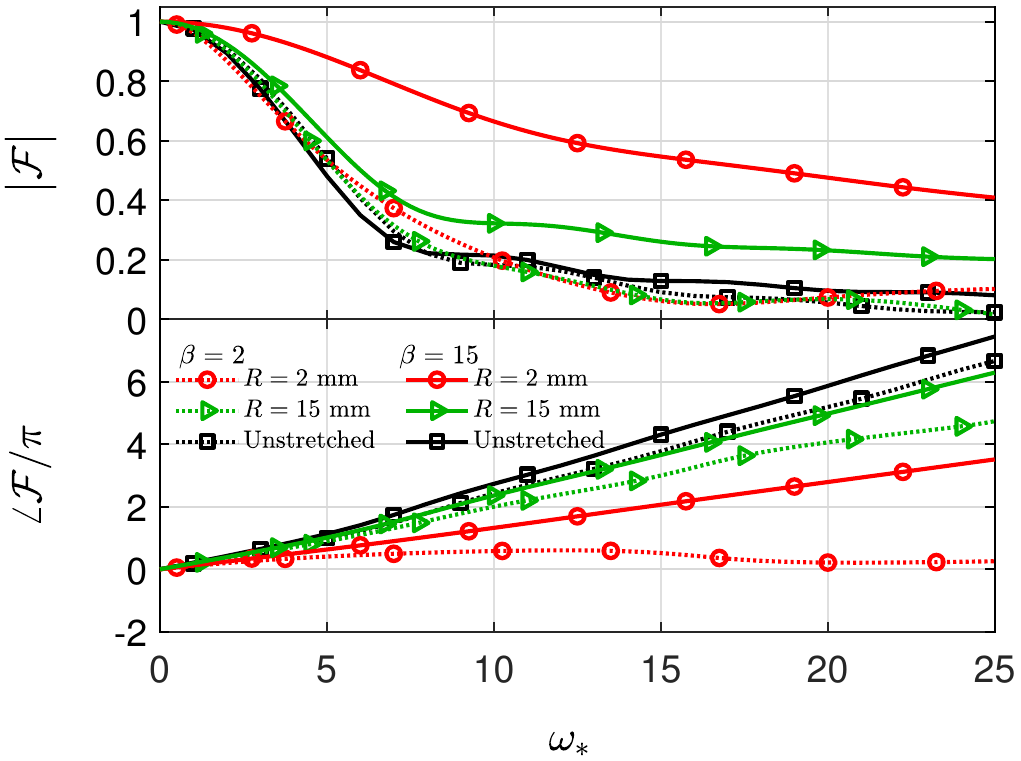}
  		}\put (-200,5) {\normalsize  $\displaystyle(a)$} 
  		\vspace*{0pt}
  		\hspace*{0pt}
  		\subfigure
  		  		{
\includegraphics[height=5.5cm]{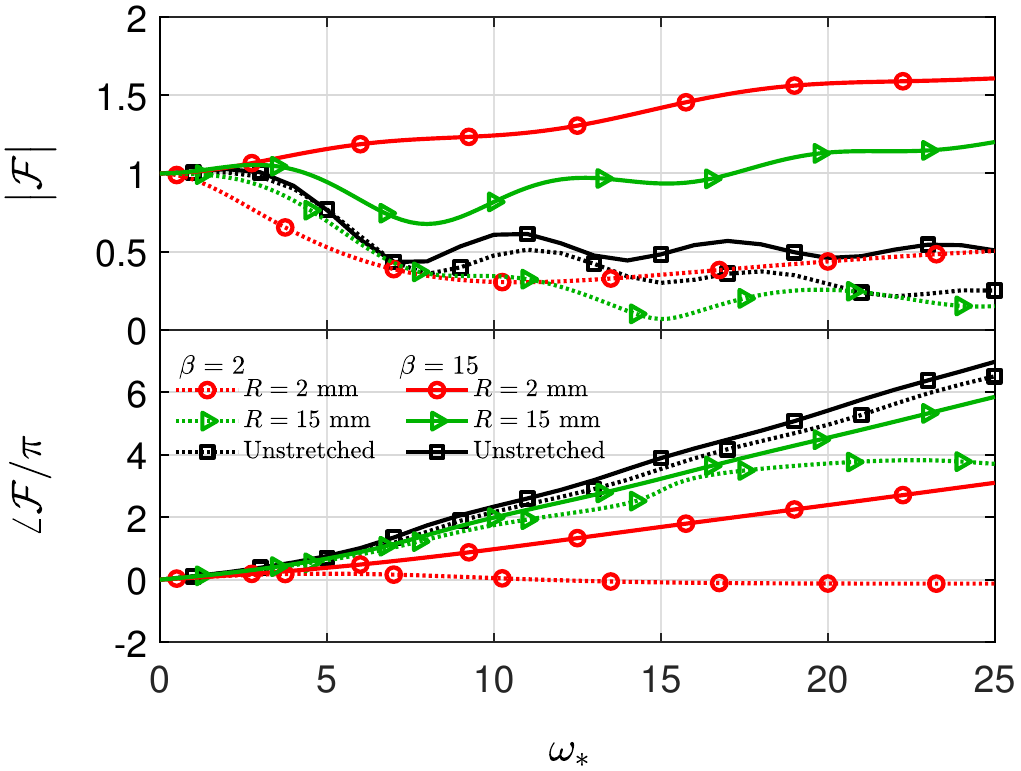}
  		}\put (-200,5) {\normalsize  $\displaystyle(b)$} 
  		\vspace*{0pt}
  		\hspace*{0pt}
  		 	 \caption{FTFs of the conical flames for different flame geometries when the flame curvature is considered or not; the (a) convective velocity perturbation model and (b) incompressible velocity perturbation model are considered, respectively.   }
	 \label{Fig:FTF_conical_curvature_C&I}
	 \vspace*{00pt}
\end{figure}

\begin{figure}[!t]
	\centering
  		\subfigure
  		{
\includegraphics[height=4.5cm]{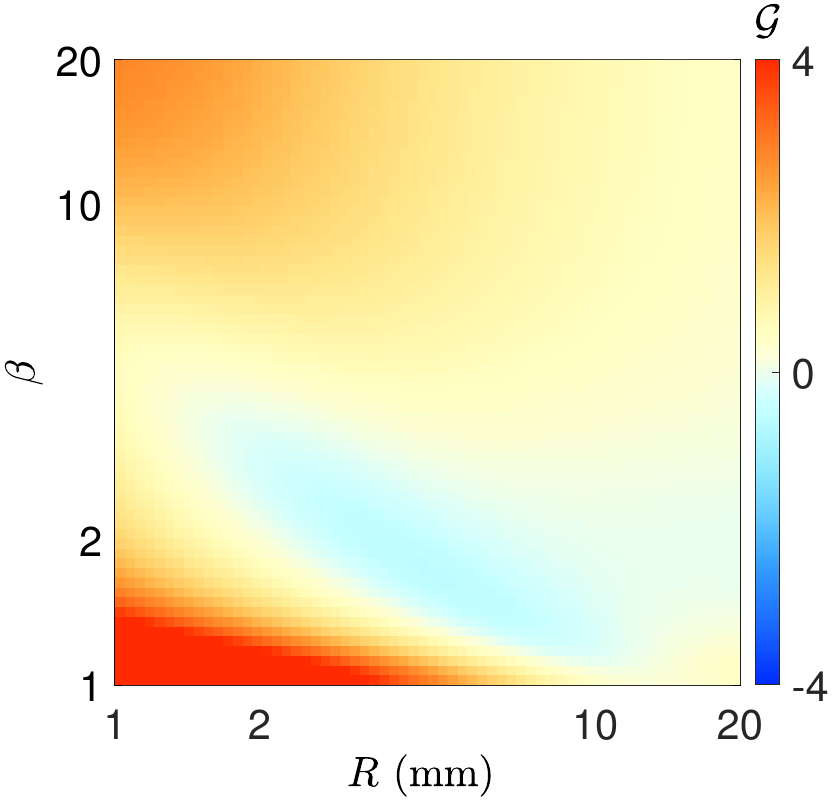}
  		}\put (-140,0) {\normalsize  $\displaystyle(a.1)$} 
  		\vspace*{0pt}
  		\hspace*{0pt}
  		\subfigure
  		  		{
\includegraphics[height=4.5cm]{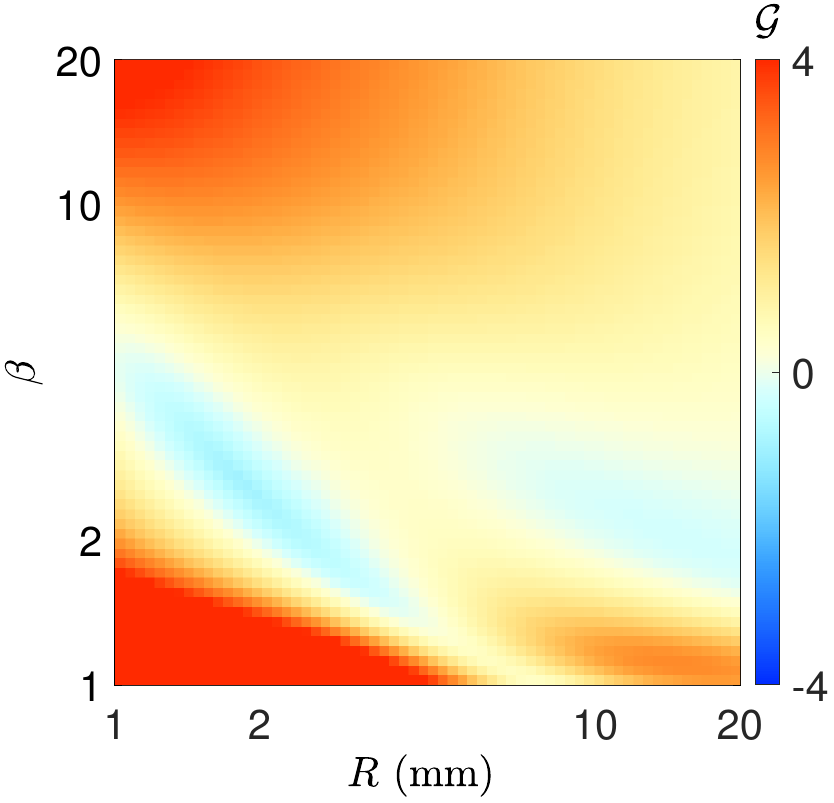}
  		}\put (-140,0) {\normalsize  $\displaystyle(a.2)$} 
  		\vspace*{0pt}
  		\hspace*{0pt}
  		\subfigure
  		  		{
\includegraphics[height=4.5cm]{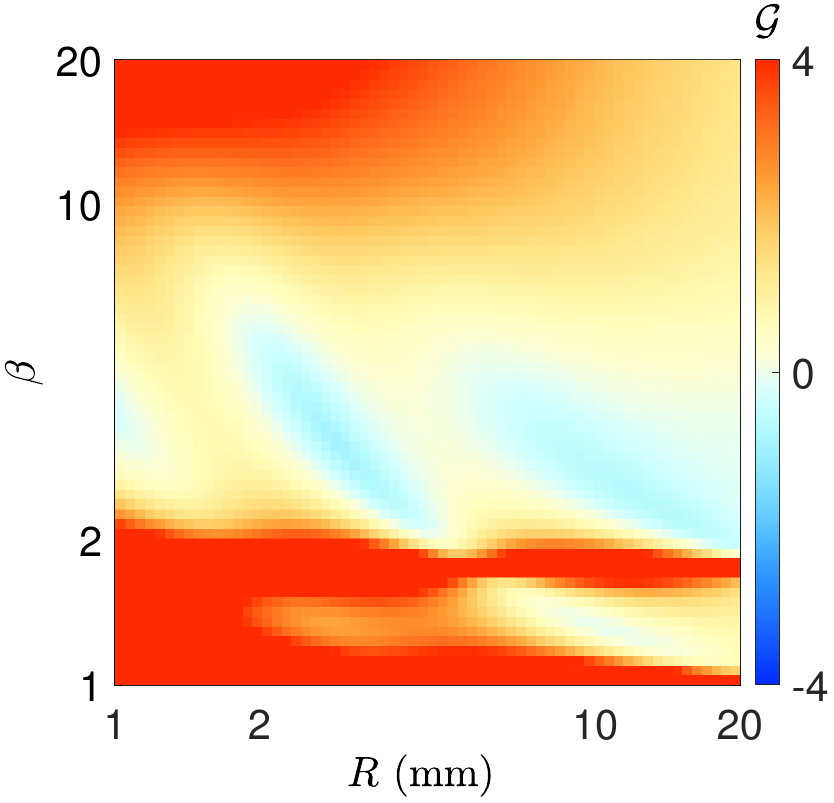}
  		}\put (-140,0) {\normalsize  $\displaystyle(a.3)$} \\
  		\vspace*{0pt}
  		\hspace*{0pt}
  		\subfigure
  		{
\includegraphics[height=4.5cm]{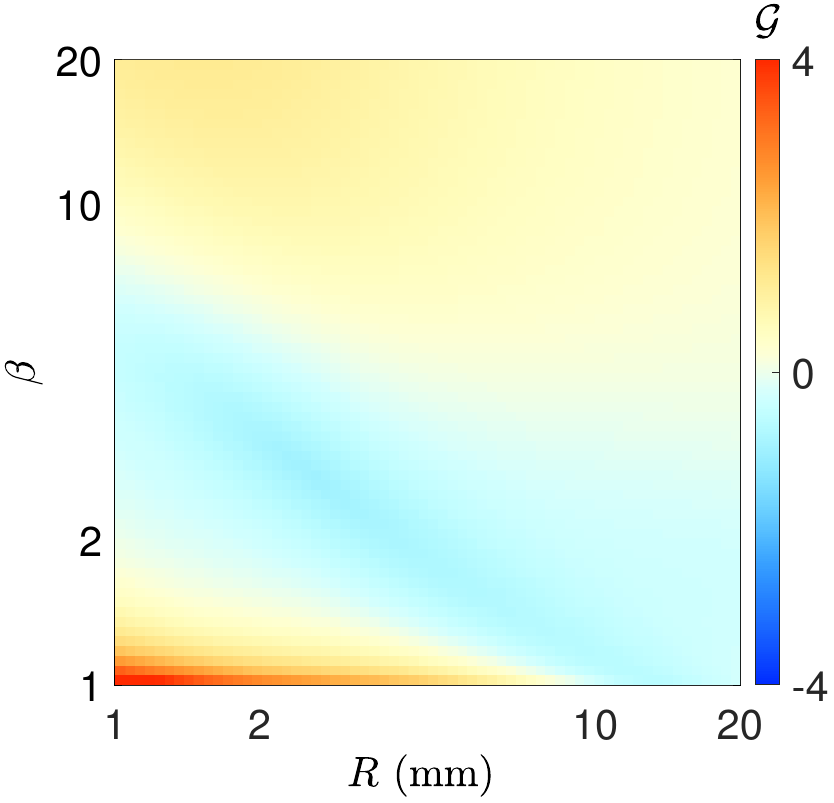}
  		}\put (-140,0) {\normalsize  $\displaystyle(b.1)$} 
  		\vspace*{0pt}
  		\hspace*{0pt}
  		\subfigure
  		  		{
\includegraphics[height=4.5cm]{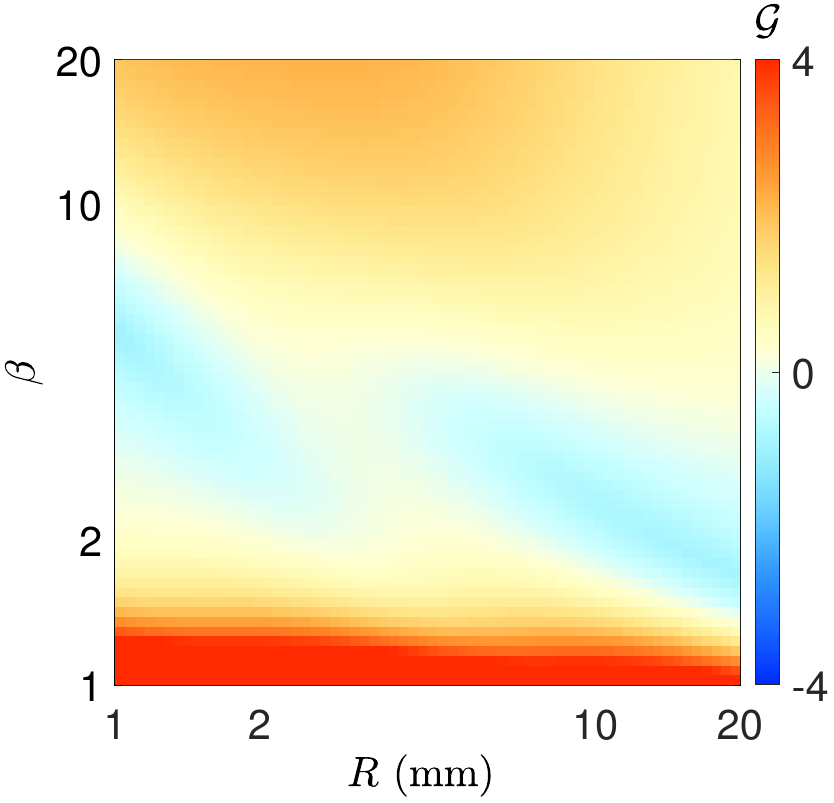}
  		}\put (-140,0) {\normalsize  $\displaystyle(b.2)$} 
  		\vspace*{0pt}
  		\hspace*{0pt}
  		\subfigure
  		  		{
\includegraphics[height=4.5cm]{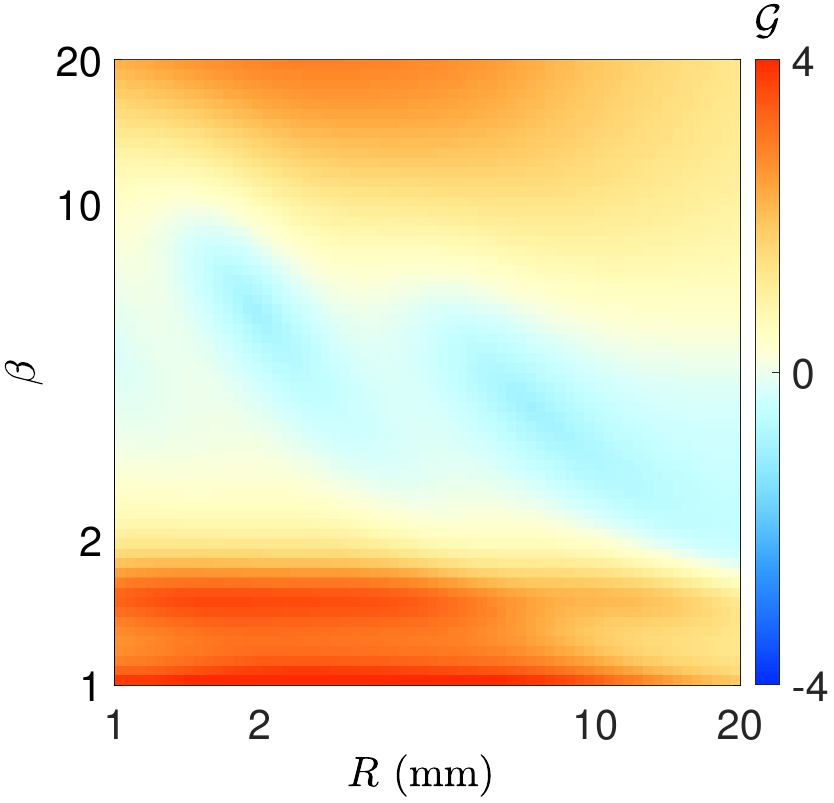}
  		}\put (-140,0) {\normalsize  $\displaystyle(b.3)$} \\
  		\vspace*{0pt}
  		\hspace*{0pt}
 	  	 \caption{Contours of the FTF gain difference ratios $\mathcal{G}$  of conical flames as functions of  $\beta$ and $R$ considering only the flame curvature effect  for the (a) convective velocity perturbation model and (b) incompressible velocity perturbation model. The normalized frequencies are (1) ${\omega}_{\ast}$ = 10, (2) ${\omega}_{\ast}$ = 15 and (3) ${\omega}_{\ast}$ = 25. }
	 \label{Fig:FTF_conical_cur_ratio}
	 \vspace*{00pt}
\end{figure}

In the case of the uniform velocity perturbation model, the flame curvature serves as a measure of the overall stretch, as demonstrated in Section \ref{subsubsec:conical} A. 
Therefore, Figs.~\ref{Fig:FTF_conical_curvature_C&I} and \ref{Fig:FTF_conical_cur_ratio} only present the FTFs and gain difference ratios $\mathcal{G}$ for different flame geometries taking into account or neglecting flame curvature under the convective and incompressible velocity perturbation models.
As the FTF neglecting the flame curvature effect is also the function of the unstretched flame aspect ratio $\beta$ (see Table \ref{TableFTF_ana} for its analytical solution), the  FTFs   for $\beta=2$ and $\beta=15$ are different. 
When $\beta$ is large, $\mathcal{G}$ is relatively high.
This can be attributed to the fact that with a longer flame length, considering curvature leads to a noticeable decrease in the absolute flame height.
Consequently, the surface area of the flame front reduces significantly, leading to a weakened flame front wrinkle counteracting effect.  As a result, the FTF gains increase.
At high frequencies, as the FTF gain $|\mathcal{F}_{none}|$ is much small, the FTF gain difference ratio  $\mathcal{G}$ can be very large. 
It is also observed that the FTFs gain for the incompressible velocity perturbation model can even exceed unity due to the radial velocity shown in Fig.~\ref{Fig:FTF_conical_curvature_C&I}(b). However, $\mathcal{G}$ does not exhibit a substantial increase since the analytical solution without considering curvature already has a relatively large gain.
Different trends are found for the shorter flames. 
The FTF gains for short flames ($\beta = 2$) decrease more rapidly compared to long flames ($\beta = 15$) and approach the case of not considering curvature.
This can be attributed to the smaller variation in flame surface area, which leads to the wrinkle counteracting effect similar to that of the analytical solution. However, it may cause the FTF gain difference ratio  $\mathcal{G}$ features periodic motions and oscillates with different $\beta$ and $R$. 
Consistent with the flame stretch results, the large $\mathcal{G}$ in the case of small size flames is still related to the significant decrease in the steady flame height.

The FTF phase for the small size flame is more sensitive to the curvature effect. 
It should be noted that flame height ratio $L_f/L_{f}^0$ shrinks more for a smaller flame, leading to a shorter time for the disturbance convected from the burner rim to the flame tip and further smaller phase lag. 
As the flame length when $\beta$ = 2 and $R$ = 2 mm greatly shrinks, the time delay nearly equals zero and an approximately zero FTF phase can be observed.

\subsubsection*{C. The effect of flow strain}
\label{sec: conical_strain}

\begin{figure}[!t]
	\centering
		\subfigure
  		{
\includegraphics[height=5.5cm]{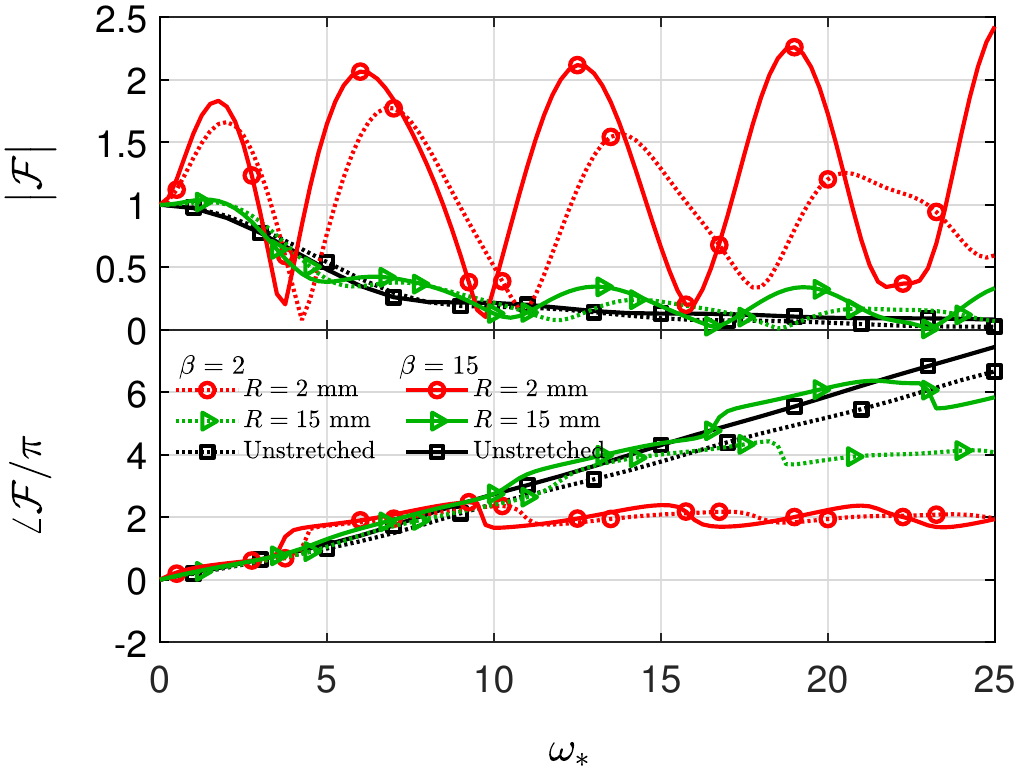}
  		}\put (-200,5) {\normalsize  $\displaystyle(a)$} 
  		\vspace*{0pt}
  		\hspace*{0pt}
  		\subfigure
  		  		{
\includegraphics[height=5.5cm]{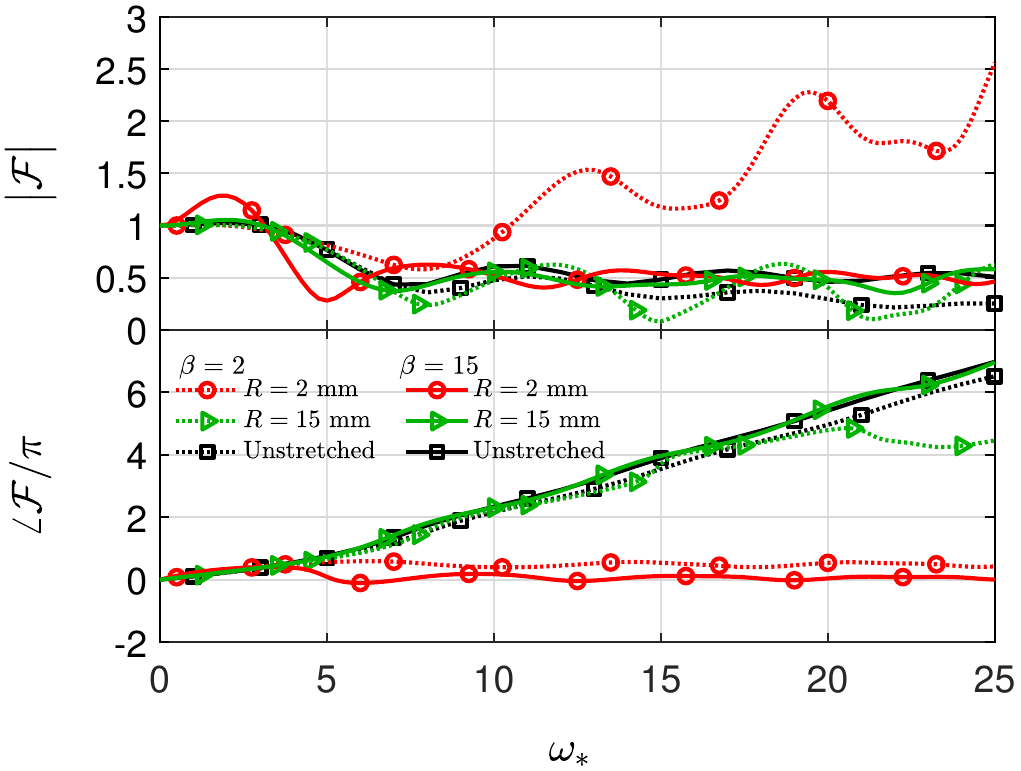}
  		}\put (-200,5) {\normalsize  $\displaystyle(b)$} 
  		\vspace*{0pt}
  		\hspace*{0pt}
 	 	 \caption{FTFs of the conical flames for different flame geometries when the flow strain is considered or not; the (a) convective velocity perturbation model and (b) incompressible velocity perturbation model are considered, respectively.   }
	 \label{Fig:FTF_conical_strain_C&I}
	 \vspace*{00pt}
\end{figure}

\begin{figure}[!t]
	\centering
  		\subfigure
  		{
\includegraphics[height=4.5cm]{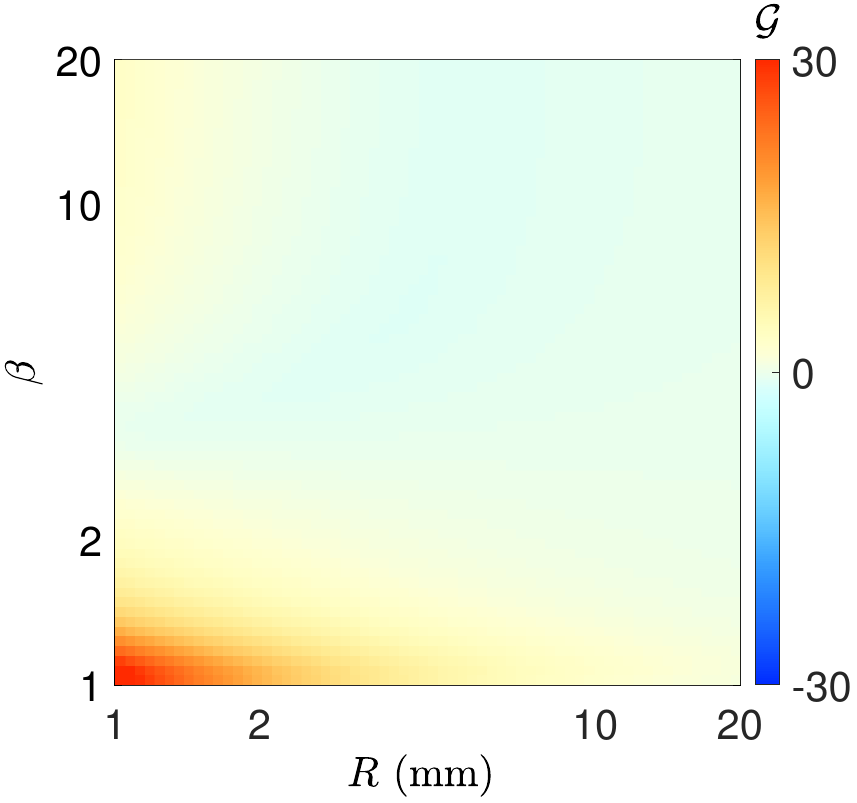}
  		}\put (-140,0) {\normalsize  $\displaystyle(a.1)$} 
  		\vspace*{0pt}
  		\hspace*{0pt}
  		\subfigure
  		  		{
\includegraphics[height=4.5cm]{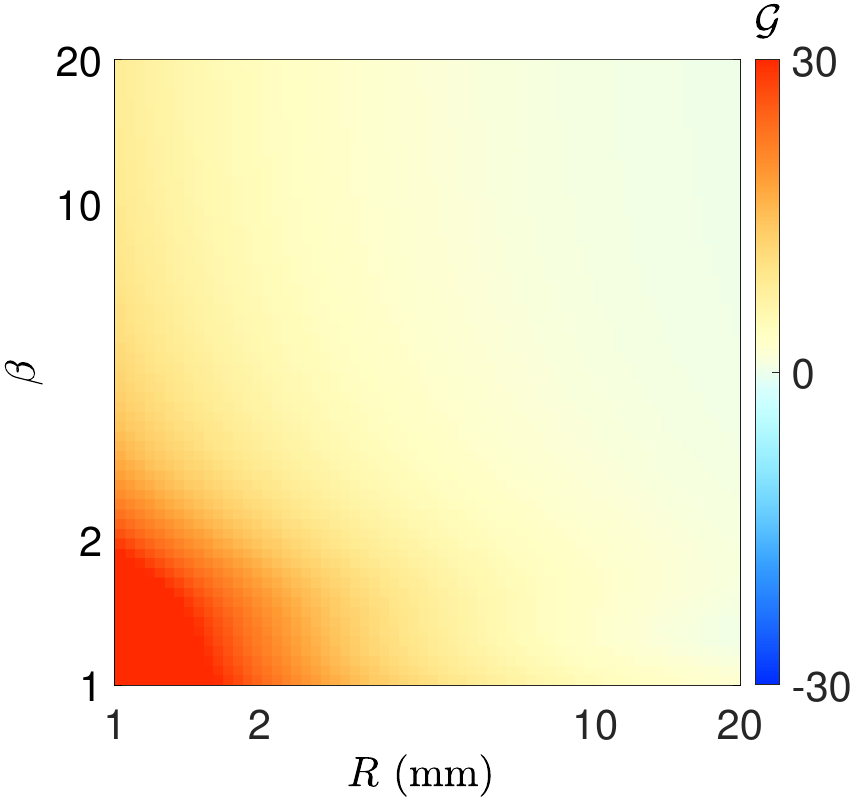}
  		}\put (-140,0) {\normalsize  $\displaystyle(a.2)$} 
  		\vspace*{0pt}
  		\hspace*{0pt}
  		\subfigure
  		  		{
\includegraphics[height=4.5cm]{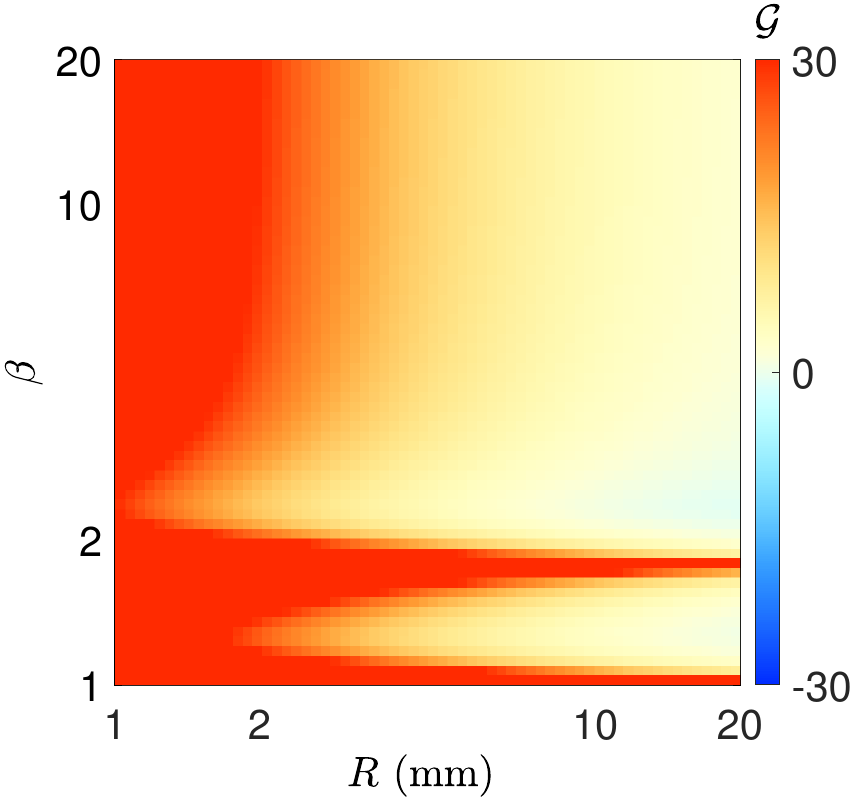}
  		}\put (-140,0) {\normalsize  $\displaystyle(a.3)$} \\
  		\vspace*{0pt}
  		\hspace*{0pt}
  		\subfigure
  		{
\includegraphics[height=4.5cm]{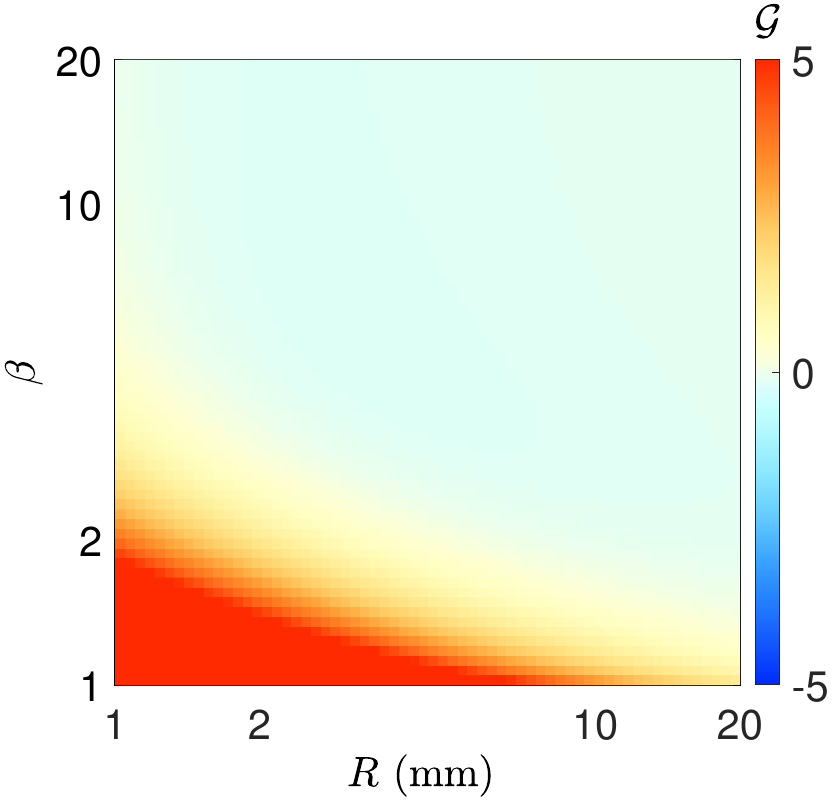}
  		}\put (-140,0) {\normalsize  $\displaystyle(b.1)$} 
  		\vspace*{0pt}
  		\hspace*{0pt}
  		\subfigure
  		  		{
\includegraphics[height=4.5cm]{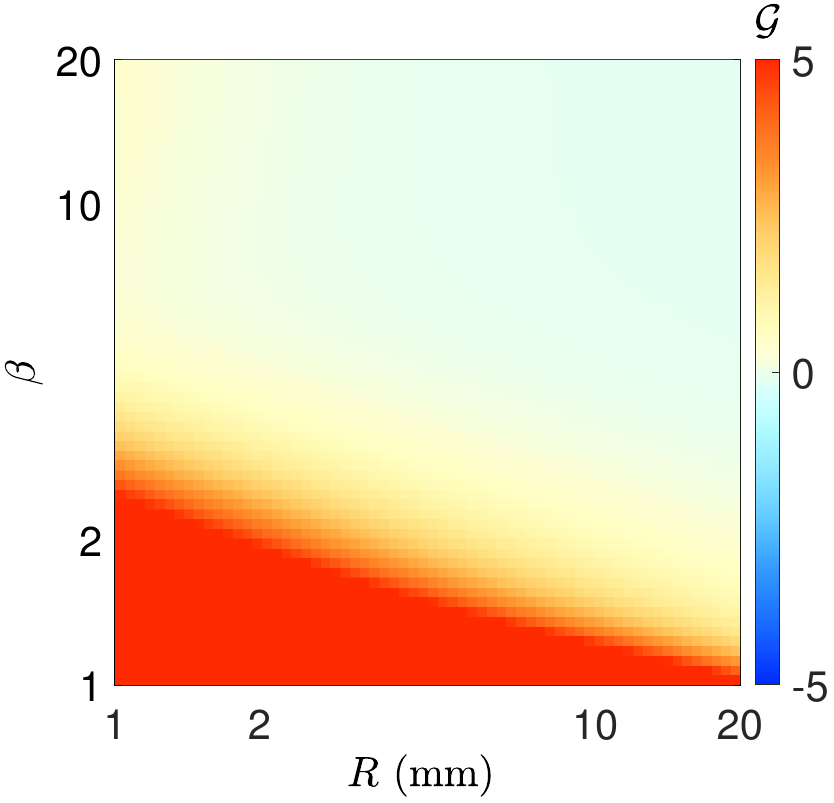}
  		}\put (-140,0) {\normalsize  $\displaystyle(b.2)$} 
  		\vspace*{0pt}
  		\hspace*{0pt}
  		\subfigure
  		  		{
\includegraphics[height=4.5cm]{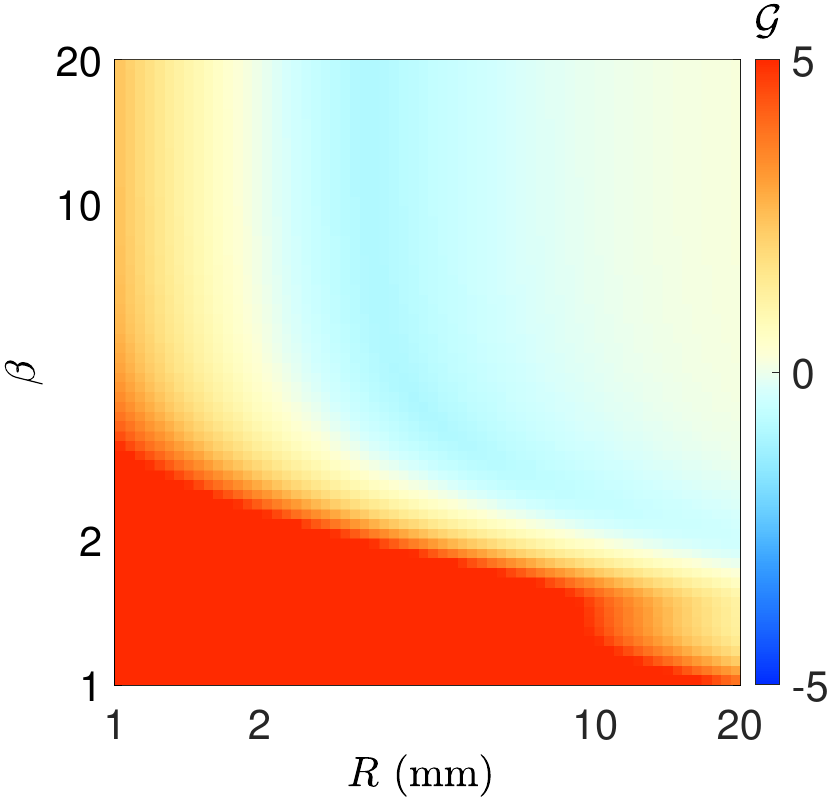}
  		}\put (-140,0) {\normalsize  $\displaystyle(b.3)$} \\
  		\vspace*{0pt}
  		\hspace*{0pt}
 	  	 \caption{Contours of the FTF gain difference ratios $\mathcal{G}$  of conical flames as functions of  $\beta$ and $R$ considering only the flow strain effect  for the (a) convective velocity perturbation model and (b) incompressible velocity perturbation model. The normalized frequencies are (1) ${\omega}_{\ast}$ = 10, (2) ${\omega}_{\ast}$ = 15 and (3) ${\omega}_{\ast}$ = 25. }
	 \label{Fig:FTF_conical_strain_ratio}
	 \vspace*{00pt}
\end{figure}

According to Eq.~(\ref{Eq:S_per_c}), the flow strain is zero for the uniform velocity perturbation model. 
One thus only discusses the results for the convective velocity perturbation model and the incompressible velocity perturbation model, e.g., as shown in Fig.~\ref{Fig:FTF_conical_strain_C&I}. 
The FTF gains for the small radius flames  ($R=2$ mm) are much larger than those for larger ones ($R=15$ mm) , indicating that the flow strain is significant for small radius flames.  
This can be further validated for more flame geometries, as shown in Fig.~\ref{Fig:FTF_conical_strain_ratio}. 
This is due to the fact that  $\hat{s}_L$ is mainly affected by the non-dimensional Markstein length ${\mathcal{L}}= \widetilde{\mathcal{L}}/R$, which is clearly inversely proportional to $R$. 
Due to the expression of the flow strain, the compensation term of the flame speed  $\mathcal{L}\hat{\mathcal{S}}$ features periodic motion with increasing normalized frequency $\omega_\ast$, leading to  similar oscillations for the FTF gain and phase, as shown in Fig.~\ref{Fig:FTF_conical_strain_C&I}. 
From the mathematical expression for $\mathcal{L}\hat{\mathcal{S}}$ (Eq.~\eqref{Eq:S_per_c}), this compensation term increases with  $\omega_\ast$. 
This is however non-physical; actually, in real flames, the velocity perturbations can be damped during their propagation and the damping effect arises with increasing the frequency \citep{Birbaud_CNF_2006}. 
Furthermore, the convection ratio $K$ also decreases with increasing frequency \citep{Birbaud_CNF_2006,Jiang_AIAA_2022}, which further reduces the value of the compensation term $\mathcal{L}\hat{\mathcal{S}}$ and the FTF gain at high frequencies. 
For the incompressible velocity perturbation model, the introduction of radial velocity perturbation  in the compensation term is more effective in counterbalancing the velocity perturbations induced by flow strain for large $\beta$, thereby resulting in a reduction in FTF gain. 
Therefore, the effect of  flow strain is thus affected by the combination of $R$ and $\beta$.

\subsubsection{V-flame }
\label{subsubsec:V}

\subsubsection*{A. The effect of flame stretch}

\begin{figure}[!t]
	\centering
		\subfigure
  		{
\includegraphics[height=5.5cm]{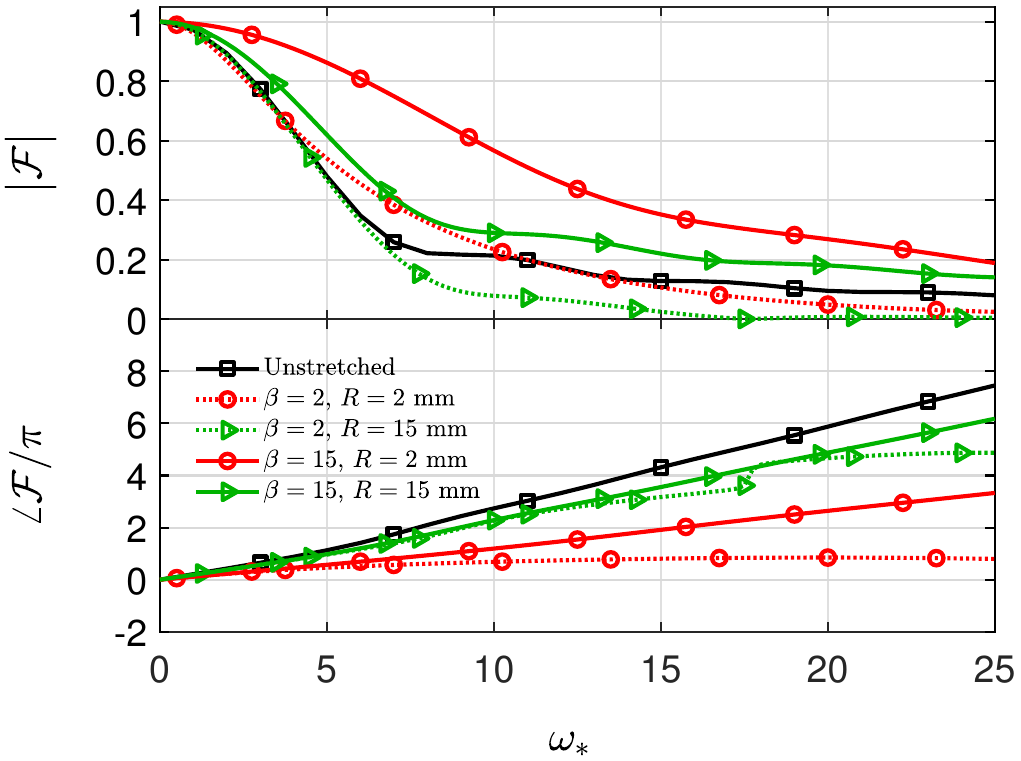}
  		}\put (-200,5) {\normalsize  $\displaystyle(a)$} \\
  		\vspace*{0pt}
  		\hspace*{0pt}
  		\subfigure
  		  		{
\includegraphics[height=5.5cm]{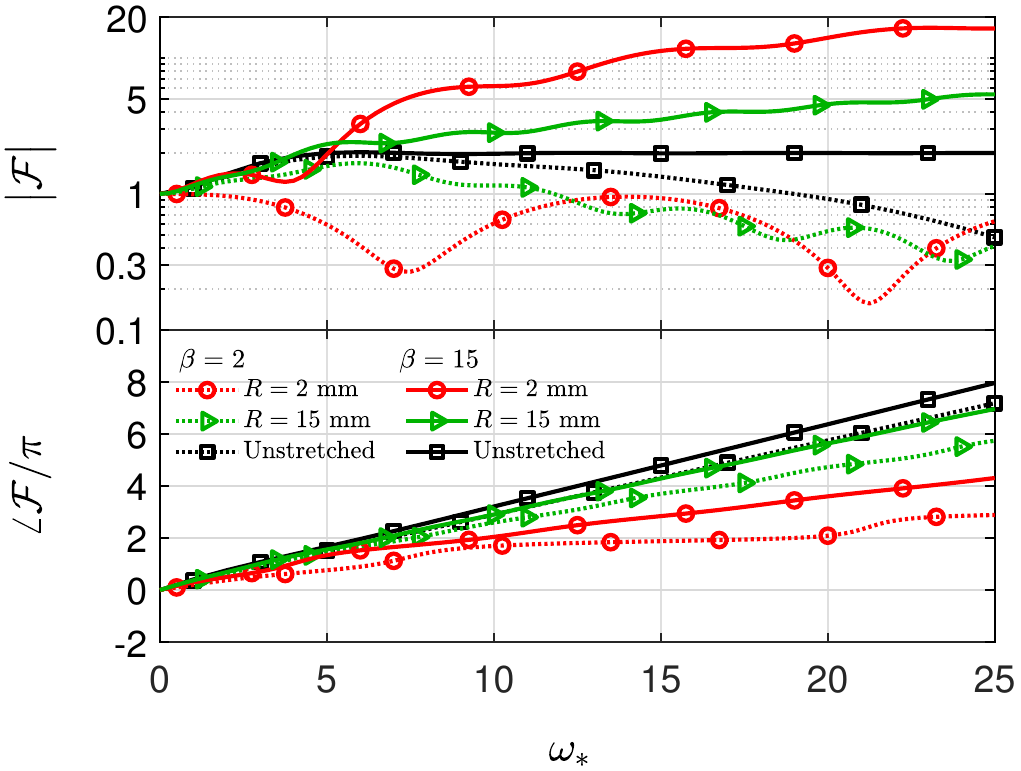}
  		}\put (-200,5) {\normalsize  $\displaystyle(b)$} 
  		\vspace*{0pt}
  		\hspace*{0pt}
  		\subfigure
  		  		{
\includegraphics[height=5.5cm]{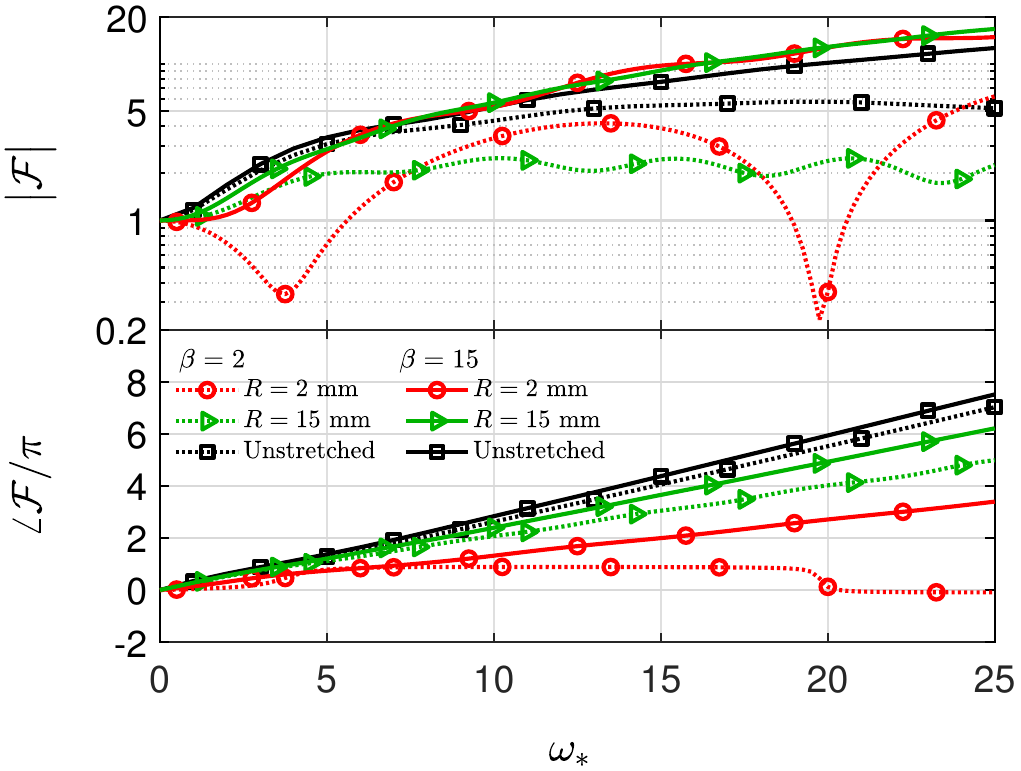}
  		}\put (-200,5) {\normalsize  $\displaystyle(c)$} \\
  		\vspace*{0pt}
  		\hspace*{0pt} 
  		 	 \caption{FTFs of the V-flames for different flame geometries when the flame stretch is considered or not; the (a) uniform velocity perturbation model, (b) convective velocity perturbation model and (c) incompressible velocity perturbation model are considered, respectively. }
	 \label{Fig:FTF_Vflame_stretch}
	 \vspace*{00pt}
\end{figure}

\begin{figure}[!t]
	\centering
		\subfigure
  		{
\includegraphics[height=4.5cm]{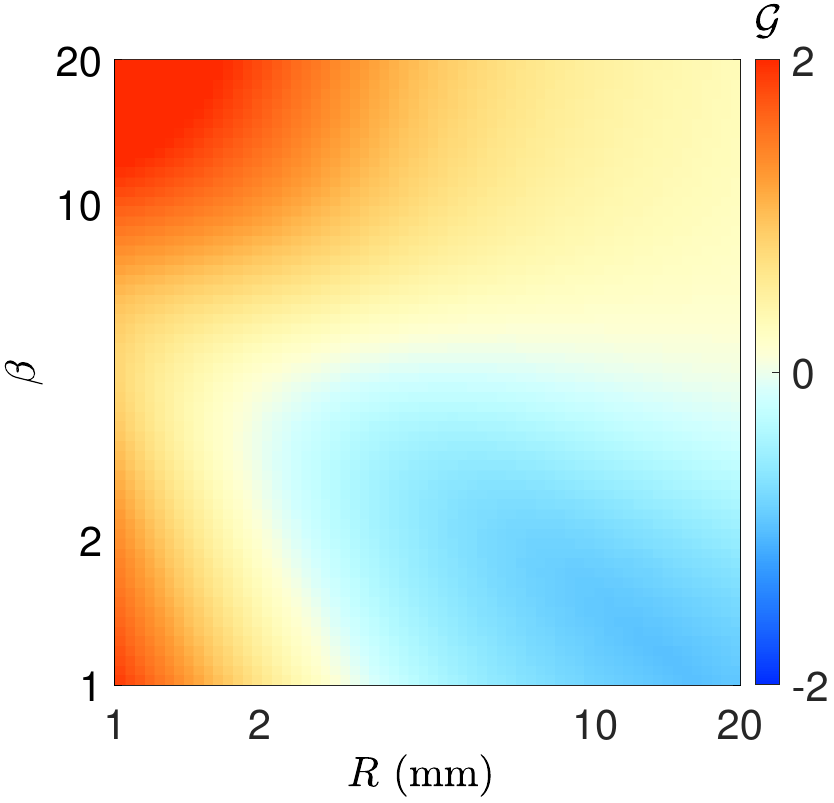}
  		}\put (-140,0) {\normalsize  $\displaystyle(a.1)$} 
  		\vspace*{0pt}
  		\hspace*{0pt}
  		\subfigure
  		  		{
\includegraphics[height=4.5cm]{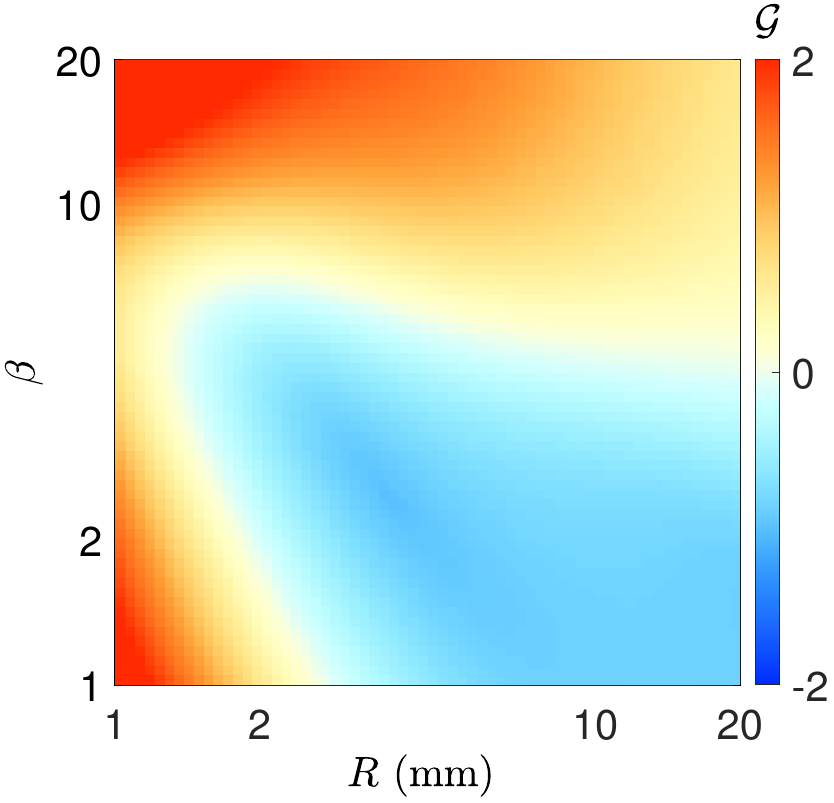}
  		}\put (-140,0) {\normalsize  $\displaystyle(a.2)$} 
  		\vspace*{0pt}
  		\hspace*{0pt}
  		\subfigure
  		  		{
\includegraphics[height=4.5cm]{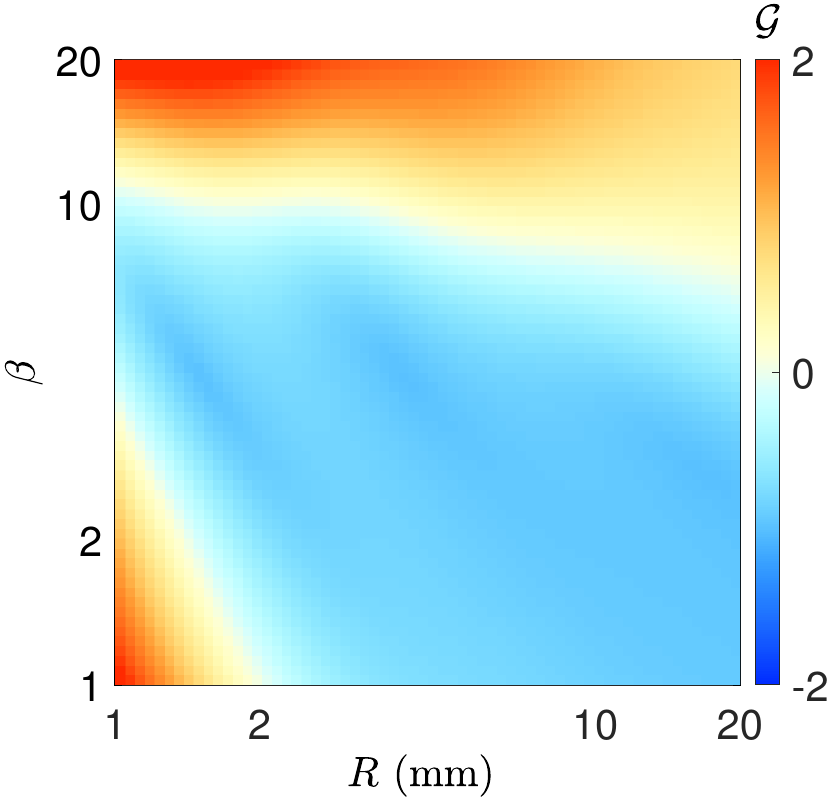}
  		}\put (-140,0) {\normalsize  $\displaystyle(a.3)$} \\
  		\vspace*{0pt}
  		\hspace*{0pt}
  		\subfigure
  		{
\includegraphics[height=4.5cm]{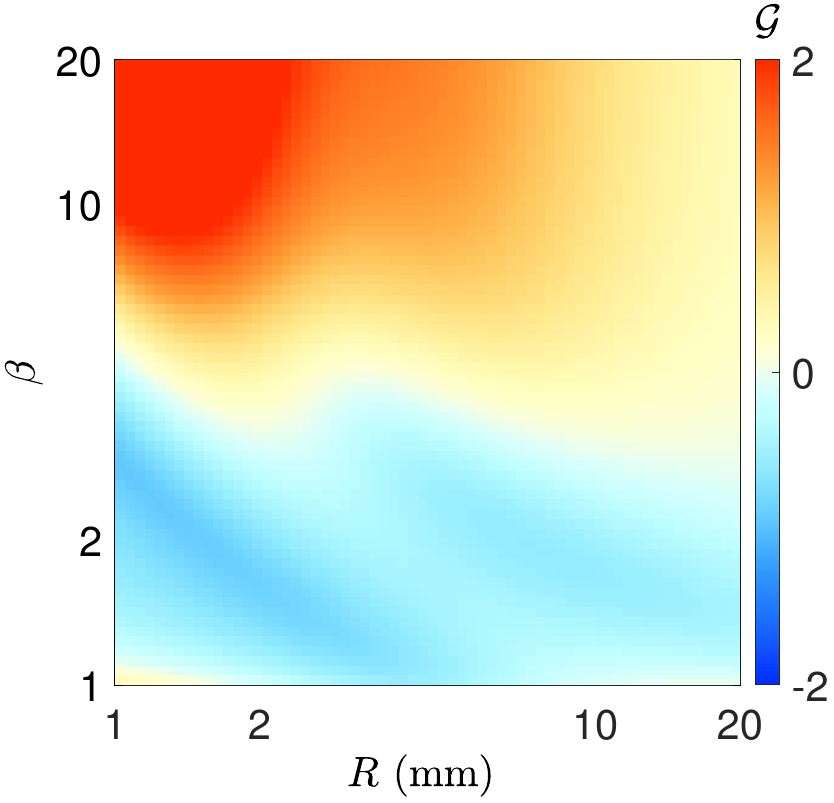}
  		}\put (-140,0) {\normalsize  $\displaystyle(b.1)$} 
  		\vspace*{0pt}
  		\hspace*{0pt}
  		\subfigure
  		  		{
\includegraphics[height=4.5cm]{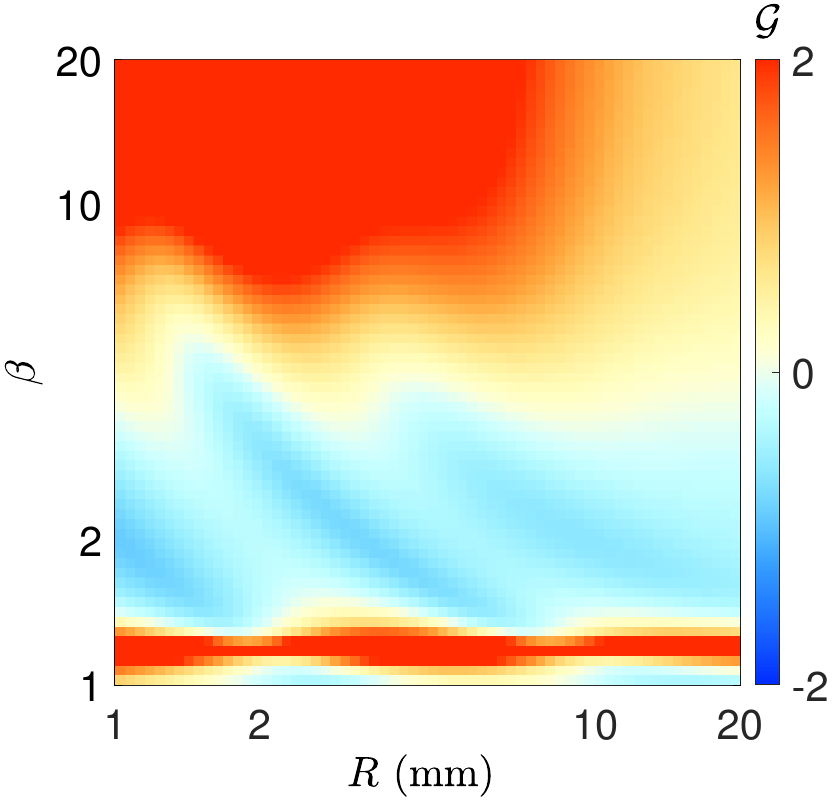}
  		}\put (-140,0) {\normalsize  $\displaystyle(b.2)$} 
  		\vspace*{0pt}
  		\hspace*{0pt}
  		\subfigure
  		  		{
\includegraphics[height=4.5cm]{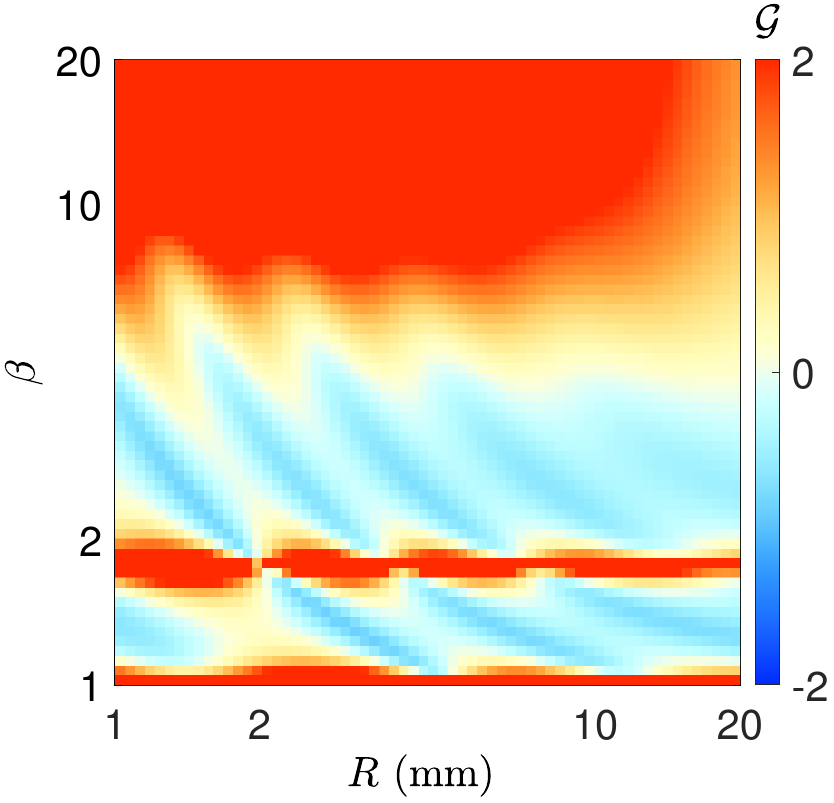}
  		}\put (-140,0) {\normalsize  $\displaystyle(b.3)$} \\
  		\vspace*{0pt}
  		\hspace*{0pt}
  		\subfigure
  		{
\includegraphics[height=4.5cm]{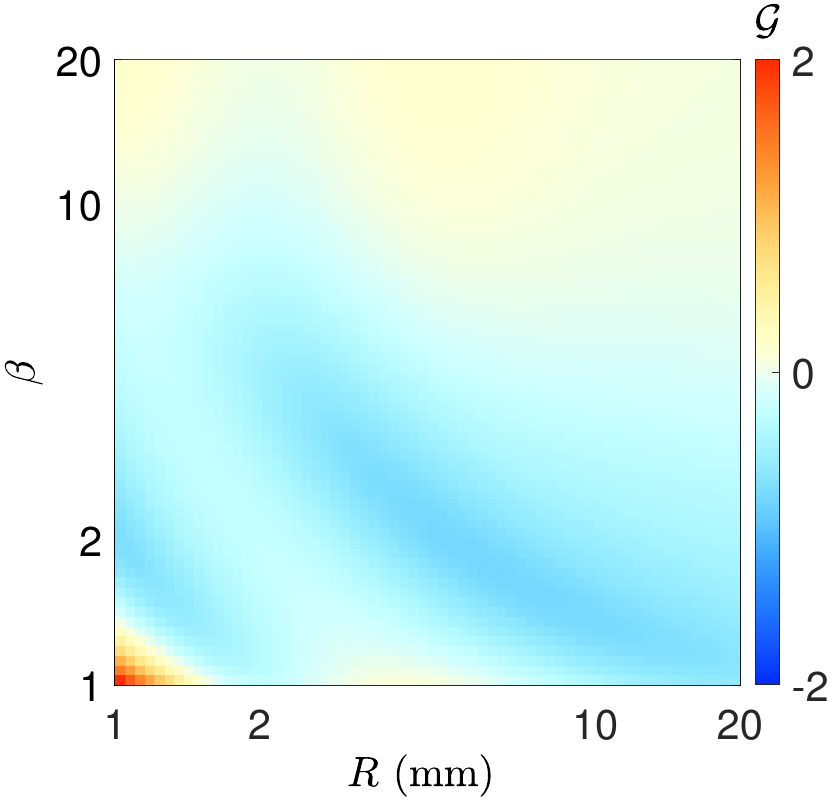}
  		}\put (-140,0) {\normalsize  $\displaystyle(c.1)$} 
  		\vspace*{0pt}
  		\hspace*{0pt}
  		\subfigure
  		  		{
\includegraphics[height=4.5cm]{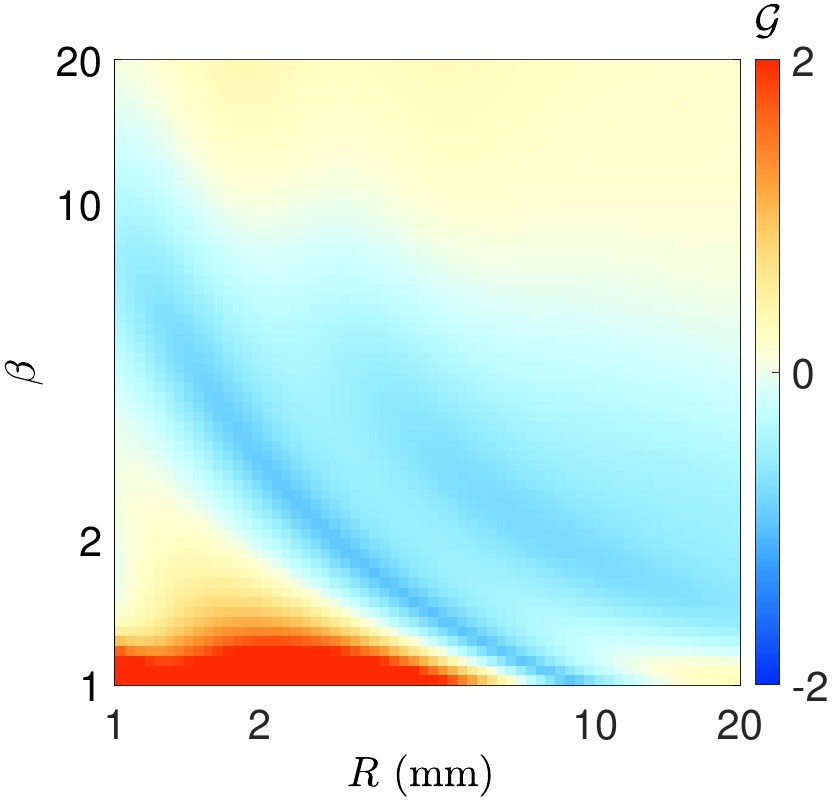}
  		}\put (-140,0) {\normalsize  $\displaystyle(c.2)$} 
  		\vspace*{0pt}
  		\hspace*{0pt}
  		\subfigure
  		  		{
\includegraphics[height=4.5cm]{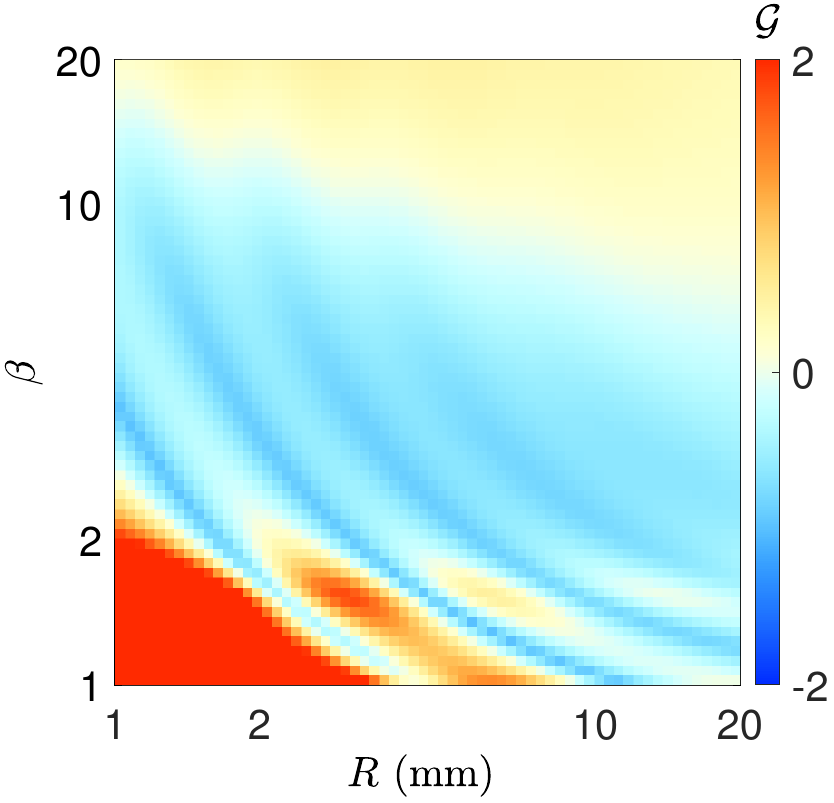}
  		}\put (-140,0) {\normalsize  $\displaystyle(c.3)$} \\
  		\vspace*{0pt}
  		\hspace*{0pt}
 	 \caption{Contours of the FTF gain difference ratios $\mathcal{G}$  of V-flames as functions of  $\beta$ and $R$ considering the flame stretch effect for the (a) uniform velocity perturbation model, (b) convective velocity perturbation model and (c) incompressible velocity perturbation model. The normalized frequencies are (1) ${\omega}_{\ast}$ = 10, (2) ${\omega}_{\ast}$ = 15 and (3) ${\omega}_{\ast}$ = 25. }
	 \label{Fig:FTF_V_curvature_ratio}
	 \vspace*{00pt}
\end{figure}

Figures~\ref{Fig:FTF_Vflame_stretch} and \ref{Fig:FTF_V_curvature_ratio} show the FTFs and gain difference ratios $\mathcal{G}$ for the V-flames considering flame stretch under three kinds of flow velocity perturbation models.
It is evident that when $\beta$ is large, the FTF gains considering stretch are larger than those of the analytical solution without stretch, whereas for small $\beta$, the opposite conclusion holds true.
The reduction in gains can be primarily attributed to the larger perturbations at the free end of the V-flame.
For the uniform velocity perturbation model shown in Fig.~\ref{Fig:FTF_Vflame_stretch}(a), the FTF gain still exhibits the characteristics of a low-pass filter.
The gain of FTF with $\beta = 2$, when considering the effect of stretch, exhibits slightly lower magnitudes than the unstretched analytical solution at low frequencies, consistent with \citep{Wang_CNF_2009} and \citep{Preetham_JPP_2010}. 
While the variation of FTF gain at high frequencies is also influenced by $R$. Furthermore, the effect of flame stretch on the FTF gain sounds for smaller $\beta$ and $R$ shown in Fig.~\ref{Fig:FTF_V_curvature_ratio}(a), which is consistent with those for the conical flame. 
For the convective and incompressible velocity perturbation model shown in Fig.~\ref{Fig:FTF_Vflame_stretch}(b) and (c), it should also be noted that, similar to the FTFs of conical flames, the FTFs (see the analytical solutions for the V-flames in Table \ref{TableFTF_ana}) are both dependent on $\xi$ and vary for different $\beta$ neglecting the stretch effect.  
Specifically, the FTF gain typically features the shape of single-hump, exceeds unity and can be very large within a certain frequency range, and when considering stretch, the FTF gain is also relatively large.  
Therefore, the magnitude of $\mathcal{G}$, as reflected in Fig.~\ref{Fig:FTF_V_curvature_ratio}(b) and (c), is similar to that of the uniform velocity perturbation model.
While the analytical solution for the incompressible velocity perturbation model is even larger, leading to a comparatively smaller gain difference ratio $\mathcal{G}$.
Furthermore, the phase variation is related to the decrease in flame height due to flame stretch.
Different from the conical flame,  the FTF phase of the V-flames typically does not saturate with increasing ${\omega}_{\ast}$. 
To explore more profound influences of the flame stretch, the effects of  flame curvature and flow strain for the V-flames are discussed separately in the following analysis. 

\subsubsection*{B. The effect of flame curvature}

\begin{figure}[!t]
	\centering
		\subfigure
  		{
\includegraphics[height=5.5cm]{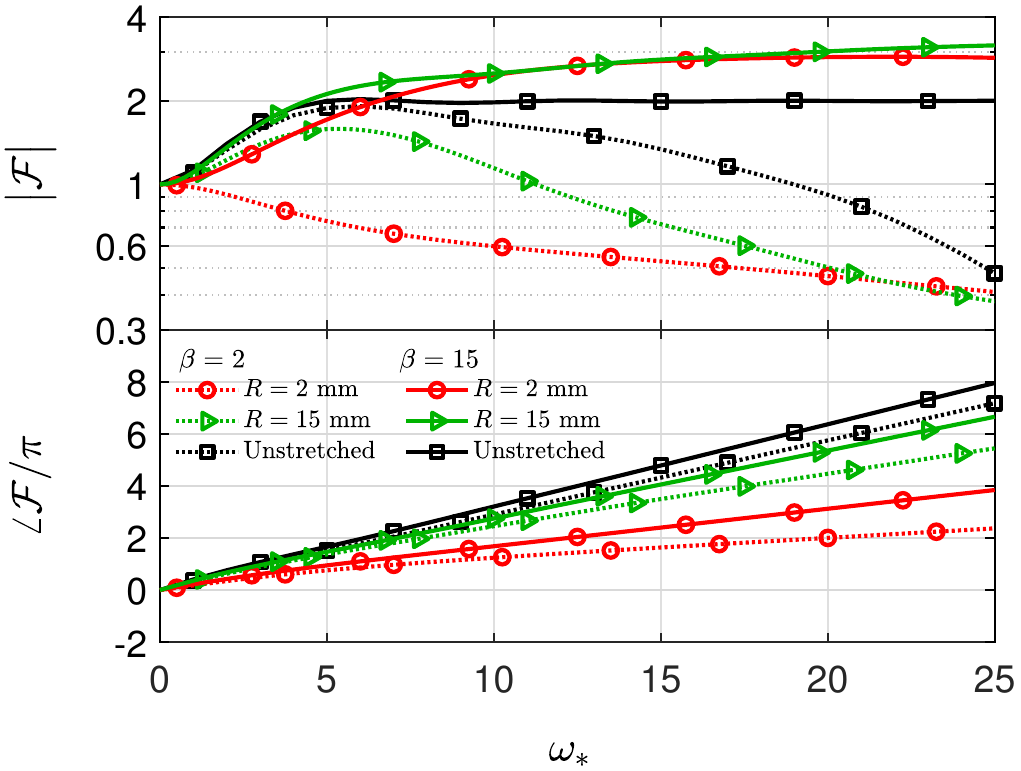}
  		}\put (-200,5) {\normalsize  $\displaystyle(a)$} 
  		\vspace*{0pt}
  		\hspace*{0pt}
  		\subfigure
  		  		{
\includegraphics[height=5.5cm]{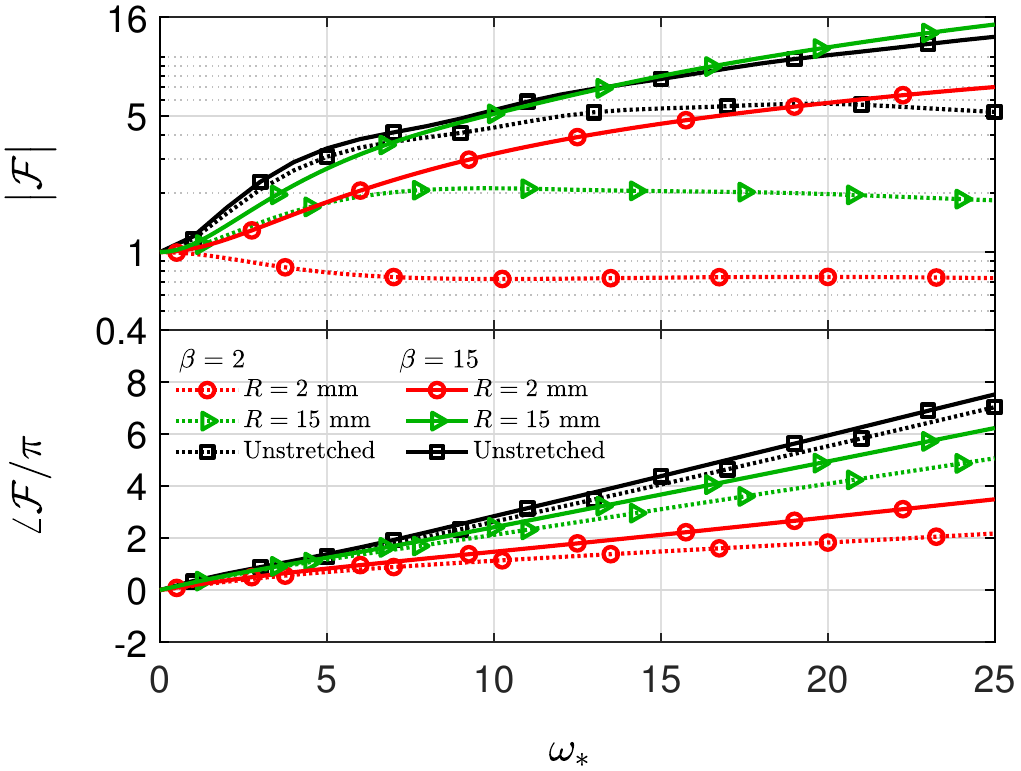}
  		}\put (-200,5) {\normalsize  $\displaystyle(b)$} 
  		\vspace*{0pt}
  		\hspace*{0pt}
  		 	 \caption{FTFs of the V-flames for different flame geometries when the flame curvature is considered or not; the (a) convective velocity perturbation model and (b) incompressible velocity perturbation model are considered, respectively.   }
	 \label{Fig:FTF_V_curvature_C&I}
	 \vspace*{00pt}
\end{figure}

\begin{figure}[!t]
	\centering
  		\subfigure
  		{
\includegraphics[height=4.5cm]{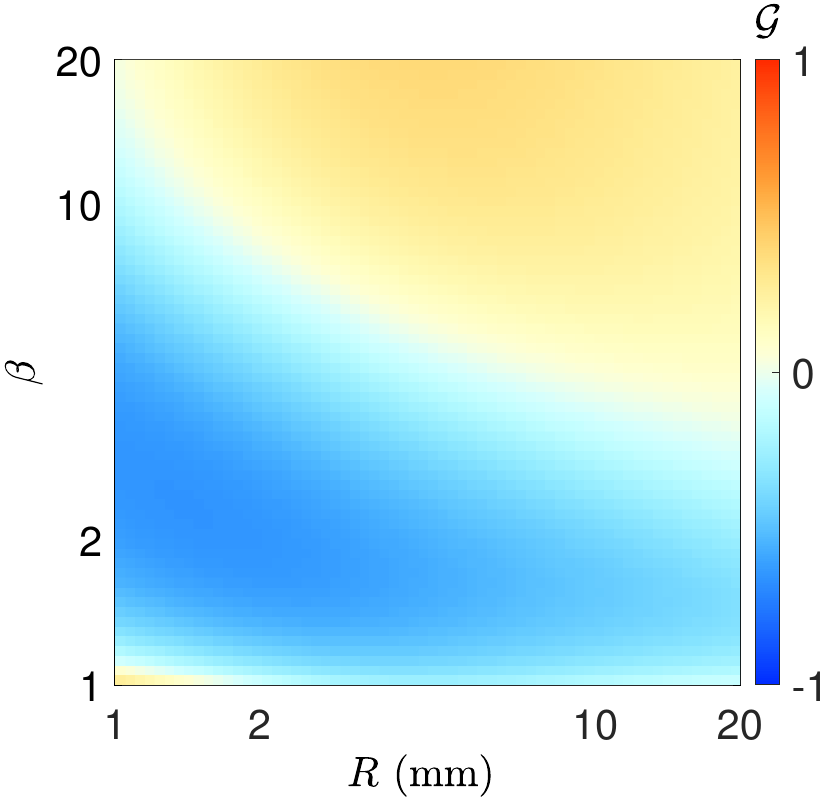}
  		}\put (-140,0) {\normalsize  $\displaystyle(a.1)$} 
  		\vspace*{0pt}
  		\hspace*{0pt}
  		\subfigure
  		  		{
\includegraphics[height=4.5cm]{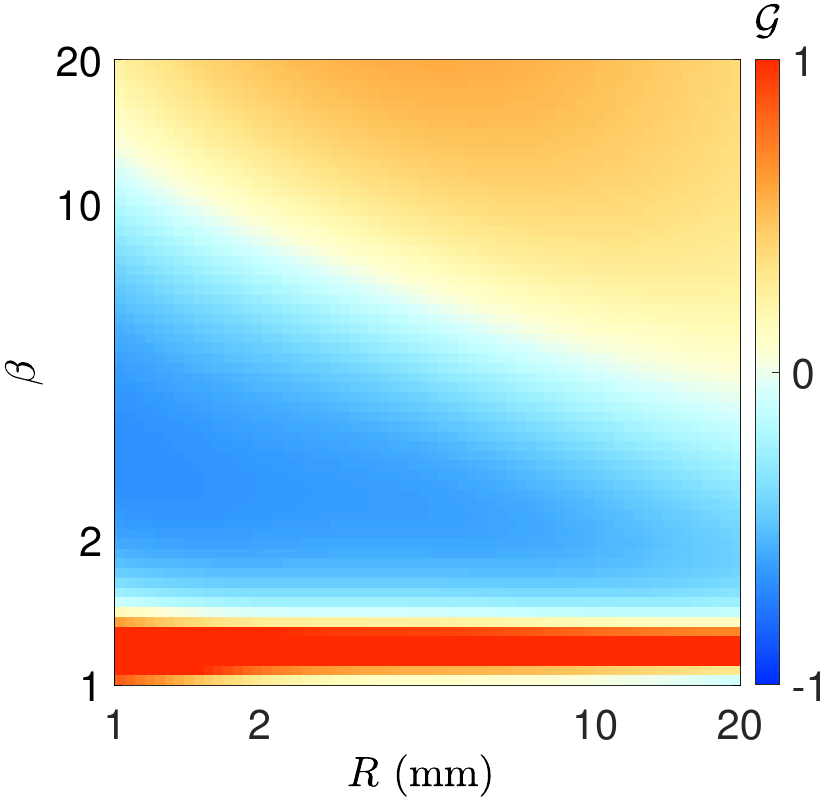}
  		}\put (-140,0) {\normalsize  $\displaystyle(a.2)$} 
  		\vspace*{0pt}
  		\hspace*{0pt}
  		\subfigure
  		  		{
\includegraphics[height=4.5cm]{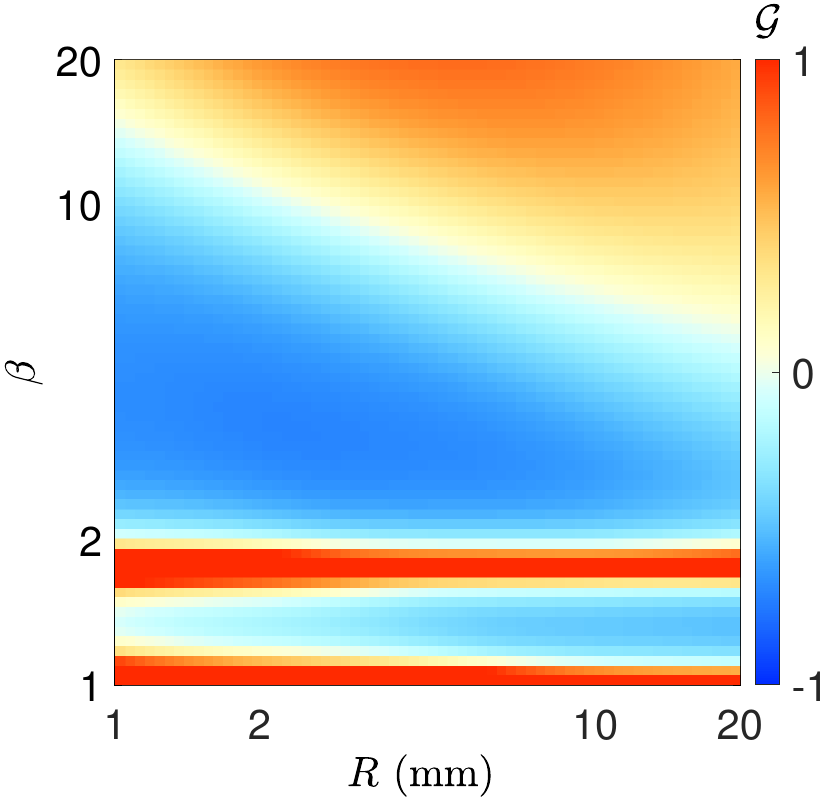}
  		}\put (-140,0) {\normalsize  $\displaystyle(a.3)$} \\
  		\vspace*{0pt}
  		\hspace*{0pt}
  		\subfigure
  		{
\includegraphics[height=4.5cm]{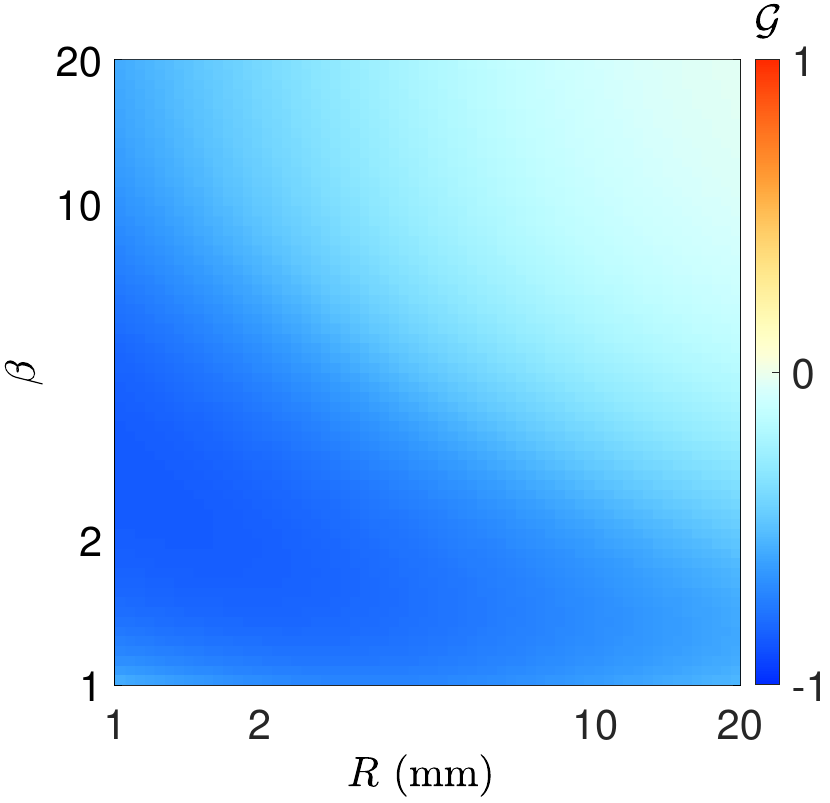}
  		}\put (-140,0) {\normalsize  $\displaystyle(b.1)$} 
  		\vspace*{0pt}
  		\hspace*{0pt}
  		\subfigure
  		  		{
\includegraphics[height=4.5cm]{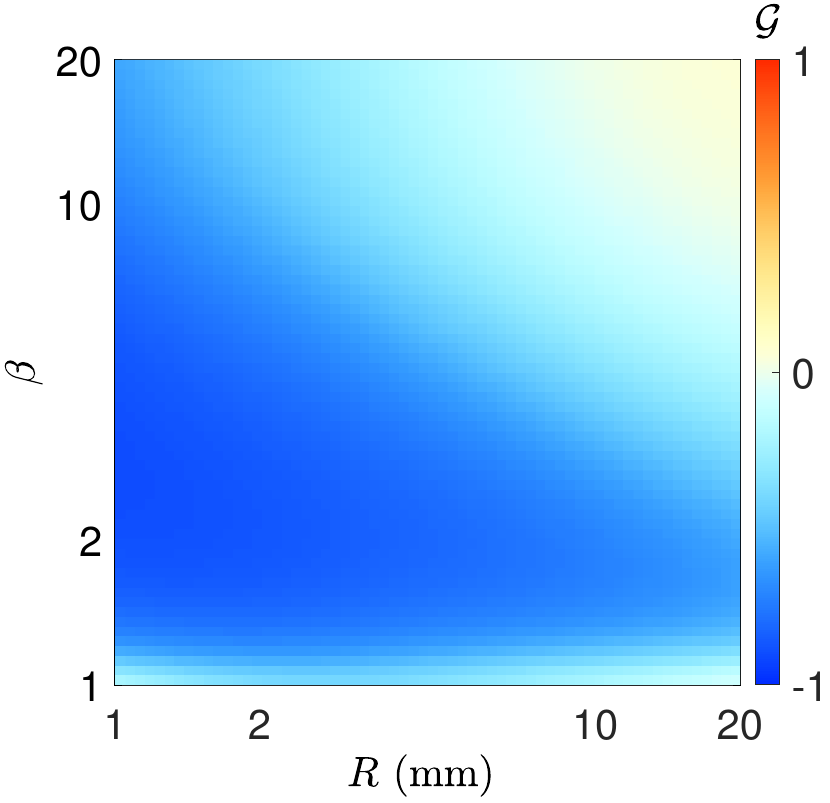}
  		}\put (-140,0) {\normalsize  $\displaystyle(b.2)$} 
  		\vspace*{0pt}
  		\hspace*{0pt}
  		\subfigure
  		  		{
\includegraphics[height=4.5cm]{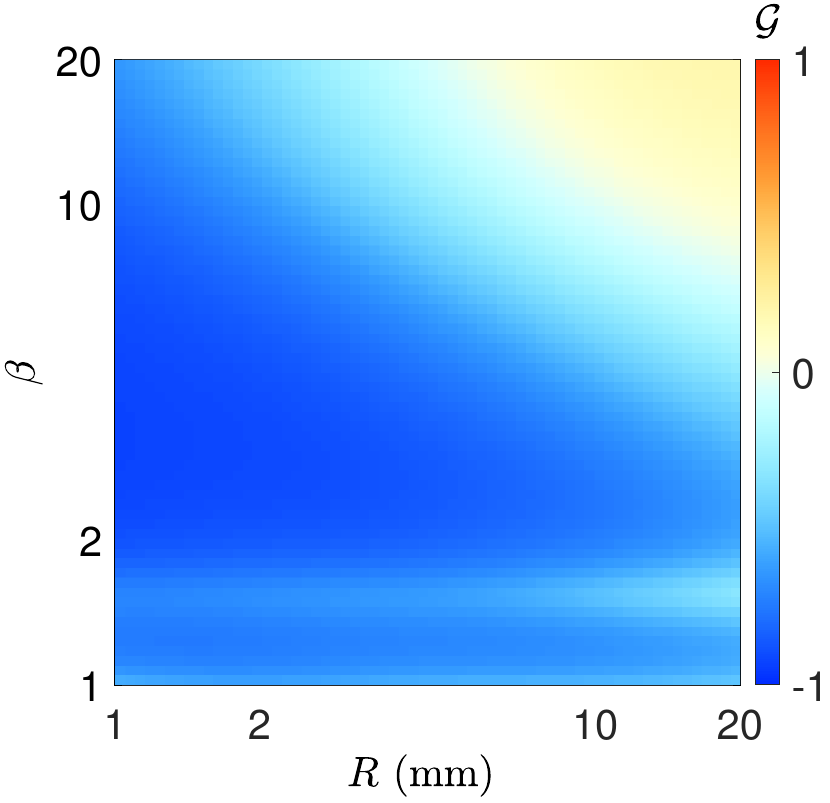}
  		}\put (-140,0) {\normalsize  $\displaystyle(b.3)$} \\
  		\vspace*{0pt}
  		\hspace*{0pt}
 	  	 \caption{Contours of the FTF gain difference ratios $\mathcal{G}$  of V-flames as functions of  $\beta$ and $R$ considering only the flame curvature effect  for the (a) convective velocity perturbation model and (b) incompressible velocity perturbation model. The normalized frequencies are (1) ${\omega}_{\ast}$ = 10, (2) ${\omega}_{\ast}$ = 15 and (3) ${\omega}_{\ast}$ = 25. }
	 \label{Fig:FTF_V_cur_ratio}
	 \vspace*{00pt}
\end{figure}

Figures~\ref{Fig:FTF_V_curvature_C&I} and \ref{Fig:FTF_V_cur_ratio} show the effect of flame curvature on the FTFs and gain difference ratios $\mathcal{G}$ for the convective and incompressible velocity perturbation models.
When the curvature effect is considered, the FTF gains for long flames ($\beta = 15$) generally exceed that without considering curvature effects, especially at high frequencies. 
However, the FTF gains are typically smaller than those neglecting the curvature effect for short flames ($\beta = 2$).
In the case of small radius flames ($R = 2$ mm, $\beta = 2$), the gains can even fall below unity.
The effect of  flame curvature can be further evaluated for more flame geometries, as shown in 
Fig.~\ref{Fig:FTF_V_cur_ratio}.  It is evident that flame curvature results in a reduction in the gain of the FTF across a significant range of flame sizes. 
This is primarily due to the fact that for V-flames, the oncoming perturbation propagates downward from the flame root to the flame edge. As the flame edge is a free boundary,  the  perturbation grows along the flame front and can attain a large perturbation amplitude when it reaches the edge. 
When the  flame curvature is considered, the flame  shrinks and the entire flame surface area  decreases. Due to the combined effect of the  large flame edge  perturbation and small flame surface area, the  wrinkle counteracting effect of the flame front is enhanced, and the FTF gain  drops fast with increasing  frequency and $\mathcal{G}$ becomes negative.
As the normalized frequency $\omega_\ast$ increases, the increased occurrence of wrinkles has strengthened the wrinkle counteracting effect. 
With gradually increasing the values of  $\beta$ and $R$, the flame becomes sufficiently long and large, leading to a weakening of the curvature effect and an increase in the surface area of the flame edge. 
However, the  perturbation at the flame edge can be further amplified. 
The effect of this amplification outweighs the wrinkle counteracting effect, resulting in an increase in the FTF gain and the $\mathcal{G}$ becoming positive.

\subsubsection*{C. The effect of flow strain}\label{sec: V_strain}
\begin{figure}[!t]
	\centering
		\subfigure
  		{
\includegraphics[height=5.5cm]{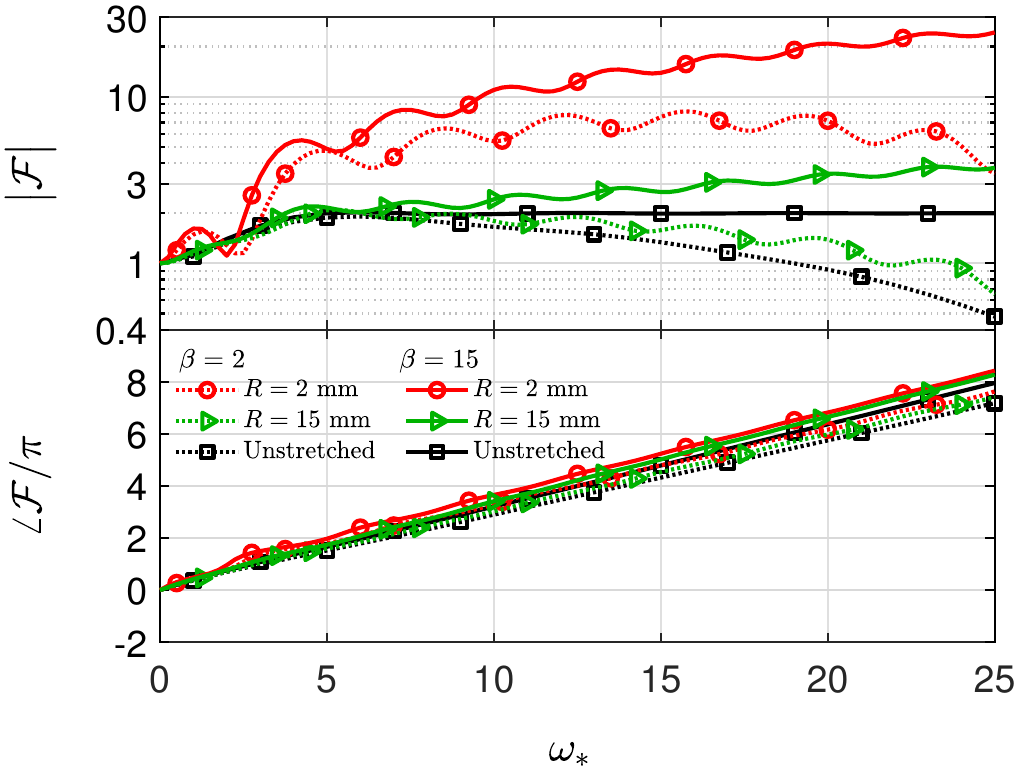}
  		}\put (-200,5) {\normalsize  $\displaystyle(a)$} 
  		\vspace*{0pt}
  		\hspace*{0pt}
  		\subfigure
  		  		{
\includegraphics[height=5.5cm]{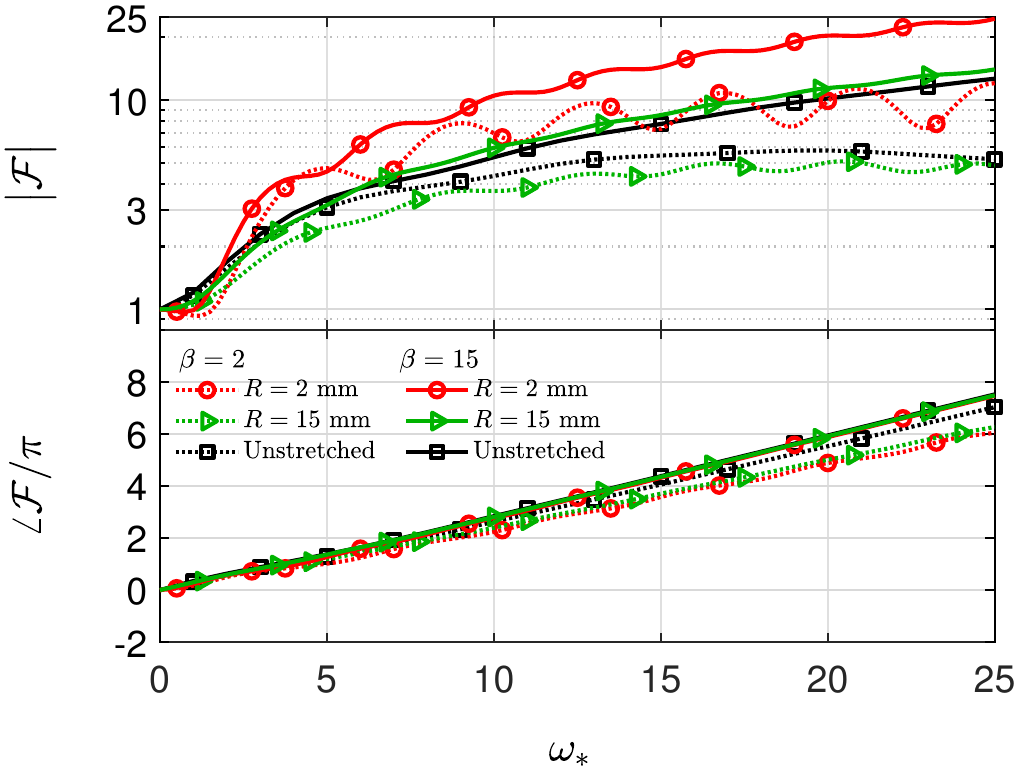}
  		}\put (-200,5) {\normalsize  $\displaystyle(b)$} 
  		\vspace*{0pt}
  		\hspace*{0pt}
 	 	 \caption{FTFs of the V-flames for different flame geometries when the flow strain is considered or not; the (a) convective velocity perturbation model and (b) incompressible velocity perturbation model are considered, respectively.   }
	 \label{Fig:FTF_V_strain_C&I}
	 \vspace*{00pt}
\end{figure}

\begin{figure}[!t]
	\centering
  		\subfigure
  		{
\includegraphics[height=4.5cm]{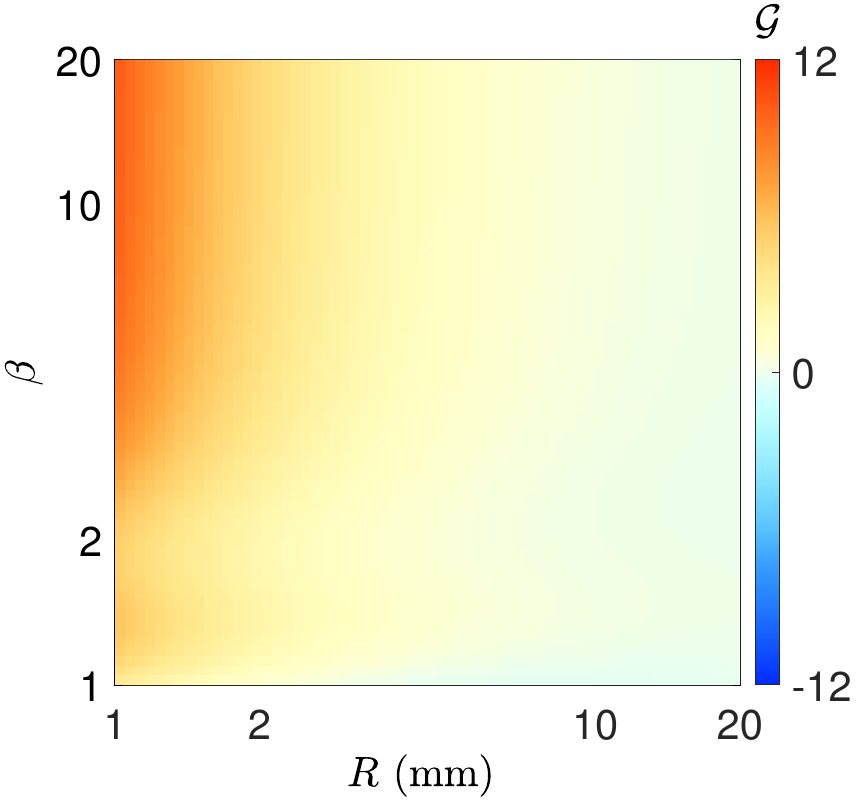}
  		}\put (-140,0) {\normalsize  $\displaystyle(a.1)$} 
  		\vspace*{0pt}
  		\hspace*{0pt}
  		\subfigure
  		  		{
\includegraphics[height=4.5cm]{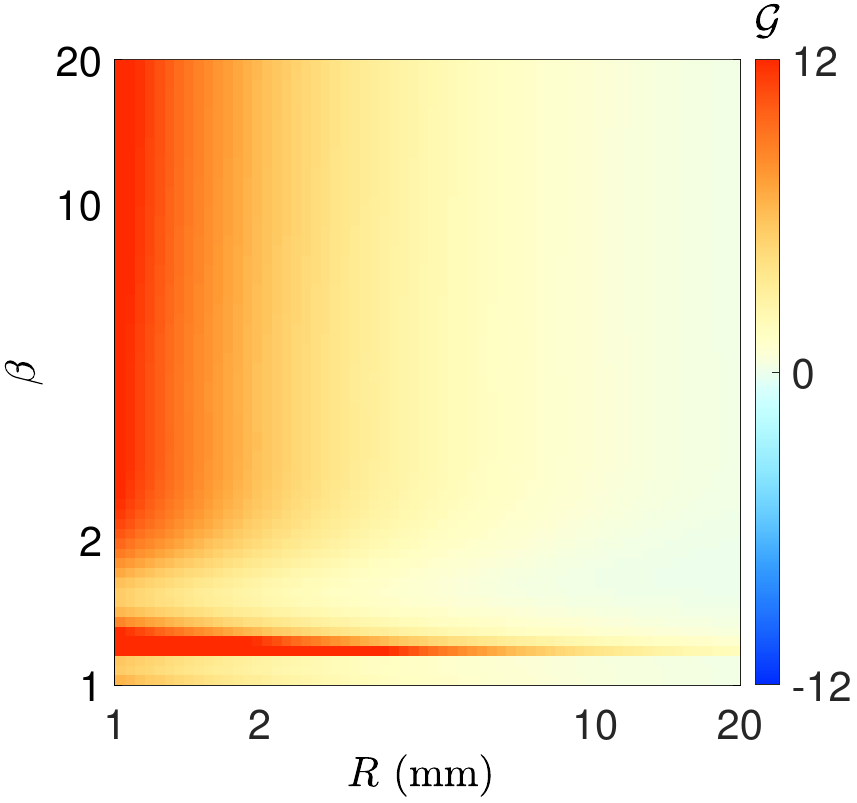}
  		}\put (-140,0) {\normalsize  $\displaystyle(a.2)$} 
  		\vspace*{0pt}
  		\hspace*{0pt}
  		\subfigure
  		  		{
\includegraphics[height=4.5cm]{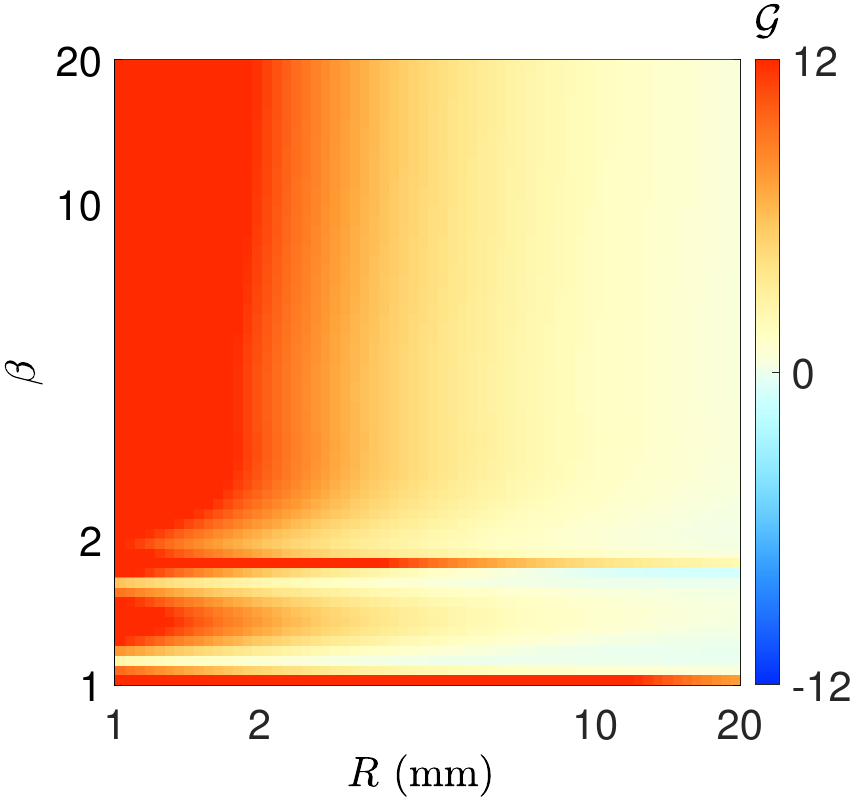}
  		}\put (-140,0) {\normalsize  $\displaystyle(a.3)$} \\
  		\vspace*{0pt}
  		\hspace*{0pt}
  		\subfigure
  		{
\includegraphics[height=4.5cm]{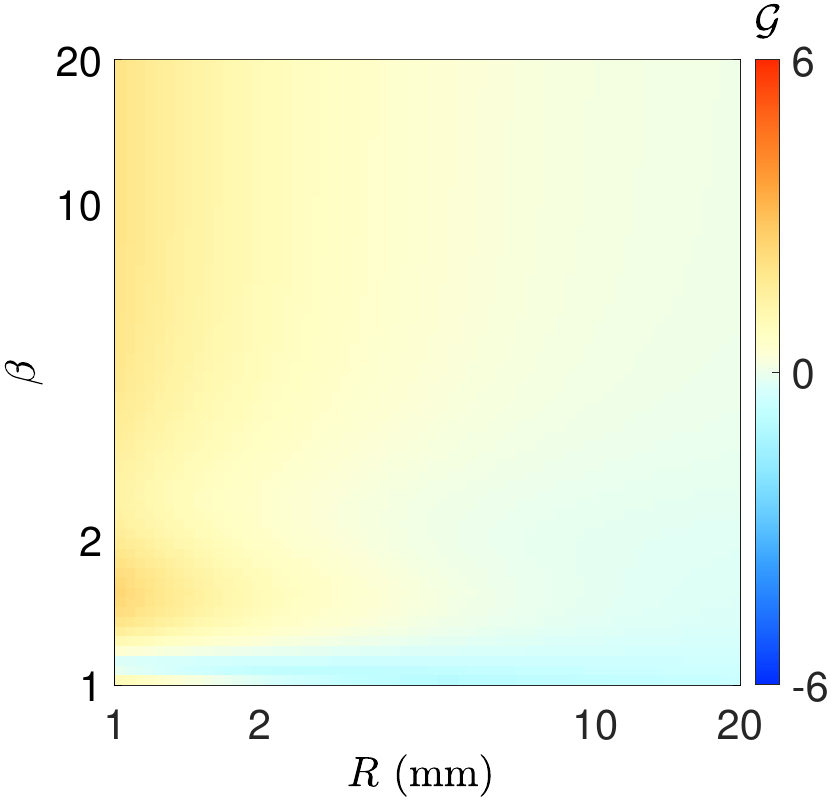}
  		}\put (-140,0) {\normalsize  $\displaystyle(b.1)$} 
  		\vspace*{0pt}
  		\hspace*{0pt}
  		\subfigure
  		  		{
\includegraphics[height=4.5cm]{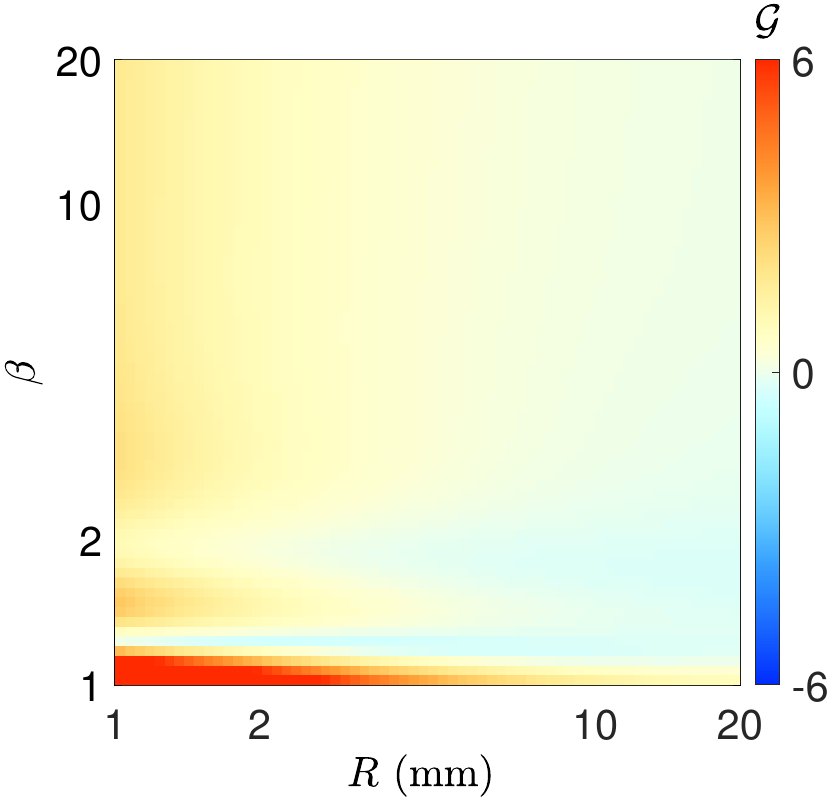}
  		}\put (-140,0) {\normalsize  $\displaystyle(b.2)$} 
  		\vspace*{0pt}
  		\hspace*{0pt}
  		\subfigure
  		  		{
\includegraphics[height=4.5cm]{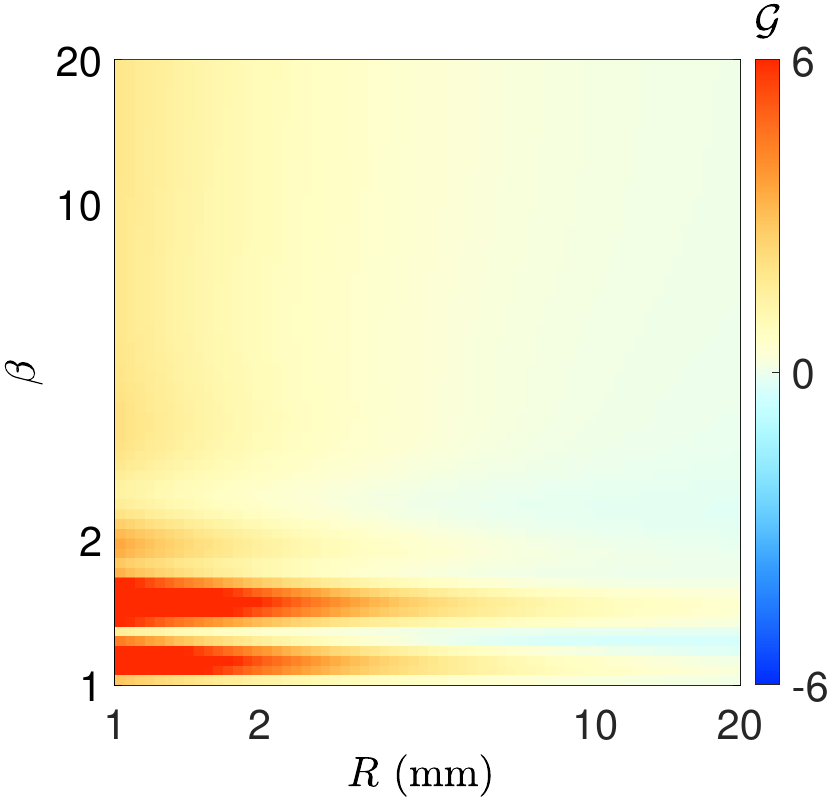}
  		}\put (-140,0) {\normalsize  $\displaystyle(b.3)$} \\
  		\vspace*{0pt}
  		\hspace*{0pt}
 	  	 \caption{Contours of the FTF gain difference ratios $\mathcal{G}$  of V-flames as functions of  $\beta$ and $R$ considering only the flow strain effect  for the (a) convective velocity perturbation model and (b) incompressible velocity perturbation model. The normalized frequencies are (1) ${\omega}_{\ast}$ = 10, (2) ${\omega}_{\ast}$ = 15 and (3) ${\omega}_{\ast}$ = 25. }
	 \label{Fig:FTF_V_strain_ratio}
	 \vspace*{00pt}
\end{figure}

Figures~\ref{Fig:FTF_V_strain_C&I} and \ref{Fig:FTF_V_strain_ratio} show the FTFs and gain difference ratios $\mathcal{G}$ of the V-flames  for different flame geometries when the flow strain is considered or not.  
In the majority of the frequency range, flow strain contributes to an increase in the FTF gain, while having negligible effect on the phase.
Similar to those for the conical flames, the effect of flow strain is mainly controlled by the factor $R$. 
For the convective velocity perturbation model, the compensation term $\mathcal{L}\hat{\mathcal{S}}$  increases with $\omega_\ast$, and the combined term ${\left(\frac{\hat{s}_L}{\sin\alpha}- \bar{s}_L  \frac{\mathrm{d} \hat{\eta}}{\mathrm{d} r} \cos\alpha\right)\left(b-r\right)}$ in the FTF definition (Eq.~\eqref{Eq:Q_ratio_non_V}) also grows along the flame front, which further amplifies the flow strain effect. 
These can be further validated for more flame geometries as shown in Fig.~\ref{Fig:FTF_V_strain_ratio}(a.1)-(a.3) at high frequencies.
For the incompressible velocity perturbation model, the effect of flow strain is controlled  by both $R$ and  $\beta$. 
Again, at high frequencies, the effect of flow strain can be restrained due to the damping effect and small convection velocity, as illustrated in Section \ref{sec: conical_strain}.

\subsection{Comparison with experiments and numerical simulations}
\label{subsec: Comparison }

\begin{figure}[!t]
	\centering
  		\subfigure
  		  		{
\includegraphics[height=5.9cm]{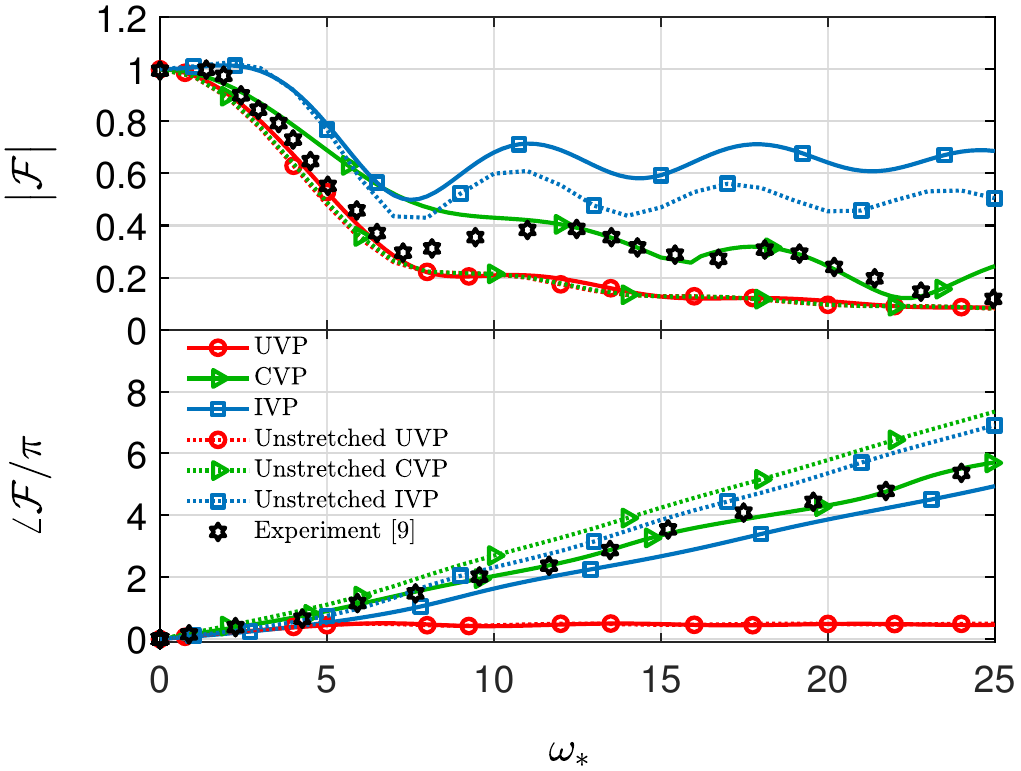}
  		}\put (-210,0) {\normalsize  $\displaystyle(a)$} 
  		\vspace*{0pt}
  		\hspace*{0pt}
  		\subfigure
  		{
\includegraphics[height=5.9cm]{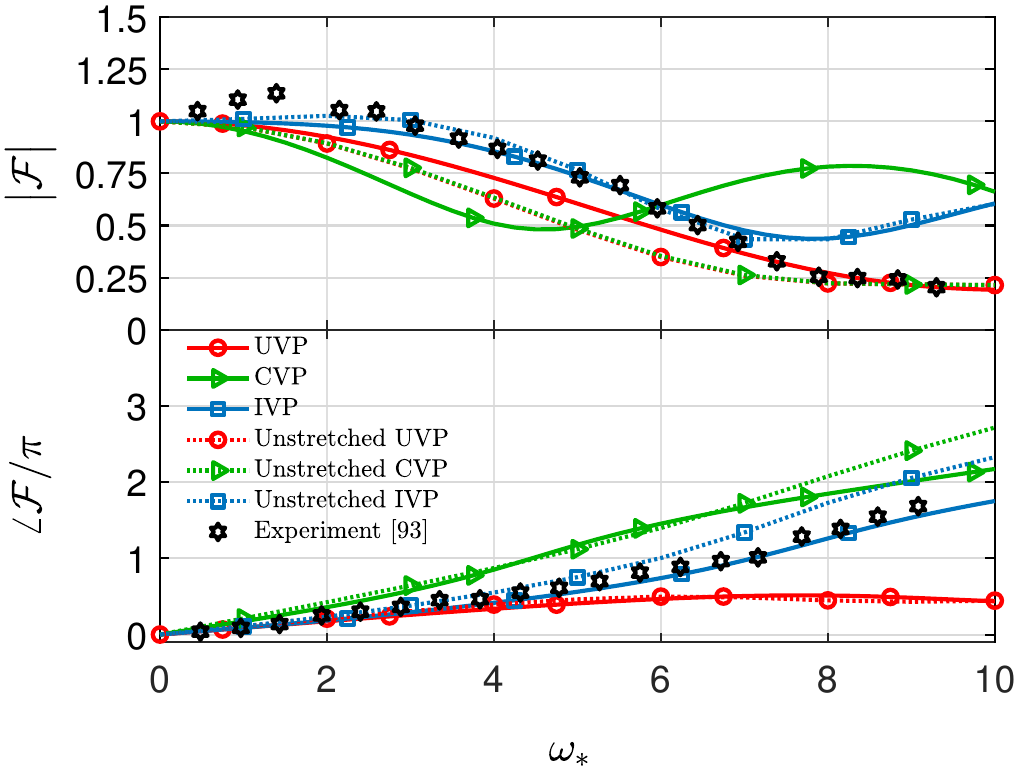}
  		}\put (-210,0) {\normalsize  $\displaystyle(b)$} 
  		\vspace*{0pt}
  		\hspace*{0pt}
 	  	 \caption{Comparisons of FTFs between experiments and $G$-equation models considering the flame stretch for the conical flames with different flame sizes. (a) $R = 11$~mm, $\beta = 6$ and (b) $R = 7$~mm, $L_f/R = 4$. In each subfigure, solid line: UVP: uniform velocity perturbation model, CVP: convective velocity perturbation model, IVP: incompressible velocity perturbation model. Dash line: analytical models without considering stretch.}
	 \label{Fig:FTF_exp}
	 \vspace*{00pt}
\end{figure}

In order to validate the proposed model, predictions from present models have been compared to experiments and high fidelity numerical simulations.
Results from the analytical models without considering the stretch are also compared.
Figure~\ref{Fig:FTF_exp} presents comparisons of different flame sizes between experimental results and the $G$-equation models considering the stretch or not.
Specifically, Figure~\ref{Fig:FTF_exp}~(a) depicts the results for $R = 11$~mm and $\beta = 6$. The hexagonal star symbols within the figure represent the experimentally measured FTF results \citep{Durox_PCI_2009}.
It can be observed that for the gain of FTF, the experimental results agree well with those obtained using the uniform velocity perturbation model at low frequencies, and the gain decreases rapidly with frequency. 
However, the gain oscillates with frequency at high frequencies, and the experimental results match well with the model using the convective velocity perturbation model considering the stretch when ${\omega}_{\ast} \textgreater 7$. 
Additionally, the FTF phase also agrees well with the result from the convective velocity perturbation model considering the  stretch.

Figure~\ref{Fig:FTF_exp}~(b) depicts the results for $R = 7$~mm, with a flame actual height-to-radius ratio of $L_f/R = 4$.
Similarly, the hexagonal star symbols represent the experimental results \citep{Gaudron_CNF_2017}.
Due to the heat transfer between the flame root and the burner rim, there is a stand-off distance between them; the experimental FTF gain thus  exhibits a slight overshoot at low frequencies.
However, since the flame root is anchored at the burner outlet in the theoretical model of this study, it is unable to adequately predict the phenomenon where this flame stand-off distance results in a gain greater than unity.
As the frequency increases, it can be observed that both the gain and  phase of the experimental results fit well with the results from the incompressible velocity perturbation model considering the stretch.
When the frequency is higher, the limitation of the larger gain using the incompressible velocity perturbation model leads to distortion in its predictions at high frequencies, and the gain of the FTF in the experiments gradually approaches that using  the uniform velocity perturbation model.

\begin{figure}[!t]
	\centering
  		\subfigure
  		{
\includegraphics[height=5.9cm]{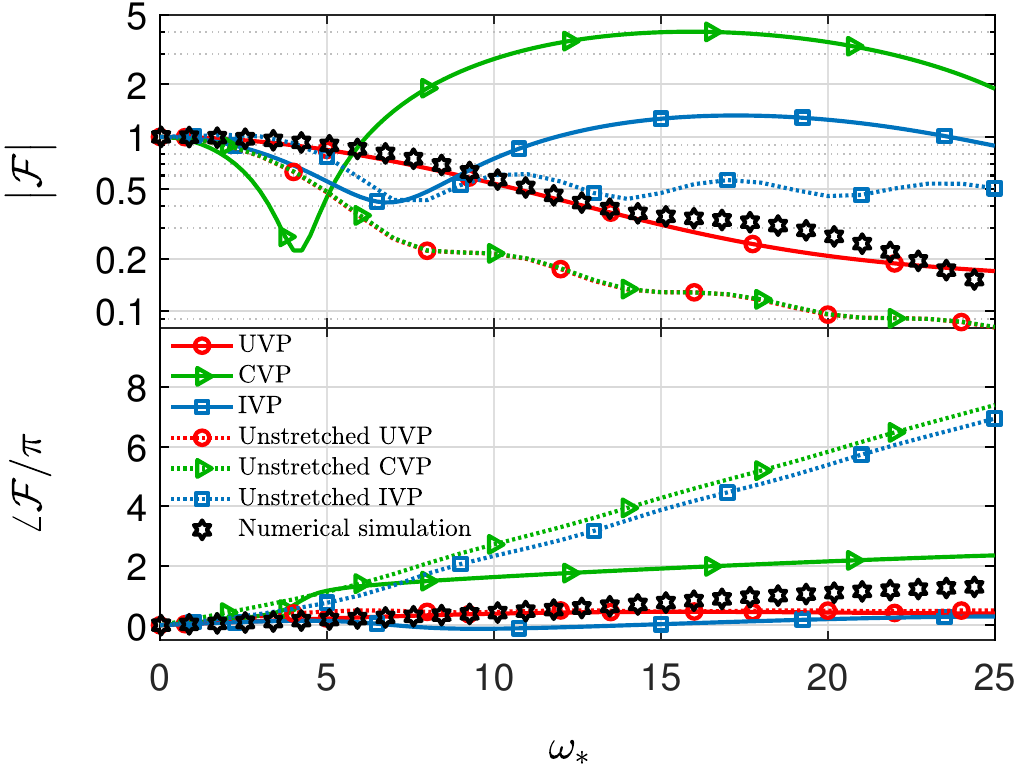}
  		}\put (-210,0) {\normalsize  $\displaystyle(a)$} 
  		\vspace*{0pt}
  		\hspace*{0pt}
  		\subfigure
  		  		{
\includegraphics[height=5.9cm]{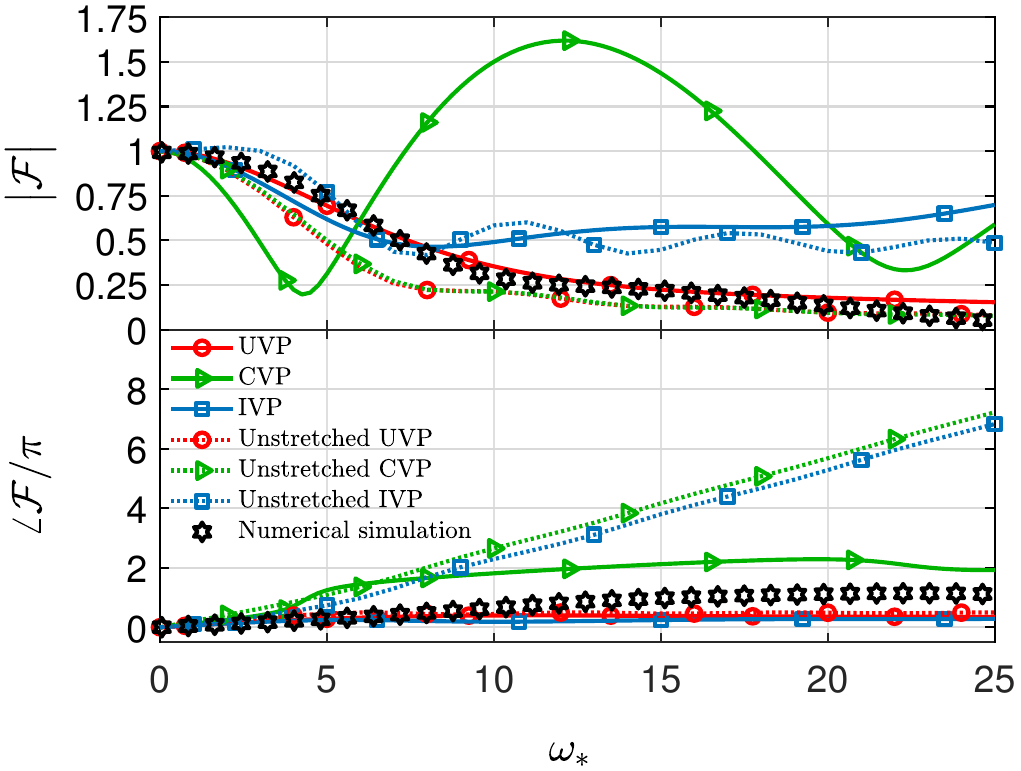}
  		}\put (-210,0) {\normalsize  $\displaystyle(b)$} \\
  		\vspace*{0pt}
  		\hspace*{0pt}
  		\caption{Comparisons of FTFs between numerical simulations and $G$-equation models considering the flame stretch for the conical flames with different flame sizes. (a) $R = 1$~mm, $\beta = 8$ and (b) $R = 2$~mm, $\beta = 4$.}
	 \label{Fig:FTF_num}
	 \vspace*{00pt}
\end{figure}

Despite the extensive experimental studies on perforated plate flames, there are limited corresponding experimental research on the single flame with  small  radii.
Therefore, Figure~\ref{Fig:FTF_num} presents comparisons between high fidelity numerical simulations and the $G$-equation models considering the stretch or not.
The two-dimensional DNS of the laminar premixed propane-air flame is conducted to numerically obtain  the FTF. System identification (SI) is employed by applying a Gaussian impulse as the oncoming flow input \citep{Polifke_ANE_2014}.
For the small size flames, corresponding to $R = 1$~mm, $\beta = 8$ and $R = 2$~mm, $\beta = 4$ in Figure~\ref{Fig:FTF_num}~(a) and (b), respectively, the prediction using the  uniform velocity perturbation model considering the stretch agrees well with the simulation results at most frequencies.
It is worth noting that the simulation results and the proposed model using the  uniform velocity model considering the stretch show larger FTF gain at low frequencies than the analytical solution model neglecting the stretch. This is primarily due to the significant influence of flame stretch on the flame shape. Therefore, the model in this study can better predict the FTF of small size flames.

In summary, when the flame size is small, the numerical simulations agree well with the results from the uniform velocity perturbation model considering the stretch. 
As the flame size increases, the FTF phase and the low-frequency gain gradually approach those obtained using the incompressible velocity perturbation model considering the stretch. 
Furthermore, for the convective velocity perturbation model considering the  stretch, good predictions are found for both  the FTF gain at higher frequencies and the FTF phase,  when the flame size becomes even larger.

\section*{Conclusions}
\label{sec:Conclusions}

This paper investigates the steady flame shape variations and the response to harmonic velocity perturbations in laminar premixed conical and V- flames considering the effects of flame stretch using the linearized $G$-equation model.
The flame speed $s_L$ in the $G$-equation is modified by introducing the flame stretch terms as the functions of both the flame curvature and the flow strain.  
It is found that even for the steady flame, the flame height significantly decreases once the stretch effect is taken into account. 
This change is more evident for small flame radius $R$ and small unstretched flame aspect ratio $\beta$.
For the perturbed flames, the present work considers three kinds of flow velocity perturbation models:  the uniform velocity perturbation model, the convective velocity perturbation model and the incompressible velocity perturbation model. 

The effects of flame stretch on the flame dynamic response of the conical and V- flames across a broad range of $\beta$ and $R$ are evaluated by comparing the  proposed  FTF models  considering the stretch effects to the analytical solutions of FTFs neglecting stretch effects,  derived in  previous research and the present work.
The FTF gain difference ratio $\mathcal{G}$ represents the influence of flame stretch on the FTF gains.
It can be partitioned into three regions based on different $\beta$ and $R$.
In the first region, where both $\beta$ and $R$ are small, $\mathcal{G}$ exhibits a large magnitude. This can be attributed to two factors: the significant reduction in the steady flame height caused by flame stretch and the flow strain, particularly when $R$ is small, which contributes to amplified FTF gains for the convective and incompressible velocity perturbation models.
As $\beta$ and $R$ gradually increase, reaching the second region, $\mathcal{G}$ oscillates periodically for the conical flame, whereas for the V-flame, $\mathcal{G}$ becomes negative. 
This behavior can be attributed to the mutual compensation between the wrinkles generated on the flame front during perturbation and the influence of flame curvature on the surface area variation.
The significant perturbations generated by the free boundary of the V-flame result in a strong wrinkle counteracting effect, leading to a decrease in the FTF gain when flame stretch is considered.
Furthermore, as the normalized frequency $\omega_\ast$ increases, the wrinkles on the flame become more pronounced. This effect plays a crucial role and influences a wider range of sizes for both conical and V- flames.
In the third region, where both $\beta$ and $R$ are large, the influence of flame curvature on surface area diminishes, and it is unable to effectively counterbalance the pronounced wrinkles induced by the flame front perturbation. Consequently, $\mathcal{G}$ exhibits a significant increase.
The FTF phase of the two kinds of flames is typically  proportional to the unstretched flame aspect ratio $\beta$. For smaller radius flames, flame stretch causes the flame height ratio $L_f/L^0_{f}$ to shrink more, resulting in shorter disturbance convection times from the burner edge to the flame tip and a further reduction in the phase lag.

The proposed models considering the flame stretch  are also compared to experiments and high fidelity  numerical simulations.
Results demonstrate that using the uniform velocity perturbation model considering the stretch can more accurately predict the FTF of small size flames, while using the incompressible and convective velocity perturbation models considering the stretch respectively agree more closely with the FTF gain of the moderate size flame at the low frequency  and the large size flame at the high frequency, as well as the phase changes.
It should be noted that the convection ratio $K$ associated with perturbation convection velocity varies for different flames and forcing frequencies. Thus, for a given velocity perturbation model, it is difficult to always find good matches between the model prediction and experimental results or numerical results.
But generally, the FTF considering the flame stretch can still enhance the agreement of the $G$-equation model approach with experiments and numerical simulations to a certain extent.

\section*{Acknowledgement}
The authors would like to gratefully acknowledge financial support from the Chinese National Natural Science Funds for National Natural Science Foundation of China (Grant no. 11927802  and U1837211). The European Research Council grant  AFIRMATIVE (2018–2023, Grant no. 772080) is also gratefully acknowledged.


\bibliographystyle{elsarticle-num}
\bibliography{ref_CNF}

\end{document}